# Finite-Dimensional ZX-Calculus
# for Loop Quantum Gravity

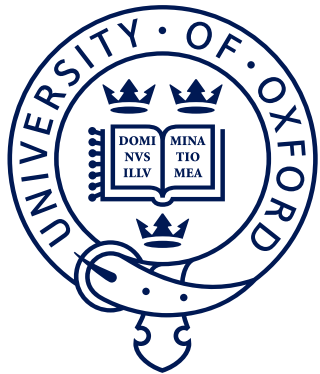

## Ben Priestley

St Hilda's College

University of Oxford

A thesis submitted for the degree of

*MSc in Advanced Computer Science*

Trinity 2025

# Acknowledgements

Upon undertaking a Master's thesis, I asked only to be interested—from start to finish—in *something or other*, so long as it was wildly different to anything I had done before. Thanks entirely to my wonderful supervisors, Razin A. Shaikh, Quanlong Wang, Richard D. P. East, and Aleks Kissinger, this request has been emphatically granted, and I do hope that my genuine excitement for researching new *things* is present in the work. My sincerest thanks goes to them, not only for introducing me to loop quantum gravity and finite-dimensional ZX-calculus (which, I suppose given the title, turned out to be quite important), but for granting me so much of their time and kindness to do so.

Also to my friends and family, and the dog above all: cheers.

# Abstract

Loop quantum gravity (LQG) attempts to unify general relativity with quantum physics to offer a complete description of the universe by quantising spacetime geometry, but the numerical calculations we encounter are extraordinarily difficult. Progress has been made in the covariant formulation of LQG, but the tools do not carry over to the canonical formulation. These tools are graphical by nature, describing space with *spin networks* to make calculations in LQG more intuitive to the human hand.

Recently, a new notation for working with spin networks has been used by Guedes et al. [1] to offer the first accurate numerical results in canonical LQG by allowing the underlying graphs to change throughout the calculation, though they are forced to concede visual intuitiveness. In this thesis, we offer a more radical rephrasing of spin network calculations by translating them into the finite-dimensional ZX-calculus [2], extending previous attempts to translate into the standard (qubit) ZX-calculus [3]. Specifically, we derive the mixed-dimensional ZX-diagrams representing the generating objects of spin networks and the rules for the forthcoming *Penrose Spin Calculus (PSC)* [4], and use these to present the ZX-form and correctness of "loop removal". We also derive the forms for several fundamental LQG objects (e.g. the Wigner $6j$-symbol) in the finite-dimensional ZX-calculus for the first time, extending Wang et al. [4] towards the domain of quantum gravity. This gives us a high-level intuitive graphical language (the ZX-calculus) that retains the flexibility of Guedes et al. [1] to handle changing graph structures, and thus we argue positions the PSC as the new definitive language for canonical LQG.

Furthermore, we investigate the possibility for a matrix-like normal form for spin networks deriving from a novel perspective in terms of W-nodes [2]. The forms we derive in this thesis are still primitive and thus limited in their scope and application, but nevertheless indicate the extraordinary potential for the PSC to offer naturally computable pathways to study LQG and spin networks.

# Contents

*If you want to learn about nature, to appreciate nature, it is necessary to understand the language that she speaks in.*

— Richard P. Feynman [5]

# 1

# Introduction

## Contents

*General relativity* is our best theory for understanding spacetime, and *quantum mechanics* provides our best understanding for matter. The former yields a good description of gravity at 'large' scales (where quantum phenomena can be safely ignored), while the latter yields an effective description of physics at 'small' scales (where gravitational forces can be safely ignored). This is fine when we are sure we only care about 'big stuff' or only about 'small stuff' at any one time, but there are some domains of our universe—those with extreme densities and/or temperatures, encountered on any worth-while trip into a black hole, or to cosmic singularities like the "Big Bang", for example—which are widely recognised to require *both* theories simultaneously [6]. Using just one somehow misses the full story, and we are yet to compile a good concoction of the two. It is in the search for an accurate (and useful) description of these domains that we find one of the driving motivators for a new theory of *quantum gravity*, satisfying the necessity for a convergence between general relativity and quantum physics [7–11]. However, quantum behaviours in gravity are yet to be observed in any recent experiments [12–14].

## 1.1 Problem and Motivation

Many proposals for a quantum theory of gravity, including the subject of this work: *loop quantum gravity (LQG)*, build from the core idea of quantising spacetime geometry [15]. In its full technical correctness, as stated by Guedes et al. [1]: it is based on a recasting of the Einstein equations in terms of holonomies in a compact gauge group and fluxes of canonically conjugate densitised triads, constructed with the so-called Ashtekar-Barbero variables [16–18]. Unfortunately, while the mathematics of LQG are pretty well-defined, they are unspeakably difficult to execute, perhaps explaining where some experimental efforts are led astray. There has been some recent progress with numerical computations [19–23] in the *covariant* formulation of the theory [see 24, for a review], which should be cause for optimism, but these numerical tools do not carry over to *canonical* LQG.

### 1.1.1 Graphical Calculi in LQG

A common basis used in LQG is spanned by eigenstates of certain geometric operators [1], namely *spin networks* [closely related to ribbon graphs and string nets; 25–27]. These objects provide a means for performing and representing the tricky calculations of LQG in a graphical manner, making them the favoured tool among researchers in the study of the quantum mechanics of angular momentum [28–30].

Guedes et al. [1] (and their accompanying letter [31]) very recently put to practice some newer notations for spin networks, deriving from Mäkinen [32], to present some of the first[1] numerical results in canonical LQG which provides a complete derivation of the action of the *Hamiltonian constraint* on spin networks—an operator defining the very dynamics of canonical LQG [34]; finding and interpreting its solutions is arguably the most important outstanding problem in the field [33]. Deriving its action has long proved to be an exceptional challenge as it changes the underlying graph and its spin assignments, yielding a large superposition of spin networks with different structures. Current graphical calculi used across

---

[1]A few months prior, Sahlmann and Sherif [33] also tackled the Hamiltonian constraint in canonical LQG, though do so via neural networks.

LQG are simply not equipped to deal with these changing graphs, leading to the need for researchers to make approximations (e.g. taking fixed graphs) that skew the accuracy of the numerics.

In an effort to remedy this issue, Guedes et al. [1] employ 'modern' conventions for $SU(2)$ recoupling theory to update older works [e.g. 35, 36] employing the outdated Temperley-Lieb algebra [30], leading to improved results. Being able to implement the action of the Hamiltonian constraint without recurring to truncation to fixed graphs is a big leap forward in our hopes for understanding the dynamics of LQG, on route to a so-called "Theory of Everything".

For all the optimism we share in Guedes et al. [1], there is a glaring issue: what this 'modern' graphical notation gains in numerical accuracy is unfortunately conceded in visual intuitiveness (as they admit). While not a full recession to cumbersome equations of symbolic mathematics, this is contrary to the mantra of a graphical language for canonical LQG.

### 1.1.2    Translation to the (Finite-Dimensional) ZX-Calculus

The *ZX-calculus* [37, 38] is a universal, sound, and complete[2] graphical calculus for qubit linear algebra [39, 40] that allows us to represent any linear map between qubits as a ZX-diagram (a type of tensor network). It has found tremendous success as a tool across the study of quantum information; e.g. in measurement-based quantum computation [41–43], topological quantum computation [44–47], quantum error correction [48], quantum circuit compilation and optimisation [49–54], etc.

The first indication of the usefulness for ZX-calculus in LQG came from East, Martin-Dussaud, and Wetering [3], which connected the ZX(H)-calculus to the representation theory of $SU(2)$ by grouping multiple qubit wires (themselves representing two-dimensional—spin-1/2—Hilbert spaces) together using a 'symmetriser' to represent higher spins, following the results of East et al. [55]. This facilitates a way to write the Yutsis diagrams of LQG [56] as ZX-diagrams, with the motivation being to provide a means to perform quantum gravity calculations directly with quantum computers [57–59].

---

[2]For a language describing some system, we have: *universal* meaning anything permissible in the system can be represented in the language; *sound* meaning everything that can be shown to be true in the language is also true in the system; and *complete* meaning everything that is true in the system can be expressed in the language.

In the years since, Wang et al. [4] have proposed[3] the *Penrose Spin Calculus (PSC)* to extend the qubit diagrams of East, Martin-Dussaud, and Wetering [3] into mixed-dimensional diagrams. This moves us from the standard ZX-calculus to the *finite-dimensional ZX-calculus* [2, 60, 61], which exponentially reduces the growth of our diagrams for LQG, and provides an intuitive means to work in the representation theory of $SU(2)$ with natively mixed-dimensional diagrams. It is the goal of this work to translate the building blocks of spin networks into the finite-dimensional ZX-calculus to recover the intuition of graphical LQG in such a way that we keep the numerical strengths that comes with the language of Mäkinen [32] and Guedes et al. [1]. In particular, we derive the relevant components and relations involved in spin networks for LQG via the PSC.

The lion's share of our efforts are spent deriving and proving the ZX-diagrammatic form for the "surgical removal" of spin network portions ("loop removal"), including the exact computation of the scalars as given by objects such as the Wigner $6j$-symbol, the $\Theta$-graph, and the tetrahedral net symbol. Proving the correctness of loop removal in the PSC enables the algorithmic approach of Guedes et al. [1] to calculate the action of the Hamiltonian constraint without the fixed-graphs approximation; i.e. we can manipulate our ZX-diagrams in the same way to handle changing graphs, and so enable their algorithm in our more intuitive language. Both the algorithmic method for loop removal, and these objects are presented in the (finite-dimensional) ZX-calculus for the first time, extending the work of Wang et al. [4]. Moreover, our work presents a higher-level language for spin networks than has previously been used for LQG [updating that of 3], giving uncontracted tensor networks, opening the door to direct diagrammatic reasoning in LQG. Indeed, abstracting the highly-specialised nature of LQG into the widely-studied ZX-calculus opens up a barrage of new techniques for studying quantum gravity, and hopefully invites new researchers to the field.

---

[3]This work is forthcoming.

To indicate the flexibility of our language (and possibly to inspire new notations for spin networks), we also propose matrix-like representations for spin networks by moving towards a novel view of the PSC in terms of W-nodes [2]. This latter work is in its earliest stages, and so has limited practical usage in its current form, but we hope that this will lead to new research with spin networks as computationally-amenable objects.

As a further remark, our work could also be of use to those generally interested in spin network calculi for purposes besides LQG; e.g. studies of non-Abelian topological error-correction codes [25–27]. There is also cause for excitement in areas like knot theory [62–64] or group theory [65], wherein we may also rely on the representation and recoupling theory of $SU(2)$.

**Problem and Motivation: Summary**

- There has been recent progress with numerical computations in the covariant formulation of LQG, but this does not carry over to the canonical formulation.

- Solving the Hamiltonian constraint—the biggest problem in canonical LQG—is difficult in present graphical languages for spin networks as the underlying graph changes. Current methods resolve this issue by assuming fixed graphs, but this leads to inaccurate numerics.

- Guedes et al. [1] employ an updated graphical language to derive the action of the Hamiltonian constraint, but admit to losing the graphical intuitiveness of the old ways to do it.

- Previously, qubit ZX-calculus has been used to provide a first attempt at LQG in the ZX-calculus, though suffers exponentially scaling diagrams. Following this, the PSC has been proposed [forethcoming in 4], which will allow us to present only polynomially-scaling diagrams for LQG.

- We use the PSC to write the necessary objects and relations, deriving the ZX-diagrammatic form for loop removal and proving its correctness. This raises us to a higher-level language, enabling direct diagrammatic reasoning and recovering intuitiveness without losing the numerical accuracy of Guedes et al. [1] (as we can handle changing graphs by the same methods, following the correctness of loop removal).

- Furthermore, we probe deeper into the use of ZX-calculus to investigate new normal forms for spin networks, using our results for loop removal and W-nodes [2] to look for a matrix-like representation.

## 1.2   Our Contributions

For clarity, we now outline the five primary contributions and novelties of this thesis.

**1. Penrose Spin Calculus Discussion.**   First and foremost, we introduce and derive many of the rules of the PSC, and explain how certain foundational objects of spin networks can be written in the finite-dimensional ZX-calculus. Our discussion reaches greater depths than in the original (forthcoming) work [4], including several new remarks and lemmas for the translation between qu*b*it and qu*d*it objects in deriving a minimal graphical form for objects such as the $3j$-symbol.

**2. The W-Node Perspective.**   Following our detailed introduction to the PSC, we demonstrate how spin network objects can be rewritten in terms of W-nodes [2], providing a novel perspective for spin networks (still within the finite-dimensional ZX-calculus). As an example of the potential utility of this notation, we demonstrate an intuitive rewrite for symmetrisers that reveals an opportunity for W-gadgets to carry the nuance of spin networks as a 'nest of W-gadgets'.

**3.  Spin Network Loop Removal.**   We perform "surgical removal" of spin network portions, as in Guedes et al. [1], in the (finite-dimensional) ZX-calculus for the first time, using the rules of the PSC, enabling the reformulation of arbitrary spin networks into loop-free normal forms. We start by deriving the ZX-diagrammatic form of 2-loop ("bubble") removal and 3-loop ("triple bubble") removal, first giving a remarkably intuitive presentation of the permutation argument for 2-loop removal, then giving an exact proof that includes the computation of the scalars. Finally, we present a novel inductive argument for the removal of $n$-loops using our results for 2- and 3-loop removal and our (novel) proof of the associativity of 3-valent spin network nodes.  In proving the correctness of these removals in the PSC, we can enable the derivation of the action of the Hamiltonian constraint via the algorithm in Guedes et al. [1] (we leave this translation itself to future work, keeping a higher-level focus in this thesis).

**4. Derivation of LQG objects in the PSC.** Within our proofs for loop removal, we further push beyond the original scope of the PSC proposal [4] to present the first mixed-dimensional ZX-diagrammatic forms for the Wigner $6j$-symbol (appearing as a scalar in 3-loop removal), the $\Theta$-graph (obtained by taking the trace of a bubble, as in Peterson [66]), and the tetrahedral net symbol (closely related to the $6j$-symbol, and again defined by a trace over a triple bubble). These objects are fundamental in LQG, apparent in their ubiquity in loop removal. Our proofs for these objects are remarkable among previous diagrammatic proofs (with lower-level languages than the ZX-calculus) for being completely self-contained—especially, they do not depend on results from Schur's lemma [67], as most other proofs do (which itself has no diagrammatic proof). Furthermore, the objects we give are high-level tensor networks, compared to the low-level scalars given in the literature.

**5. Investigating a Matrix-like Normal Form for Spin Networks.** Finally, as an extension to the loop-free normal form we obtain from our algorithmic loop-removal, we investigate the potential for additional normal forms building on our novel view of spin networks in the W-node perspective. This includes an intriguing rewrite of arbitrary spin networks as a nest of W-gadgets among a fully-connected structure between the open links of the network, which we write as an induced hypergraph. Our primitive normal forms, among their present limitations, aim to be computationally efficient by writing spin networks as dense matrices. Our hope is that, with a matrix-like representation for spin networks, a plethora of new computational research may be undertaken with spin networks, further simplifying the numerical difficulties faced in experimental tests of quantum gravity.

## 1.3 Structure of the Thesis

**Chapter 2.** Our background covers the essential $SU(2)$ representation and recoupling theory that underpins the mathematics of spin networks (section 2.1.1), then presents the foundations of the Yutsis calculus (section 2.1.2) and Penrose calculus (section 2.1.3) for graphically representing these concepts. This culminates in the definition of the Clebsch-Gordan coefficients and conditions, and their relationship with the Wigner $3jm$-symbol, which acts as the generating object of graphical calculi for spin networks. We then introduce the notation of Mäkinen [32] and Guedes et al. [1] (section 2.2), giving the new 'modern' take on a graphical calculus for spin networks in LQG. Within this discussion, we also present the bubble identity (G13) and triple bubble (G28) relations, and meet the $6j$-symbol and $\Theta$-graph. Finally, this chapter also introduces the qubit ZX-calculus (section 2.3), then presents the (relevant parts of the) finite-dimensional ZX-calculus (section 2.4) as an extension to mixed dimensions.

**Chapter 3.** This chapter presents the complete PSC, first offering a quick overview (section 3.1) that presents the generators of the calculus in table 3.1 and the complete rule set in Fig. 3.1. We then move on to a detailed derivation (section 3.2) of the generators and any rules used for proofs in this work. The discussion with our derivations are broken into 'qubit' (sections 3.2.1 and 3.2.3) and 'qudit' (sections 3.2.2 and 3.2.4) to offer greater detail in our rewrites and later present an explicit dissection of the ZX-diagram for the $3j$-symbol (section 3.2.5). The chapter closes out with a re-framing of the aforementioned discussions in terms of W-nodes (section 3.3), deriving their corresponding diagrams rigorously.

**Chapter 4.** Our most significant contributions can be found in this chapter. We begin with 2-loop removal (section 4.1), where we present the analogue to the bubble identity (G13) in the PSC and prove its correctness, then follow this with a presentation of an analogue of the $\Theta$-graph (section 4.1.2) by taking the trace of a bubble. We then define the tetrahedral net symbol and $6j$-symbol in the PSC (section 4.2), again giving a relationship between these objects with a tracing operation. Using these objects, we can then move on to 3-loop removal (section 4.3), presenting the analogue of the triple bubble (G28) relation in the PSC and prove its correctness. Finally, following a novel proof of the associativity of spin network nodes (section 4.4), we conclude the chapter with an intuitive inductive proof of $n$-loop removal (section 4.5), acting as an analogue to surgical removal of spin network portions (G27).

**Chapter 5.** This chapter offers an extension to our algorithmic method for loop removal in chapter 4 by offering a view of spin networks with W-nodes. We first show how the W-node notation can be used to form a fully-connected graph between the open links of the network (section 5.1.1), then demonstrate how the resulting W-gadget nest on this fully-connected structure can be viewed as an induced hypergraph (section 5.1.2). We then discuss the withstanding open problem following assumptions made about wire dimensions (section 5.1.3). Furthermore, this chapter also presents several examples (section 5.2) of this potential normal form for different spin network structures, hopefully indicating how such structures can be read in matrices for the first time.

**Chapter 6.** To conclude the thesis, we present some further high-level discussion and contextualise our work within LQG. This chapter also offers our thoughts for future works building on our new language, theorems, and normal forms both inside and outside the study of quantum gravity and spin networks.

*I don't like it, and I'm sorry I ever had anything to do with it.*

— Erwin Schrodinger

# 2

# Background

## Contents

## 2.1   Traditional Spin Networks

The quantum theory of angular momentum has been graphically reasoned about since the introduction of the *Yutsis calculus* [56]. The diagrams of this calculus can be embedded into the more general *Penrose graphical calculus* [68]; specifically, into the *binor calculus* [see e.g. 69]: the subcase of Penrose diagrams which are decomposed into qubits (that is, where diagrams are made up of spin-1/2 strands). It is with this binor calculus that a proposal for a fundamentally discrete model of quantum spacetime—*spin networks*—was put forward [70, 71]. Spin networks gives us a basis for the states of quantum geometry [62]; i.e. the kinematic states within the Hamiltonian study of quantum gravity, but they are also generally useful across many areas of mathematics and physics, such as knot theory [62–64] and group theory [65].

In this section, we give a brief review of the $SU(2)$ representation and recoupling theory that underlies these spin networks (section 2.1.1), in which we introduce the Clebsch-Gordan coefficients and the Wigner-$3jm$ symbol. We then present the original graphical language due to Yutsis, Levinson, and Vanagas [56] that permits the diagrammatic reasoning of $SU(2)$ recoupling theory (section 2.1.2). Finally, we draw the $3j$-symbol as a Penrose diagram (section 2.1.3), forming a foundation for the Penrose Spin Calculus (PSC) that this work later uses to reframe the manipulation of spin networks.

The structure and content of this background largely follows Major [69]—to this day, an exceptional introductory resource for learning the intuition behind spin networks, aimed at an undergraduate level. We also share an approach with East, Martin-Dussaud, and Wetering [3], from which the PSC of Wang et al. [4] is generalised, and in particular, our section 2.1.1 is in close concordance with their section 2.1 since our work has similar motivations and roots.

### 2.1.1   Review of $SU(2)$ Representation/Recoupling Theory

$SU(2)$ is the group of $2 \times 2$ unitary matrices with unit determinant. Of course, this group is central to the theory of quantum angular momentum, as we discuss in this work. But it can also be seen further afield in qubit quantum computation, or in the studies of isospins or electroweak theory.

A *unitary representation* of $SU(2)$ is a group homomorphism $SU(2) \to U(\mathcal{H})$ to the set of unitaries on a finite-dimensional Hilbert space $\mathcal{H}$. Given such a representation, we can always decompose into a direct sum of irreducible representations (irreps). For any integer dimension, there is a unique irrep (up to isomorphism), which we label by "spins" $j \in \mathbb{N}/2$ (such that the spin-$j$ irrep has dimension $2j + 1$).

**Definition 1** (Fundamental Irrep). *The fundamental irrep (spin-1/2 representation) $SU(2) \to U(\mathbb{C}^2)$ is the $2 \times 2$ matrix multiplication of $SU(2)$ over $\mathbb{C}^2$.*

Using symmetrised copies of the spin-1/2 representation, we can define arbitrary $j$-spin irreps. For this, we need the following projector.

**Definition 2** (Symmetrisation Projection). *The symmetrisation projector $\mathcal{S}_{2j}$ is defined to be the endomorphism over $(\mathbb{C}^2)^{\otimes 2j}$ such that,*

$$\mathcal{S}_{2j}(v_1 \otimes \cdots \otimes v_{2j}) = \frac{1}{(2j)!} \sum_{\sigma \in \mathfrak{G}_{2j}} U_\sigma(v_1 \otimes \cdots \otimes v_{2j}) \quad,$$

*where $\mathfrak{G}_{2j}$ is the $2j$-element permutation group, and hence $U_\sigma$ is the unitary performing permutation $\sigma \in \mathfrak{G}_{2j}$ on the given tensor product. $\mathcal{S}_{2j}$ can be seen to be both self-adjoint and idempotent, hence is a well-defined projector.*

The name is apt, since $\mathcal{S}_{2j}$ takes any $v_1 \otimes \cdots \otimes v_{2j}$ to a uniform superposition of all possible permutations of the $v_i$ such that it becomes invariant under the action of the permutation unitaries—it 'symmetrises' the given vector. $\mathcal{S}_{2j}$ thus defines a subspace $\mathcal{H}_j \subset (\mathbb{C}^2)^{\otimes 2j}$ consisting of symmetric vectors with $\dim(\mathcal{H}_j) = 2j + 1$.

With the usual canonical basis of $\mathbb{C}^2$, we can write the canonical orthonormal basis of $\mathcal{H}_j$ as,

$$|j; m\rangle := \sqrt{\frac{(2j)!}{(j+m)!(j-m)!}} \; \mathcal{S}_{2j}(\underbrace{|0\rangle \otimes \cdots \otimes |1\rangle}_{j+m \text{ times } 0}) \quad,$$

with integer $m \in [-j, j]$. Here we are using the physicist's notation to characterise a single spin state by its intrinsic angular momentum $j$ (denoting that we are in $\mathcal{H}_j$) and its azimuthal component $m$ (labelling the actual basis vector).

Since any finite-dimensional representation of $SU(2)$ can be decomposed as a direct sum of irreps, it follows that any tensor product of irreps can be written as a direct sum of irreps. This implies the existence of a bijective equivariant map—an *intertwiner*—from the tensor product to a direct sum of irreps.

**Definition 3** (Intertwiner)**.** *Given two representations $V, W$ of a group $G$, an intertwiner is a linear map $\iota : V \to W$ which commutes with the group action; i.e. for all $g \in G$ and $v \in V$, we have that $\iota(g \cdot v) = g \cdot \iota(v)$.*

*The set of intertwiners forms a vector space $\mathrm{Hom}_G(V, W)$ that is isomorphic to $\mathrm{Inv}_G(V \otimes W^*)$, where $W^*$ is the dual vector space of $W$ with the dual representation.*

The most fundamental intertwiner, which is of particular importance as we move on to constructing spin networks, is the *Clebsch-Gordan intertwiner*. Given the Hilbert spaces $\mathcal{H}_{j_1}$ and $\mathcal{H}_{j_2}$, we can decompose their tensor product $\mathcal{H}_{j_1} \otimes \mathcal{H}_{j_2}$ into a direct sum of irreps with the following equivalence of representations,

$$C : \mathcal{H}_{j_1} \otimes \mathcal{H}_{j_2} \to \bigoplus_{j=|j_1-j_2|}^{j_1+j_2} \mathcal{H}_j \quad .$$

Specifically, for any state $|j; m\rangle \in \mathcal{H}_j$, we have that,

$$|j; m\rangle = \sum_{m_1=-j_1}^{j_1} \sum_{m_2=-j_2}^{j_2} |j_1; m_1. j_2; m_2\rangle \langle j_1; m_1. j_2; m_2 \mid j; m\rangle$$

$$= \sum_{m_1=-j_1}^{j_1} \sum_{m_2=-j_2}^{j_2} C^{jm}_{j_1 m_1 j_2 m_2} |j_1; m_1. j_2; m_2\rangle$$

where $C^{jm}_{j_1 m_1 j_2 m_2}$ are called the *Clebsch-Gordan coefficients*, and are defined by,

$$C^{jm}_{j_1 m_1 j_2 m_2} := \langle jm | C | j_1 m_1; j_2 m_2\rangle \in \mathbb{R} \quad ,$$

If $m \neq m_1 + m_2$, then $C^{jm}_{j_1 m_1 j_2 m_2} = 0$. Furthermore, we set $C^{jm}_{j_1 m_1 j_2 m_2} = 0$ whenever the following *Clebsch-Gordan conditions* are not satisfied:

$$j_1 + j_2 + j \in \mathbb{N} \qquad \text{and} \qquad |j_1 - j_2| \leq j \leq j_1 + j_2 \quad . \tag{CG}$$

This has an implicit asymmetry that may not be preferable for setting up a graphical calculus, so we can instead give a description of a single state composed of three Hilbert spaces (without privileging any one of them).

**Definition 4** (Wigner $3jm$-Symbol). *Suppose the Clebsch-Gordan coefficients are satisfied (and write $j = j_3$). Then,*

$$\dim\left(\mathrm{Inv}_{SU(2)}(\mathcal{H}_{j_1} \otimes \mathcal{H}_{j_2} \otimes \mathcal{H}_{j_3})\right) = 1 \ .$$

*Hence, there is a unique unit vector in this subspace, up to phase, which we can write as,*

$$|j_1, j_2, j_3\rangle = \sum_{m_1, m_2, m_3} \begin{pmatrix} j_1 & j_2 & j_3 \\ m_1 & m_2 & m_3 \end{pmatrix} |j_1 m_1; j_2 m_2; j_3 m_3\rangle \quad ,$$

*where*

$$\begin{pmatrix} j_1 & j_2 & j_3 \\ m_1 & m_2 & m_3 \end{pmatrix} := \frac{(-1)^{j_1 - j_2 - m_3}}{\sqrt{2j_3 + 1}} C_{j_1 m_1 j_2 m_2}^{j_3, -m_3} \quad .$$

*This collection of coefficients is called the Wigner $3jm$-symbol (or, for brevity, just the "3j-symbol").*

## 2.1.2 Original Conventions for Yutsis Diagrams

Now we can see the graphical calculus introduced in Yutsis, Levinson, and Vanagas [56] for the recoupling theory of $SU(2)$. In this section, we use the original conventions, and later in section 2.2, we will explore the current 'modern' conventions.

**Definition 5** (Yutsis 3-Valent Node). *The basic generator of Yutsis diagrams is the 3-valent node, which represents the $3j$-symbol,*

$$\begin{pmatrix} j_1 & j_2 & j_3 \\ m_1 & m_2 & m_3 \end{pmatrix} \quad = \quad$$  $$\quad = \quad$$  $$\quad ,$$

*where the sign here denotes the cyclicity of the node; clockwise ($-$) or anti-clockwise ($+$). We can omit the sign by prescribing clockwise as the default cyclicity.*

3-Valent nodes are well-behaved under topological deformations; e.g.

 $$\quad = \quad$$  $$\quad = \quad$$  $$\quad .$$

The orientation indicated by the arrows drawn on each link corresponds to a sign on the magnetic index; flipping the sign reverses the orientation,

$$
\begin{pmatrix} j_1 & j_2 & j_3 \\ -m_1 & m_2 & m_3 \end{pmatrix} \quad = \quad
$$

An important special case—so important, in fact, that we may prefer to have set as a generator—is a 3j-symbol where one of the wires has spin 0 (denoted below by a dotted line),

$$
\quad = \quad \frac{(-1)^{j_1+m_1}}{\sqrt{2j_1+1}} \delta_{m_1,-m_3} \delta_{j_1,j_3} \quad .
$$

The importance of this special case will become clear when we arrive at the PSC in chapter 3, where it will translate into the cup that gives us the ability to bend input states into output states and vice versa. This then lets us be far more deliberate in how we go about manipulating diagrams through having the freedom to bend legs around as we see fit. For instance, we can note that the symmetric Wigner $3jm$-symbol (as above) can be bent into the asymmetric design of the Clebsch-Gordan intertwiner using this special case:

$$
\quad = \quad \quad ,
$$

but we are getting ahead of ourselves. To appreciate this, we need to understand how to perform operations on Yutsis diagrams (e.g. the above is connecting the wires of two 3j-symbols).

Lastly, but not least-ly, it is convenient to also consider a single wire as a generating element of the calculus:

$$
\quad = \quad (-1)^{j_2-m_2} \delta_{j_1,j_2} \delta_{m_1,m_2} \quad \text{and} \quad \quad = \quad (-1)^{j-n} \delta_{m,n} \quad ,
$$

where the phase serves to guarantee that the composition of two singles wires is also a single wire.

## Operations on diagrams

Graphically speaking, multiplication is implemented by juxtaposing diagrams (no operator symbols are necessary here),

$$
\vcenter{\hbox{}} \quad = \quad \begin{pmatrix} j_1 & j_2 & j_3 \\ m_1 & m_2 & m_3 \end{pmatrix} \begin{pmatrix} j_4 & j_5 & j_6 \\ m_4 & m_5 & m_6 \end{pmatrix} \quad,
$$

and summation by connecting wires labelled by the same spin (summing over the magnetic index $m$ on the corresponding free ends of the joining links),

$$
\vcenter{\hbox{}} \quad = \quad \sum_{m=-j}^{j} (-1)^{j-m} \vcenter{\hbox{}} \quad .
$$

## Closed diagrams as invariant functions

Before we move on to translate these into Penrose diagrams, we should note that not all diagrams have open (or "external") links. All of the above diagrams have at least one open link, and hence represent tensors with a number of indices equal to the number of open links. However, this may not be the case, and whenever we have a digram with no open links, it encodes a number, since all magnetic indices are summed over leaving only a function of the internal spins. We call such an expression an *invariant function*.

For example, take the trace of a single wire by connecting the ends of the identity wire to each other. This encodes its dimension,

$$
\vcenter{\hbox{}}^{\,j} \quad = \quad \sum_m (-1)^{j-m} \,\,{}_{n}^{m}\!\!\bigg|_j \quad = \quad \sum_m (-1)^{j-m}(-1)^{j-m} \quad = \quad 2j+1 \quad .
$$

A prominent example that will crop up when we come to perform loop extraction in chapter 4 is the following $6j$-symbol:

$$
\begin{Bmatrix} j_1 & j_2 & j_3 \\ j_4 & j_5 & j_6 \end{Bmatrix} \quad = \quad \vcenter{\hbox{}} \quad .
$$

Of course, the story of Wigner symbols goes on like a Tolstoy novel for more and more internal spins. The above is all we will need explicitly for the derivations in this work, but the ideas all hold for other invariant functions.

**What does it look like as an explicit expression?**

Perhaps by this point you are asking, *why do we actually want a graphical calculus; can't we just write all this out in equations?* It is a reasonable question, and yes, you could just write these things out as explicit expressions. In fact, let us go ahead and do that for just one coefficient of the $3j$-symbol:

$$
\left( \begin{smallmatrix} j_1 & j_2 & j_3 \\ m_1 & m_2 & m_3 \end{smallmatrix} \right) \equiv
$$

$$
\delta_{m_1+m_2+m_3}(-1)^{j_1-j_2-m_3}\sqrt{\frac{(j_1+j_2-j_3)!\,(j_1-j_2+j_3)!\,(-j_1+j_2+j_3)!}{(j_1+j_2+j_3+1)!}}\times
$$

$$
\sqrt{(j_1-m_1)!\,(j_1+m_1)!\,(j_2-m_2)!\,(j_2+m_2)!\,(j_3-m_3)!\,(j_3+m_3)!}\times
$$

$$
\sum_{k=K}^{N}\frac{(-1)^k}{k!\,(j_1+j_2-j_3-k)!\,(j_1-m_1-k)!\,(j_2+m_2-k)!\,(j_3-j_2+m_1+k)!\,(j_3-j_1-m_2+k)!}\;.
$$

Oh dear, it seems that the simplest of generators translates as several lines of mathematical sick. It barely fits on the page. It should make you even more nauseous to contemplate that the $6j$-symbol above is a 6-fold summation over products of four $3j$-symbols.

The graphical approach allows us to abstract this difficult-to-follow detail away and focus only on what is important. We should never have to resort to vomiting out the explicit expression (at least, not in full). Indeed, it is a primary motivation of this work to express spin networks in as clean and intuitive a diagrammatic language as we can muster.

### 2.1.3  Into Penrose Diagrams

**Definition 6** (Penrose Identity). *The identity over $\mathbb{C}^2$ is an undirected wire,*

$$
\Big|\quad \cong \quad |0\rangle\langle 0| + |1\rangle\langle 1|\quad,
$$

*with the open ends carrying implicit labels of copies of $\mathbb{C}^2$.*

**Definition 7** (Penrose Cap/Cup). *The cap and cup respectively are given by,*

$$
\frown\quad \cong \quad i\langle 01| - i\langle 10|\qquad and \qquad \smile\quad \cong \quad i|01\rangle - i|10\rangle\quad.
$$

The imaginary phase imparted on each of the cap and cup is necessary to obtain well-behaved diagrams under topological deformation,

$$\bigcap \bigcup \;=\; \Big| \quad \text{and} \quad \overgroup{\bigcap\bigcap} \;=\; \bigcap \quad.$$

The RHS to both of these relations would pick up a negative phase without the global $i$ in Def. 7. For a similar reason, we are forced to associate, with each crossing of links (a "swap"), a further negative phase factor.

**Definition 8** (Penrose Swap)**.** *The swap is the standard definition, with a global negative phase,*

$$\bigtimes \;=\; -|00\rangle\langle00| - |10\rangle\langle01| - |01\rangle\langle10| - |11\rangle\langle11| \quad.$$

In addition to the above, we now also have that the following relation holds,

$$\bigotimes \;=\; \bigcap \quad,$$

which encompasses all we need for our calculus to have *planar isotopy*; that is, any diagram may be continuously deformed in the plane while preserving its interpretation as a linear map.

With these definitions in place, we can immediately see the two fundamental equations of the binor calculus. First, we can yield the dimension of the tensor calculus [negative, in this case 68] with a loop,

$$\bigcirc \;=\; -2 \quad.$$

Second, we have the so-called "*binor identity*" (or "*skein relation*"),

$$\Big) \Big( \;+\; \bigtimes \;+\; \smile \;=\; 0 \quad.$$

Now we can make clear the embedding of Yutsis diagrams into Penrose diagrams. The spin-$j$ irrep is composed by (anti-)symmetrising strands as,

$$\left|\overset{2j}{\cdots}\right| \;:=\; \frac{1}{(2j)!} \sum_{\sigma\in\mathfrak{S}_{2j}} (-1)^{|\sigma|} \;\boxed{\overset{\cdots}{\underset{\sigma}{\phantom{x}}}} \quad,$$

where $|\sigma|$ is the parity of the permutation $\sigma \in \mathfrak{G}_{2j}$, and the $\sigma$-box represents the corresponding permutation of the $2j$ strands. Due to the global negation in Def. 8, this operator is a projector from $\mathbb{C}^2 \otimes \cdots \otimes \mathbb{C}^2$ to $\mathcal{S}_{2j}(\mathbb{C}^2 \otimes \cdots \otimes \mathbb{C}^2)$. Hence,

$$
\left|\begin{array}{c} \overset{2j}{\cdots} \\ \underbrace{0 \ldots 1}_{j+m \text{ times } 0} \end{array}\right|
\quad = \quad
\mathcal{S}_{2j}(\underbrace{|0\rangle \otimes \cdots \otimes |1\rangle}_{j+m \text{ times } 0})
\quad = \quad
\sqrt{\frac{(j-m)!(j+m)!}{(2j)!}} \; |j; m\rangle \quad .
$$

Together with the important invariance property of the cup under the action of $SU(2)$; i.e. for any $u \in SU(2)$,

$$
\boxed{u}\;\boxed{u} \quad \cong \quad i \; u \cdot (|01\rangle - |10\rangle) \quad = \quad i \; (|01\rangle - \|10\rangle) \quad \cong \quad \bigcup \quad ,
$$

we can draw the (analogue of) the $3j$-symbol in the binor calculus: the diagram,

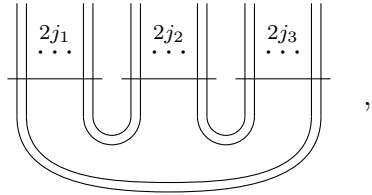

depicts a vector in the invariant subspace $\mathrm{Inv}_{SU(2)}(\mathcal{H}_{j_1} \otimes \mathcal{H}_{j_2} \otimes \mathcal{H}_{j_3})$, and is thus proportional to $|j_1, j_2, j_3\rangle$. Since we are spaces that are symmetrised, we only care about how many links connect any two $2j_i$ bundles. And, in fact, there is a unique way to connect the links (when it is possible), and the Clebsch-Gordan coefficients precisely specify when such a recoupling is possible.

It is with these objects that we can build up arbitrary spin networks.

**Definition 9** (3-Valent Spin Network). *A 3-valent spin network is an open graph wherein each node has exactly 3 associated links and wherein each link is associated with a spin $j_i \in \mathbb{N}/2$ such that the three spins involved in any given node satisfy the Clebsch-Gordan coefficients.*

The definition easily generalises to higher valencies, but this work will restrict to 3-valent spin networks only. Indeed, in a crude way, a 4-valent spin network secretly masquerades as superpositions of 3-valent spin networks, and similarly moving up the valencies inductively.

## 2.2  'Modern' $SU(2)$ Recoupling Theory

The now 'old-fashioned' Temperley–Lieb algebra [30, 72] can been seen prominently across much of the early LQG literature [35, 36, 73, 74], and it does have a certain graphical intuitiveness. Temperley-Lieb tangles [closely related to knots 75] are only proportional to spin networks [74], and the conversion factors between them are particularly complex. This then leads to the need for normalisations of states, 3j-symbols, and so on (e.g. 6j- and 9j-symbols that also commonly appear), which makes the numerics involved in calculations on even simple spin networks inescapably difficult.

Recently, Guedes et al. [1] put to use a more 'modern' convention for $SU(2)$ recoupling theory, following from Mäkinen [32]. This is put forward as a more favourable representation for robust numerical accuracy, at the self-admitted cost to graphical intuition (which this work hopes to rectify via the PSC). In this section, we introduce their notation, recovering several definitions from the Yutsis calculus in the up-to-date graphical language, and present useful rewrite rules for spin networks (introduced in section 2.2.3, then used for rewrites in section 2.2.4) that will be the subject of our translations in chapter 4. Equations in this section are labelled as (GX) to correspond to equations labelled as (X) in Guedes et al. [1] to smoothen our later discussions—names for graphical relations are not the most consistent in the literature.

### 2.2.1  $SU(2)$ Elements as Graphical Links

**Definition 10** (Identity)**.** *The identity is written as an undirected wire labelled with a spin $j$,*

$$\delta_{m,n}^{(j)} \quad = \quad m \; \underrightarrow{\hspace{0.6cm} j \hspace{0.6cm}} \; n \quad , \tag{G2}$$

*where each end is carrying one of the $d_j := 2j + 1$ possible indices (half-integer steps from $-j$ to $j$, including $0$) of the identity matrix.*

As before, the contraction of two matrices looks, graphically, like joining the wires. By this, we imply a summation over the indices at the corresponding connecting ends.

**Definition 11** (*j*-Representational Tensor)**.** *The j-representational tensor $\epsilon_{m,n}^{(j)} = (-1)^{j-m}\delta_{m,-n}^{(j)} = (-1)^{2j}\epsilon_{n,m}^{(j)}$ is written as a directed wire (m to n), with $m, n, j$ playing the same roles as in Def. 10,*

$$\epsilon_{m,n}^{(j)} \quad = \quad m \xrightarrow{\quad j \quad} n \quad .$$ (G3)

This tensor of Def. 11 satisfies the following relations:

$$\epsilon_{m,n}^{(j)}\epsilon_{n,k}^{(j)} \quad = \quad m \xrightarrow{\quad j \quad\quad} k \quad = \quad (-1)^{2j}\delta_{m,k}^{(j)} \quad ,$$ (G4)

$$\epsilon_{m,n}^{(j)}\epsilon_{k,n}^{(j)} \quad = \quad m \xrightarrow{\quad j \quad}\!\!\xleftarrow{\quad\quad} k \quad = \quad \delta_{m,k}^{(j)} \quad .$$ (G5)

**Definition 12** (Wigner Matrix)**.** *Given an element g of the SU(2) group, the Wigner matrix $D_{m,n}^{(j)}(g)$ for g is graphically written as a wire directed (m to n) by a g-triangle,*

$$D_{m,n}^{(j)}(g) \quad = \quad m \xrightarrow{\quad j \quad}\!\!\triangleright_{\!g}\!\!\longrightarrow n \quad ,$$ (G6)

We can now observe that $\epsilon_{m,n}^{(j)}$ is invariant under $SU(2)$ transformations; i.e. given a Wigner matrix $D_{m,n}^{(j)}(g)$ for $g \in SU(2)$, we have that,

$$D_{m,n}^{(j)}(g)\epsilon_{n,p}^{(j)}D_{q,p}^{(j)}(g) = \epsilon_{m,q}^{(j)} = D_{n,m}^{(j)}(g)\epsilon_{n,p}^{(j)}D_{p,q}^{(j)}(g) \quad ,$$

which, graphically, takes the form,

$$m \xrightarrow{\ g\ }\!\!\triangleright\!\xrightarrow{\quad j \quad}\!\triangleleft\!\xleftarrow{\ g\ } q \quad = \quad m \xrightarrow{\quad j \quad} q \quad = \quad m \xleftarrow{\ g\ }\!\triangleleft\!\xrightarrow{\quad j \quad}\!\triangleright\!\xleftarrow{\ g\ } q \quad . $$ (G7)

## 2.2.2   Coupling Links into Nodes

Now that we have basic relations for representing elements of $SU(2)$ as *links*, we graduate to *nodes* by coupling several links according to the $SU(2)$ decomposition rule into irreps. Following this decomposition enforces the Clebsch-Gordan coefficients on the spins meeting at a node (see section 2.1.1).

The smallest valency with a non-trivial character is three coupled spins, which mathematically presents itself as the Wigner $3jm$-symbol of def,

$$\begin{pmatrix} j_1 & j_2 & j_3 \\ m_1 & m_2 & m_3 \end{pmatrix} \quad = \quad \underset{j_2}{\overset{j_1}{\bigvee_{\!\!+\!\!}}}_{\,j_3} \quad = \quad \underset{j_3}{\overset{j_1}{\bigvee_{\!\!-\!\!}}}_{\,j_2} \quad ,$$ (G9)

where the sign denotes the cyclicity; counter-clockwise $(+)$ or clockwise $(-)$. The matrix-like form of the $3j$-symbol presents an upper row of spins and a lower row of associated spin projections (a.k.a. "magnetic numbers"; i.e. matrix components $m_i$ in a given irrep $j_i$).

Using a technique often referred to as *"braiding"*, we can introduce a phase factor $(-1)^{j_1+j_2+j_3}$ to flip the cyclicity of a node[1].

Similarly to (G7), the $3j$-symbol is invariant under $SU(2)$ transformations;

$$\tag{G10}$$

We can also note the following correspondence between the $3j$-symbol and the tensor $\epsilon_{m,n}^{(j)}$,

$$\tag{G11}$$

### 2.2.3 Important Graphical Relations

The overwhelming majority of the spin network simplifications we can make at any time either directly use, or derive from, the following two relations (respectively, sometimes called the "Θ-*graph*" and the "*bubble identity*").

**Definition 13** (Θ-Graph).

$$\tag{G12}$$

**Definition 14** (Bubble Identity).

$$\tag{G13}$$

---

[1]Notice that, because we restrict $j_1 + j_2 + j_3 \in \mathbb{N}$ (to fulfil the Clebsch-Gordan conditions), a double braid leaves the node invariant, since the introduced phase cancels as $(-1)^{2(j_1+j_2+j_3)} = 1$.

Revealing opportunities for simplifications using the above relations sometimes takes a little creative license. For example, any two links (of arbitrary spin) can be coupled making use of the following sum,

$$
j_1 \Big| \ \Big| j_2 \quad = \quad \sum_j d_j \ \vcenter{\hbox{}} \quad , \tag{G14}
$$

which runs, in principle, over $j \in \mathbb{N}/2$, but in effect, from $|j_1 - j_2|$ to $j_1 + j_2$, by implication of the Clebsch-Gordan conditions that $j, j_1, j_2$ are forced to satisfy amongst each other. This kind of manipulation is facilitated by,

$$
\vcenter{\hbox{}} \quad = \quad \sum_k d_k (-1)^{j_2+j_3+k+\ell} \begin{Bmatrix} j_1 & j_2 & k \\ j_4 & j_3 & \ell \end{Bmatrix} \vcenter{\hbox{}} \quad , \tag{G17}
$$

making use the 6j-symbol, written in curly braces above (recall section 2.1.2).

### 2.2.4 "Surgical Removal" of Spin Network Portions

In addition to the bubble identity of (G13), there is a possibility to "surgically remove" more substantial portions of a spin network, which can then be contracted in some careful way to form disjoint, closed networks—secondary to the main network—that give invariant functions. This section discusses the pieces that puzzle together to permit this removal [discussion summarises section V of 1], which will be translated to the PSC in this work.

Recall that intertwiners (Def. 3) are invariant tensors of the $SU(2)$ group. Given a collection of representations $j_i$ acting on Hilbert spaces $\mathcal{H}_{j_i}$, the intertwiners are the elements of the space of spin singlets $\mathrm{Inv}_{SU(2)}(\bigotimes_i \mathcal{H}_{j_i})$.

We have already met the trivial intertwiner—the tensor $\epsilon_{n,q}^{(j)}$—which is the basis element of the 1-dimensional space $\mathrm{Inv}_{SU(2)}(\mathcal{H}_j \otimes \mathcal{H}_j)$ for a given $j$. Again as we have seen in the discussion surrounding Def. 4, the foundational (non-trivial) intertwiner is the 3j-symbol, and this is the sole basis element of the space $\mathrm{Inv}_{SU(2)}(\mathcal{H}_{j_1} \otimes \mathcal{H}_{j_2} \otimes \mathcal{H}_{j_3})$ for given $j_1, j_2, j_3$.

It turns out that this is all we need; 2- and 3-valent intertwiners can be suitably contracted[2] to build up intertwiners of arbitrary valency. The corresponding space is $\mathrm{Inv}_{SU(2)}(\bigotimes_i \mathcal{H}_{j_i})$, and it may generally have many basis elements corresponding to all possible choices of inner-link spins. The identity of such spaces can then be resolved as the weighted sum over all inner spins of the non-contracted double $n$-valent intertwiners:

$$\mathbb{I} \;=\; \sum_{\{i\}} \left( \sum_{\ell=1}^{n-3} d_{i_\ell} \right) \quad \text{,} \qquad \text{(G27)}$$

where open links drawn in opposing directions, in the above, are denoting bras from kets. Introducing the intertwiner-space identity of (G27) graphically accounts for "breaking" links in a spin network, in an idea Guedes et al. [1] likened to lattice surgery [26, 76, 77].

Using the resolution of the identity, the "surgical removal" becomes possible; for example, another useful example is that which generalises the bubble identity from a 2-loop to a 3-loop,

$$ \qquad = \qquad \qquad . \qquad \text{(G28)}$$

---

## 2.3   (Qubit) ZX-Calculus

The ZX-calculus is a graphical language comprising ZX-diagrams and a set of rewrite rules to transform between them [37, 38], giving us a tool with which to diagrammatically reason about linear maps [78]. ZX-diagrams are themselves built up from (or 'generated by') *spiders*, representing particular linear maps, composing to yield a network of spiders represents a tensor network.

Certain rule sets for the ZX-calculus are *complete*; that is, if two diagrams represent equivalent linear maps, then there always exists a sequence of diagrammatic rewrites to transform one into the other [39, 79, 80]. Hence, anything we could want to reason about quantum computation can be done entirely within the ZX-calculus.

### 2.3.1   Spiders as Tensors

The standard ZX-calculus is entirely generated by both the Z- and X-spider, each of which can have an arbitrary number of input wires and output wires[3].

**Definition 15** (Z-Spider). *The Z-spider is a linear map given by, under the standard interpretation $[\![\cdot]\!]$,*

$$\overset{\overbrace{\vdots}^{m}}{\underset{\underbrace{\vdots}_{n}}{\alpha}} \quad \overset{[\![\cdot]\!]}{\longmapsto} \quad |0,\overset{n}{\cdots},0\rangle\langle0,\overset{m}{\cdots},0| + e^{i\alpha}|1,\overset{n}{\cdots},1\rangle\langle1,\overset{m}{\cdots},1| \quad,$$

*with a phase $\alpha \in \mathbb{R}$ (although, by the periodicity of $e^{i\alpha}$, we can take $\alpha \mod 2\pi$).*

It is extremely common to see $\alpha = 0$, since this gives $e^{i\alpha} = 1$ to input and output the same computational basis state on each wire, so we denote this by an unlabelled spider,

$$\overset{\cdots}{\underset{\cdots}{\bullet}} \quad := \quad \overset{\cdots}{\underset{\cdots}{0}} \quad.$$

The Z-spider here has been defined with respect to the eigenbasis of the Z-gate ($|0\rangle$ and $|1\rangle$), hence the name; similarly we can define the X-spider with respect to the eigenbasis of the X-gate ($|+\rangle$ and $|-\rangle$).

---

[3]All diagrams are drawn top to bottom in this work, to better align with the physicists' Yutsis and Penrose calculi (sections 2.1.2 and 2.1.3).

**Definition 16** (X-Spider)**.** *The X-spider is a linear map given by*

$$\overset{\overbrace{\cdots}^{m}}{\underset{\underset{n}{\cdots}}{\alpha}} \quad \overset{[\![\cdot]\!]}{\mapsto} \quad |+,\cdot\overset{n}{\cdots}\cdot,+\rangle\langle+,\cdot\overset{m}{\cdots}\cdot,+| + e^{i\alpha}|-,\cdot\overset{n}{\cdots}\cdot,-\rangle\langle-,\cdot\overset{m}{\cdots}\cdot,-| \quad,$$

*again with a phase $\alpha \in \mathbb{R}$ taken mod $2\pi$.*

Similarly, a phase-free X-spider omits the label,

$$\overset{\cdots}{\underset{\cdots}{\bullet}} \quad := \quad \overset{\cdots}{\underset{\cdots}{0}} \quad .$$

Some interesting forms of spider are those with only a single leg, corresponding to Fourier/computational basis states/costates, up to scalar,

$$\overset{\circ}{|} = \sqrt{2}\,|+\rangle \qquad \underset{\circ}{|} = \sqrt{2}\,\langle+| \qquad \overset{\pi}{|} = \sqrt{2}\,|-\rangle \qquad \underset{\pi}{|} = \sqrt{2}\,\langle-|$$

$$\overset{\bullet}{|} = \sqrt{2}\,|0\rangle \qquad \underset{\bullet}{|} = \sqrt{2}\,\langle0| \qquad \overset{\pi}{|} = \sqrt{2}\,|1\rangle \qquad \underset{\pi}{|} = \sqrt{2}\,\langle1| \quad .$$

## 2.3.2 Compositions of Spiders as Tensor Networks

Spiders can be composed in parallel or in sequence to build up a structure representing a tensor network.

*Parallel composition* ($\otimes$) is given by a tensor product of the corresponding linear maps, graphically drawn as juxtaposed diagrams,

$$\overset{\curlywedge}{} \quad \otimes \quad \overset{\curlyvee}{} \quad = \quad \overset{\curlywedge}{}\,\overset{\curlyvee}{} \quad .$$

More interestingly, *sequential composition* ($\circ$) is given by multiplication of the corresponding matrices (implying, for one thing, that the inputs of one line up with the outputs of the other), graphically drawn by connecting pairs of wires,

$$\overset{\curlyvee}{} \quad \circ \quad \overset{\curlywedge}{} \quad = \quad \overset{\bigcirc}{} \quad .$$

The ZX-calculus has some pleasing topological behaviour: *only connectivity matters*. Diagrams can be deformed on the plane without changing the linear map they represent, so long as the connectivity between the spiders is not changed; e.g. the diagrammatic representation of the CNOT-gate can be written in any number of equivalent ways,

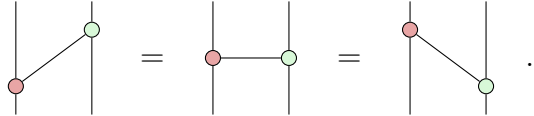

### 2.3.3   Closed Diagrams as Scalars

Diagrams with no inputs or outputs, corresponding to contracting all the indices of the represented tensor network, are scalars. This, on its own, is a powerful idea; any complex number can be graphically presented using only the following closed diagrams,

$$
\bigcirc = 2 \qquad\qquad \alpha\!-\!\!\bigcirc = \sqrt{2}
$$

$$
\pi = 0 \qquad\qquad \alpha\!-\!\pi = e^{i\alpha}
$$

$$
\alpha = 1 + e^{i\alpha} \qquad\qquad \bigcirc\!\!\!\!=\!\!\!\!\bullet = \frac{1}{\sqrt{2}} \quad .
$$

It can often be inconvenient to collect lots of closed sub-diagrams—representing global phase factors—within a larger diagram when rewriting. We simply do not need the precision in many cases! So, it is not uncommon to ignore scalars. Although, we should be careful to check whether the scalars we are ignoring are not actually zero, since this would obviously change the character of our diagram beyond global phase.

### 2.3.4   Rules of the ZX-Calculus

To simplify the rules presented here, denote the Hadamard-gate by,

$$
\begin{array}{c}\blacksquare\end{array} \overset{\llbracket\cdot\rrbracket}{\longmapsto} \frac{1}{\sqrt{2}}\begin{pmatrix}1 & 1\\ 1 & -1\end{pmatrix} \quad .
$$

This H-box is not strictly a generator, since it can be decomposed into Z- and X-spiders as follows,

$$ \square \quad \propto \quad \begin{array}{c} \pi/2 \\ \pi/2 \\ \pi/2 \end{array} \quad , $$

but it makes our lives easier to treat it like one. With this last piece, we can now see the rules of the ZX-calculus in Fig. 2.1. All these rules also hold true with the colours flipped, making use of $(h)$ and $(hh)$.

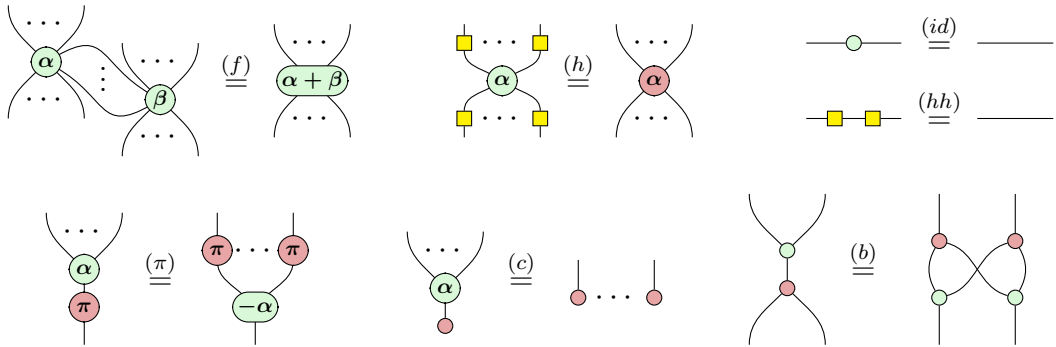

**Figure 2.1:** A convenient set of rules for the ZX-calculus, correct up to non-zero scalar, as seen in Wetering [78]. These rules are commonly called: spider-$(f)$usion, $(h)$adamard, $(id)$entity, $(hh)$-cancellation, $(\pi)$-commute, $(c)$opy, and $(b)$i-algebra.

From these axioms, we can go on to define more useful rules that make the process of rewriting diagrams very intuitive. As a prominent example, we can see the qubit Hopf rule.

**Lemma 17** (Qubit Hopf Rule)**.**

$$ \propto \quad . $$

# 2.4   (Qudit) Finite-Dimensional ZX-Calculus

The ZX-calculus can—and has—been extended in a great many ways to help us take a diagrammatic approach to different fields of quantum computation and theory with greater ease. Broadly speaking, we can extend the language in two ways:

- We can change the set of generators, as in the ZW-calculus [81, 82] or the ZXW-calculus [83, 84]; and/or

- We can give the generators a different interpretation, which may involve extending the language to higher-dimensional systems, such as qudits [60, 85, 86] or the finite-dimensional setting [87–89].

In particular, the ZW- and ZXW-calculi enable complete reasoning for both qudits and finite-dimensional Hilbert spaces, and it has recently been shown that we can open up some novel and interesting applications of the ZX-calculus itself by defining it in the same generality as these other languages [2]. Indeed, a suitable framework for studying representation theory—including its proclivity to interact the same group in different dimensional Hilbert spaces—is a mixed-dimensional one. With this in mind, we now introduce the *finite-dimensional ZX-calculus* [2, 60, 61] as a mixed-dimensional extension to the standard ZX-calculus [37, 38].

## 2.4.1   Generators

We define the generators of the mixed-dimensional ZX-calculus, together with their standard interpretations $[\![\cdot]\!]$, in accordance with Poór, Shaikh, and Wang [2].

**Definition 18** (Z-spider). *The Z-spider, sometimes called a 'box', is given by,*

$$
\xrightarrow{[\![\cdot]\!]} \quad \sum_{j=0}^{\min\{d_i\}_i-1} a_j |j,\ldots,j\rangle\langle j,\ldots,j| \quad ,
$$

*where we set $a_0 := 1$ and write $\overrightarrow{a} = (a_1,\ldots,a_{\min\{d_i\}_i-1})$.*

**Definition 19** (X-spider)**.** *The X-spider, for which all wires have dimension $d$, is given by,*

$$\vcenter{\hbox{}} \;=\; \vcenter{\hbox{}} \quad \overset{[\![\cdot]\!]}{\longmapsto} \quad \sum_{\substack{i+j_1+\cdots+j_m \\ \equiv k_1+\cdots+k_n \;\mathrm{mod}\; d}} |j_1,\ldots,j_m\rangle\langle k_1,\ldots,k_n| \quad,$$

*for some $i \in \mathbb{Z}$.*

Any two X-spiders with the same connectivity that are labelled by $i$ and $i'$ are exactly equal, by definition, if $i \equiv i' \mod d$. It often makes for clearer figures to simply write the dimension label alongside the spider in this case. Single-legged X-spiders labelled by $0 \leq i < d$ correspond to the computational basis state/costate;

$$\vcenter{\hbox{}} \quad \overset{[\![\cdot]\!]}{\longmapsto} \quad \langle i| \qquad \text{and} \qquad \vcenter{\hbox{}} \quad \overset{[\![\cdot]\!]}{\longmapsto} \quad |d-i\rangle \quad.$$

**Definition 20** (Hadamard Box)**.** *The Hadamard box, which we optionally[4] define as a generator for ease of use, is given by*

$$\vcenter{\hbox{}} \quad \overset{[\![\cdot]\!]}{\longmapsto} \quad \frac{1}{\sqrt{d}}\sum_{k,j=0}^{d-1} \omega^{jk}|j\rangle\langle k| \quad,$$

*where $\omega = \exp(2\pi i/d)$ is the $d$-th root of unity.*

**Definition 21** (Identity Wire)**.** *The identity wire is given by,*

$$\vcenter{\hbox{}} \quad \overset{[\![\cdot]\!]}{\longmapsto} \quad I_d = \sum_{j=0}^{d-1} |j\rangle\langle j| \quad.$$

**Definition 22** (Swap)**.** *The swap generator is given by,*

$$\vcenter{\hbox{}} \quad \overset{[\![\cdot]\!]}{\longmapsto} \quad \sum_{i=0}^{d_1-1}\sum_{j=0}^{d_2-1} |j,i\rangle\langle i,j| \quad.$$

## 2.4.2 Convenient Notation and Basic Rules

An extremely useful Z-spider is that labelled by the all-ones vector $\vec{1} = (1,\ldots,1)$. We will use this box so often, e.g. when changing the dimension of a wire, that we will omit the label;

$$\vcenter{\hbox{}} \;:=\; \vcenter{\hbox{}} \quad.$$

---

[4]Alternatively (though strenuously), it can instead be derived from Z- and X-spiders [see (HD) of 2].

There is also a special X-spider: the phase-free X-spider is denoted,

Similarly, the case wherein only the last/first component of the vector is non-zero, say $a \in \mathbb{C}$, is also special enough to earn its own notation:

although note that the 0-th component remains fixed as unit.

More generally, the idea of 'padding with zeros' will crop up a lot; most often when up-casting a wire's dimension. To this end, it is convenient to have notation in place to indicate how much padding we have for an arbitrary Z-spider in $d$ dimensions;

$$\overrightarrow{a_k} := (a_1, \dots, a_{k-1}, 0, \overset{d-k}{\dots}, 0) \quad .$$

The last special case we should note is the zero box as both a state and a scalar:

**Rules for fusing, bending, and connecting**

The fusion of two Z-spiders takes the element-wise product of the fusing labels (padding as necessary), and preserves the external connectivity;

with $M = \min\{m_1, \dots, m_j, n_1, \dots, n_\ell, r_1, \dots, r_s\}$ so that,

$$\overrightarrow{a \odot b} = (a_1 b_1, \dots, a_{M-1} b_{M-1}, 0, \dots, 0) \quad .$$

The fusion of two X-spiders preserves connectivity and sums up phases;

$$\tag{S4}$$

Since our qudit calculus extends the qubit calculus, many rules carry over seamlessly; namely, the *identity* and oh-so useful *bi-algebra* rules hold when all wires have the same dimension.

$$\tag{S2}$$

$$\tag{B2}$$

We can also write Bell states $\sum_{i=0}^{d-1} |i, i\rangle$ (and costates) as caps (and cups) as in the qubit case,

$$\cap \;:=\; \text{and} \quad \cup \;:=\; \tag{S3}$$

however, we differ from the qubit calculus when it comes to X-spiders (as we so often will),

$$\neq$$

The common practice of "bending wires" is slightly more deliberate in the finite-dimensional ZX-calculus. We introduce the *dualiser* [as defined in 38] to make this practice possible.

**Definition 23** (Dualiser)**.** *The dualiser, which we denote by a yellow D-box, is given by,*

$$\boxed{D} \;:=\; \xmapsto{\;[\![\cdot]\!]\;} \sum_{i=0}^{d-1} |d-i\rangle\langle i| \quad .$$

We can now see how a dualiser may be used to arbitrarily change the connectivity of X-spiders,

**Lemma 24.**

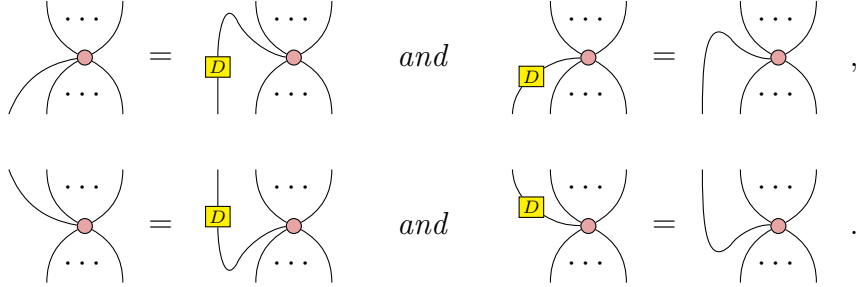

*Proof.* This can easily be seen using only the (S1-4) rules,

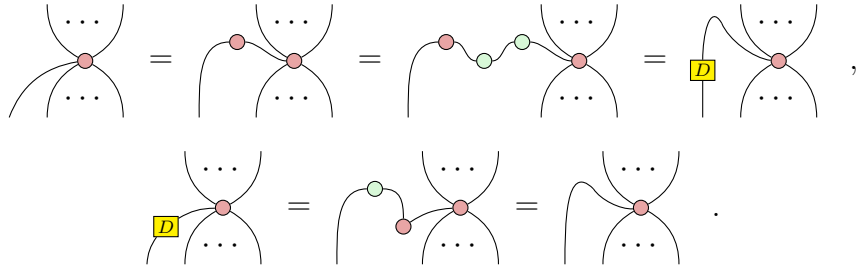

Similarly for the vertically flipped cases.                                      □

Generally speaking, dualisers are very fluid within diagrams. For example, we can slide them along wires, caps, and cups at whim, so long as the connectivity for the spiders whose legs it occupies are preserved. More importantly, they have the freedom to push through boxes, with the consequence of reversing the order of elements in their wake,

$$
\boxed{(a_1,\ldots,a_{d-1})} \;\; = \;\; \boxed{(a_{d-1},\ldots,a_1)} \; . \tag{D1}
$$

Another point of difference from the qubit calculus is an allowance for spiders of opposite colours to be connected by multiple edges (which cannot be further simplified). We can define some syntactic sugar to represent such multi-edges [see e.g. 90–92].

**Definition 25** (Multiplier). *A multiplier indicates the number of edges connecting a Z-spider to an X-spider (and vice versa),*

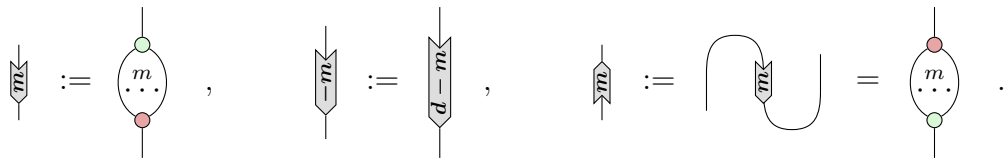

This gives us an alternative definition for the dualiser as a multiplier;

$$\boxed{D} \quad = \quad \text{[diagram]} \quad . \tag{P1}$$

A further special case multiplier is the one labelled by the dimension $d$, which serves as the generalised (mixed-dimensional) Hopf rule.

**Lemma 26.** *The qubit Hopf rule extends to the qudit case as,*

$$\text{[diagram]} \quad = \quad \text{[diagram]} \quad = \quad \text{[diagram]} \quad , \qquad \text{[diagram]} \quad = \quad \text{[diagram]} \quad = \quad \text{[diagram]} \quad . \tag{Hopf}$$

**Rules for pushing spiders**

X-spiders can be 'copied' through Z-spiders (if they are acting like states/costates), or simply 'pushed through' an arbitrary box by either of the following K-rules,

$$\text{[diagram]} \quad = \quad \text{[diagram]} \quad , \tag{K0}$$

where $0 \leq j < m$ and $N = \min\{m, n_1, n_2\}$.

$$\text{[diagram]} \quad = \quad \text{[diagram]} \quad , \tag{K2}$$

where $k_j(\overrightarrow{a}) = (a_{1-j}/a_{d-j}, \ldots, a_{d-1-j}/a_{d-j})$.

The exchange of positions between an X- and mixed-dimensional Z-spider can be made cleaner if there happens to be a dualiser between them to keep things diplomatic;

$$\text{[diagram]} \quad = \quad \text{[diagram]} \quad , \tag{PK}$$

where $2 \leq n \leq m$.

On the other hand, boxes can be copied through X-spiders (again, acting as states/costates), and can *sometimes* be symmetrically split as they push through an X-spider:

where $r_0 = \sum_{i=0}^{d-1} a_i b_{d-i \mod d}$, and the $k$-th element of $\overrightarrow{c}$ is given by,

$$c_k = \frac{1}{r_0} \sum_{i=0}^{d-1} a_i b_{k-i \mod d},$$

that is, $\overrightarrow{c}$ is the discrete convolution of $\overrightarrow{a}$ and $\overrightarrow{b}$.

if the elements satisfy $p_i p_j = q_{i+j \mod d}$ for all $1 \leq i < m$ and $1 \leq j < n$.

### Rules for manipulating dimensions

Changing the dimensions within a diagram is only difficult when we encounter X-spiders, since Z-spiders can be padded with zeros to up-cast seamlessly. For our first rule, we can increase the dimension of an X-spider arbitrarily if we can observe no compression of information; that is, if all of the information flowing into the X-spider can already be described, then we can further inflate the dimension of the space without changing the quality of that information:

for some non-negative integer $k$.

More generally, one can employ multipliers to introduce a *larger* space, or to account for X-spiders that split/combine wires,

where $m \geq n$.

We can now extend the bi-algebra (B2) to arbitrary mixed dimensions.

**Lemma 27.**

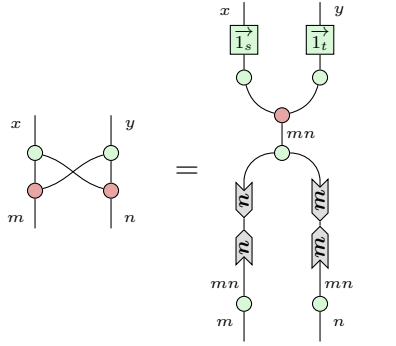

where $s = \min\{x, m, n\}$ and $t = \min\{y, m, n\}$.

*Proof.* Immediately follows (S1), (DZX), and (B2) [see 4, lemma 4]. □

### 2.4.3 Including the W-node

Within this work, we will encounter a great deal of combinatorics to deal with permutations and symmetries. Part of the deal, then, is that the coefficients of our Z-spiders (and whatever else) will often contain factorials, binomials, and maybe even multinomials. This is not a problem on its own, but it can make diagrammatic reasoning less straightforward to such an extent that the 'visual intuitiveness' that makes graphical calculi such appealing tools could become lost. To reinstate that intuition, we make use of a few nibbles of the ZW-calculus [81, 82], particularly in the finite-dimensional setting [2, 89].

**Definition 28** (W-node). *The W-node[5], and its interpretation, are given by,*

$$\text{W} \quad \xmapsto{\;[\![\cdot]\!]\;} \quad \sum_{\substack{0 \le k_i \le d_i, \\ k_1 + \cdots + k_n \le x}} \sqrt{\binom{k_1 + \cdots + k_n}{k_1, \ldots, k_n}} |k_1, \ldots, k_n\rangle\langle k_1 + \cdots + k_n| \quad ,$$

*where $x \ge \max_i d_i$, $d = \sum_i d_i$, and we use the functional notation of Z-boxes as functions of $N$; i.e. $\sqrt{N}$ denotes that the $i$-th coefficient is $\sqrt{i}$.*

Under this definition, there are several helpful rules from the literature with these W-nodes that we will use throughout this work. Perhaps most importantly,

---

[5]This definition aligns with Poór, Shaikh, and Wang [2], and will not necessarily be equivalent to alternatives [most notably 60, defines a different W-node].

the bi-algebra rule of (finite-dimensional) ZW-calculus [89] can be seen in the following lemma.

**Lemma 29.**

$$\text{(WW)}$$

where $x \geq \min\{\sum a_i, \sum b_i\}$, $\ell_{i,j} = \min\{a_i, b_j\}$, and the internal dimensions of the W-nodes are, on the LHS $A = \sum a_i$ on the top and $B = \sum b_i$ on the bottom, and on the RHS $A_i = \sum_k \ell_{i,k}$ along the top and $B_i = \sum_k \ell_{k,i}$ along the bottom.

Similarly, we can note some other bi-algebra variants.

**Lemma 30.**

$$\text{(WZ)}$$

for some $r \in \mathbb{C}$, with the same notation and restrictions as in lemma 29.

## 2.4.4   Rules Summary

Acknowledging that we have introduced many rules, all with abstract labels deriving from an increasingly complex web of ZX research, we give a summary in Fig. 2.2 for quick reference.

**Figure 2.2:** Summary of the rules of the finite-dimensional ZX-calculus that are relevant for this work. See Poór, Shaikh, and Wang [2] for a complete set, including further discussion and rules with Hadamard boxes (Def. 20).

# 3

# Penrose Spin Calculus

## Contents

In this chapter, we introduce the *Penrose Spin Calculus (PSC)*, following
upcoming work by Wang et al. [4], which we elevate to the level of a formal
diagrammatic language by embedding it into the (mixed-dimensional) ZX-calculus
as the fragment describing the recoupling theory of $SU(2)$. In this way, the
language encompasses all Yutsis diagrams and their compositions; i.e. arbitrary
spin networks can be naturally expressed by this fragment. Compared to previous
works constrained to qubit ZX-diagrams [3] which grow exponentially or worse[1] in
the size of the spin network, our qudit ZX-diagrams have only polynomial growth,

---

[1]The rapid growth in East, Martin-Dussaud, and Wetering [3] can be seen in the complex
structure of the "crown", algorithmically constructed in section 4.1, whose number of graphical
elements increases exponentially.

making them the natural choice for scaling to larger networks.

To begin with, we will provide a glancing view of the components needed to use our language; section 3.1 defines the generators of PSC, together with their standard interpretation, in direct analogue with the Yutsis diagrams to which they correspond. We then see the rules of our calculus in the familiar Yutsis style, which serves to provide an accessible view for the behaviour of these diagrams. Section 3.2 spends time embedding the generators through detailed derivations and graphical rewrites in the ZX-calculus. We will further see the translation of the Wigner matrix into the PSC (together with a diagrammatic proof), as well as the diagrammatic techniques used to move between the symmetric Wigner $3jm$-symbol and Clebsch-Gordan coefficients. We leave most rule derivations to Wang et al. [4], although for completeness we will provide proofs for the rules that this work directly uses. Finally, section 3.3 then takes an entirely novel view for the generators in terms of W-nodes, which will facilitate incredibly intuitive rewrites towards the normal forms of chapter 5.

# 3.1 Glancing View: Generators and Rules

Recall from Yutsis (section 2.1.2) and Penrose diagrams (section 2.1.3) the following objects: the *identity* element, the 3-valent node corresponding to the *3j-symbol*, and the special case 3-valent node with a single spin-0 link (an object we call the "*cup/cap*"). As before, these make a nice choice for the generators of our calculus. Hence, the generators of the PSC are given by a suitable translation of these objects into the (mixed-dimensional) ZX-calculus: see table 3.1. The normalisation factor $N(j_1, j_2, j_3)$, seen in the 3j-symbol, is defined by

$$N(j_1, j_2, j_3) = \sqrt{\frac{(j_1 + j_2 + j_3 + 1)! \, (-j_1 + j_2 + j_3)! \, (j_1 - j_2 + j_3)! \, (j_1 + j_2 - j_3)!}{(2j_1)! \, (2j_2)! \, (2j_3)!}} \quad .$$

Using these generators, we can obtain all other orientations of the 3-valent vertex; we can bend one or two links with the cup to reverse their direction (taking them from outputs to inputs), thereby revealing the asymmetric arrangements,

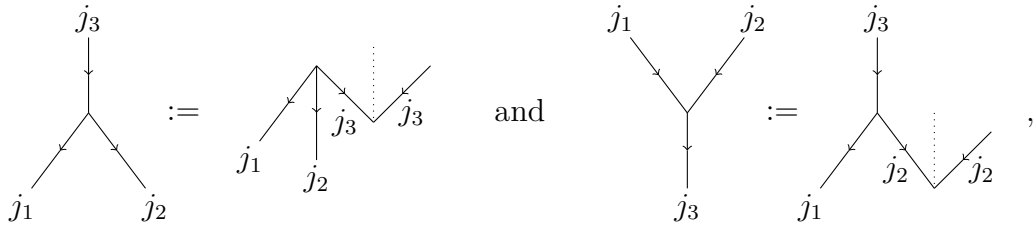

or we could go the whole hog and bend all three to recover a symmetric arrangement with the opposing flow,

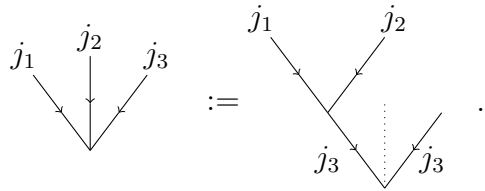

By stacking 3-valent vertices, we can arrive at any $(n+1)$-valent vertex embedding a spin-$n/2$ into $n$ copies of the fundamental spin-1/2 irrep,

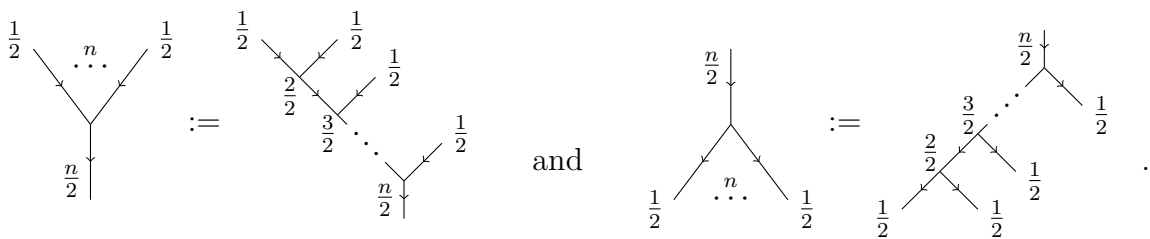

| Yutsis Diagram | Finite-Dimensional ZX-Diagram | Interpretation |
| --- | --- | --- |
| 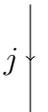 | 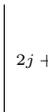 | $\sum_m |j;m\rangle\langle j;m|$ |
| 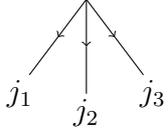 | 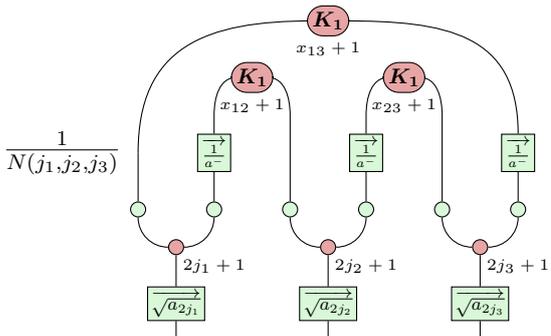 | $\sum_{m_1,m_2,m_3}$ $\begin{pmatrix} j_1 & j_2 & j_3 \\ m_1 & m_2 & m_3 \end{pmatrix}$ $|j_1;m_1,j_2;m_2,j_3;m_3\rangle$ |
| 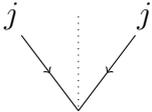 | 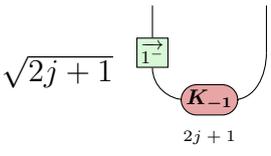 | $\sqrt{2j+1}\sum_m$ $(-1)^{j+m+1}\langle j;m,j;-m|$ |

**Table 3.1:** The generators of the PSC, alongside the corresponding Yutsis diagrams and standard interpretations. We have spins $j, j_1, j_2, j_3 \in \mathbb{N}/2$ where $j_1, j_2, j_3$ are such that they satisfy the Clebsch-Gordan conditions. We use the notation $x_{ab} := j_a + j_b - j_c$, with $c$ being inferred from context (in this case, just the other spin), and $\overrightarrow{1} := ((-1)^1, \dots, (-1)^{2j})$. We also have $\overrightarrow{\sqrt{a_n}} = \left( \frac{1}{\sqrt{\binom{n}{1}}}, \dots, \frac{1}{\sqrt{\binom{n}{n}}} \right)$ and $\overrightarrow{\frac{1}{a_n}} := (\binom{n}{1}, \dots, \binom{n}{n})$. Then use $\overrightarrow{\frac{1}{a^-}} = \overrightarrow{1} \odot \overrightarrow{\frac{1}{a}}$. Where $n$ is omitted in the declaration of an $a$-box (as above), it defaults as $n = d - 1$.

**Definition 31** (Symmetriser). *Taking this embedding vertex of $n$ spin-1/2 irreps together with its adjoint (corresponding to decoding back into spin-1/2s) defines the symmetriser graphically,*

$$
\begin{array}{c}
\text{(diagram)}
\end{array}
$$

**Definition 32** (Qubit Cap/Cup). *A pair of spin-1/2 links emanating from a cap*

*(i.e. coupled with a spin-0 link) takes the following (qubit) ZX-form,*

*and similarly for the conjugate case.*

It is worth explicitly defining the following Z-boxes involved in table 3.1, as we will use them a lot!

**Definition 33** (*a*-Box and its Variants)**.** *The a-box, $\sqrt{a}$-box, $(1/a)$-box, and $(1/\sqrt{a})$-box are each Z-spiders in $d+1$ dimensions, with coefficients given by the vectors,*

$$\overrightarrow{a} \quad := \quad \left( \frac{1}{\binom{d}{1}}, \frac{1}{\binom{d}{2}}, \dots, \frac{1}{\binom{d}{d}} \right) , \quad \overrightarrow{\sqrt{a}} \quad := \quad \left( \frac{1}{\sqrt{\binom{d}{1}}}, \frac{1}{\sqrt{\binom{d}{2}}}, \dots, \frac{1}{\sqrt{\binom{d}{d}}} \right) ,$$

$$\overrightarrow{\frac{1}{a}} \quad := \quad \left( \binom{d}{1}, \binom{d}{2}, \dots, \binom{d}{d} \right) , \quad \overrightarrow{\frac{1}{\sqrt{a}}} \quad := \quad \left( \sqrt{\binom{d}{1}}, \sqrt{\binom{d}{2}}, \dots, \sqrt{\binom{d}{d}} \right) .$$

*We may further write a subscript $n$ on these boxes, either for clarity (e.g. writing $n = d$) or to denote that the box is embedded into a larger space (when $n < d$); e.g.*

$$\overrightarrow{a_n} \quad := \quad \left( \frac{1}{\binom{n}{1}}, \overset{n}{\dots}, \frac{1}{\binom{n}{n}}, 0, \overset{d-n}{\dots}, 0 \right) .$$

**Definition 34** (($1^-$)-Box)**.** *The $(1^-)$-box is a Z-spider in $d+1$ dimensions, with coefficients given by the vectors,*

$$\overrightarrow{1^-} \quad := \quad \left( (-1)^1, (-1)^2, \dots, (-1)^d \right) .$$

*Combining this with the above definition allows us to define the $(1/a^-)$-box (and its variants in a similar manner),*

$$\overrightarrow{\frac{1}{a^-}} \quad = \quad \overrightarrow{1^-} \odot \overrightarrow{\frac{1}{a}} \quad = \quad \left( (-1)^1 \binom{d}{1}, (-1)^2 \binom{d}{2}, \dots, (-1)^d \binom{d}{d} \right) .$$

Finally, to complete the overview, we can see the full rule set for the PSC, derived via the (mixed-dimensional) ZX-calculus, in Fig. 3.1.

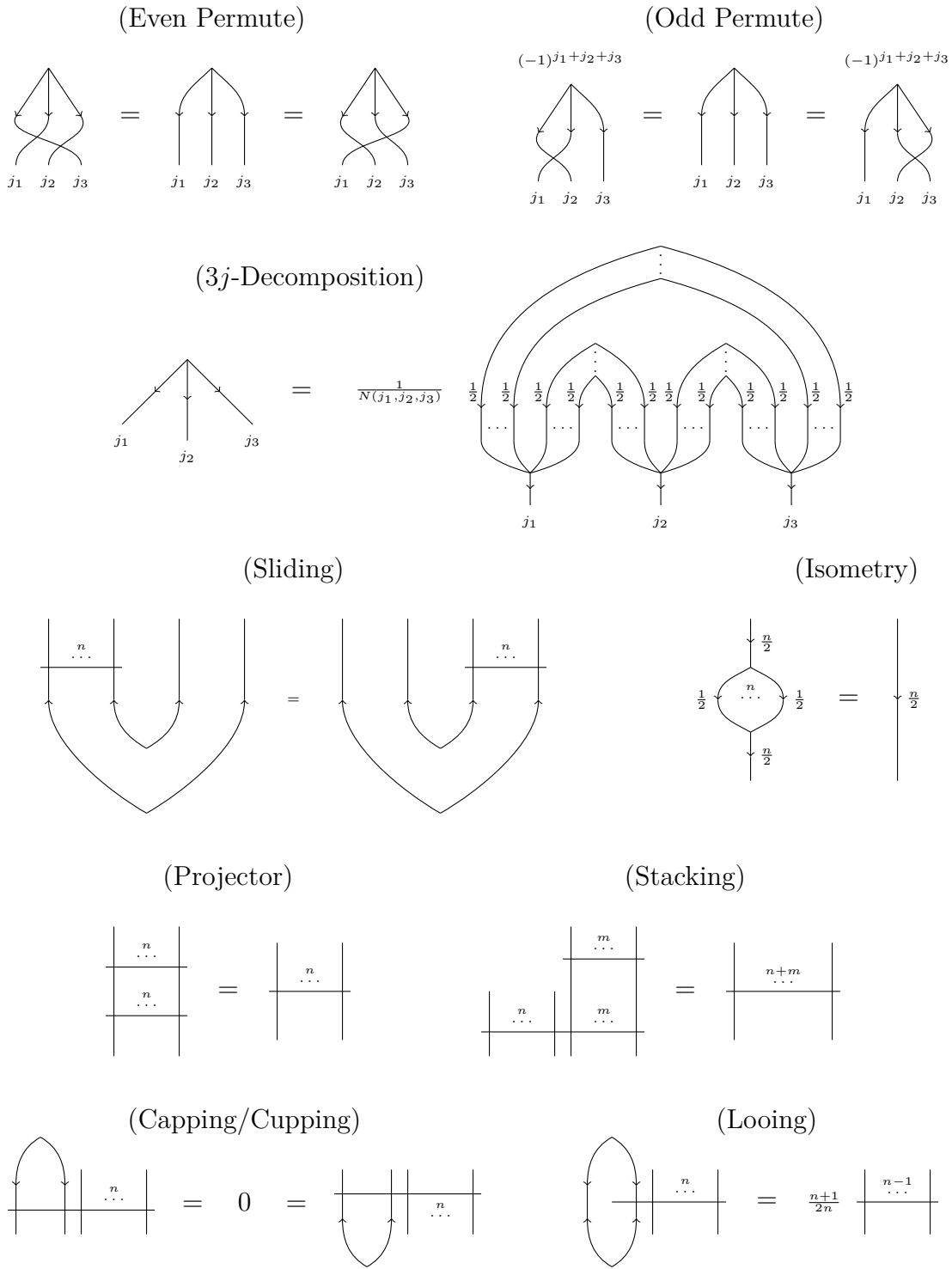

**Figure 3.1:** Rules of the Penrose Spin Calculus (PSC), as proposed in Wang et al. [4].

## 3.2   Detailed Derivations

In this section, we carefully derive the generators of table 3.1, starting with the isometry $V_j$ and symmetriser $\mathcal{S}_n$ in subsections 3.2.2 and 3.2.2, then composing the isometry with cups/caps in subsections 3.2.3 and 3.2.4 to build the $3j$-symbol generator. In each case, we start with groupings of qubit wires (giving the "qu*bit*" isometry, etc.), then merge these into qudit wires with the $a$-box (giving the "qu*dit*" isometry, etc.). In subsection 3.2.5, we make clear the structure of the qudit $3j$-symbol, with all objects properly defined.

### 3.2.1   The Qu*bit* Isometry and Symmetriser

Recall the symmetrisation projection $\mathcal{S}_n$ of Def. 2, where we are setting $n = 2j$, and the subspace $\mathcal{H}_j$ of $(\mathbb{C}^2)^{\otimes n}$ that it defines. There exists an isometry $V_j : \mathcal{H}_j \to (\mathbb{C}^2)^{\otimes n}$ which embeds the canonical basis of $\mathcal{H}_j$ into $(\mathbb{C}^2)^{\otimes n}$, given in the following proposition.

**Proposition 35** (Qubit Isometry)**.** *The isometry $V_j$ is given by the following diagram,*

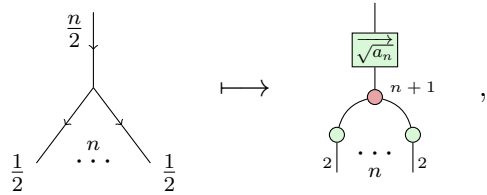

*where* $\overrightarrow{\sqrt{a_n}} = \left( \frac{1}{\sqrt{\binom{n}{1}}}, \dots, \frac{1}{\sqrt{\binom{n}{n}}} \right).$

*Proof.* We prove this by induction.

For the base case, let $n = 2$, which is to take the $3j$-symbol with a spin-1 input link splitting into two spin-1/2 output links. This base case holds immediately via the $3j$-symbol generator of table 3.1 (composed with a cup to bend the spin-1 link). So, suppose it is true for some $n$.

For the inductive step with $n + 1$, we have,

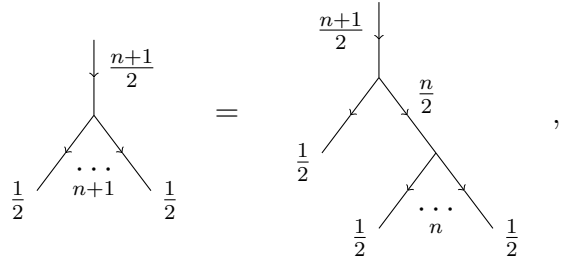

which holds in the PSC by the following diagrammatic argument,

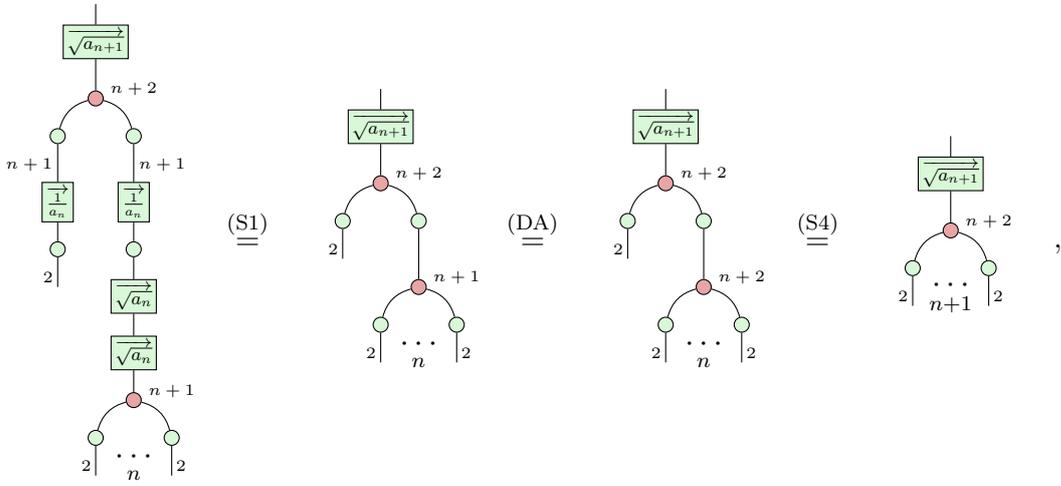

where $\overrightarrow{\frac{1}{a_n}} := \left( \binom{n}{1}, \ldots, \binom{n}{n} \right)$, hence $\overrightarrow{\frac{1}{a_n}} \odot \overrightarrow{a_n} = \overrightarrow{1}_n$ as in the first equality. $\qquad \square$

To verify that $V_j$ is indeed an isometry, we will be needing the following helpful lemmas that give us a way to 'break open' a $(1/a)$-box into a grouping of qubit wires.

**Lemma 36.**

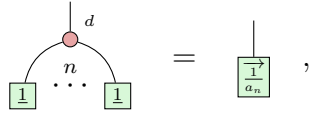

where $n < d$, and the dimension itself is at least qubit $d \geq 2$.

*Proof.* Immediately follows from (PA). $\qquad \square$

**Lemma 37.**

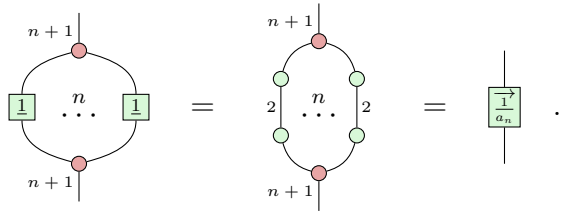

*Proof.* The first equality is a trivial application of (S1). For the second equality,

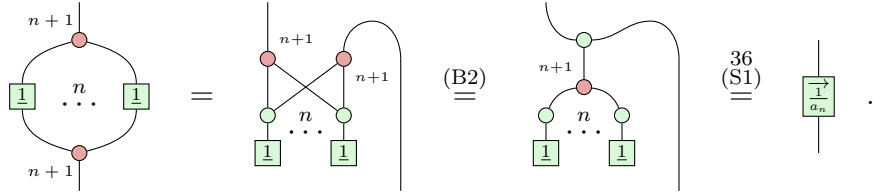

□

**Proposition 38.** $V_j$ *is an isometry; i.e.* $V_j^\dagger V_j = I$*, graphically written,*

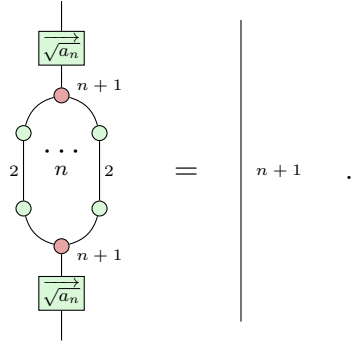

*Proof.* By lemma 37, we can form a $(1/a)$-box from the $n$ grouped qubit wires, which then fuses with the two $\sqrt{a}$-boxes by (S1). Then (S2) to finish. □

With the reverse composition, we can diagrammatically represent the symmetriser $\mathcal{S}_n$ of Def. 31 in the ZX.

**Definition 39** (Qubit Symmetriser).

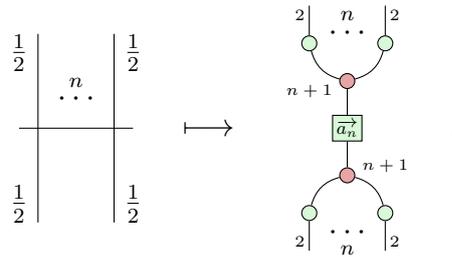

Using lemma 37, we can show that it is indeed a projector, proving the eponymous rule in Fig. 3.1.

**Proposition 40** (Projector). $\mathcal{S}_n = V_j V_j^\dagger$ *is a projector; i.e.* $\mathcal{S}_n^2 = \mathcal{S}_n$*, graphically written,*

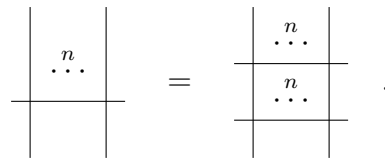

*Proof.*

$$\mathcal{S}_n^2 \quad = \quad \overset{37}{=} \quad = \quad = \quad \mathcal{S}_n \quad .$$

□

Also, following the qubit definitions of the symmetriser and the (spin-1/2) cap, we can prove the 'capping' rule of Fig. 3.1 that will be very useful to us in loop removal (chapter 4). Similarly for the cup and 'cupping'.

**Proposition 41** (Capping / Cupping)**.**

$$= \quad 0 \quad = \quad .$$

*Proof.* We can demonstrate the first equality by showing that the LHS is equal to its own negative, as follows,

$$\overset{(K2)}{=} \qquad = \qquad = \qquad .$$

Similarly for the second equality.                                                □

Using the above, we can now write the canonical orthonormal basis of $\mathcal{H}_j$ diagrammatically,

where, in addition to merging the qubit representation of the input state into the qudit representation, the first equality uses the Def. 39 to expose the isometry of Prop. 35.

## 3.2.2 The Qu*d*it Isometry and Symmetriser

In fact, leveraging the power of the qudit calculus, we can write the isometry in a specialised form that natively captures the embedding into two disjoint spaces. In particular, recalling our discussion of the $3j$-symbol in section 2.1.3, we can express "how many links connect any two $2j_i$ bundles" directly in the dimensions carried by the wires of our isometry.

Before seeing this isometry, we specialise lemma 36 (and, by extension, lemma 37). Suppose, for this discussion, that we have spins $j, j_k, j_\ell$, and we are looking to write the isometry from $\mathcal{H}_j$ to $\mathbb{C}^{2j_k} \otimes \mathbb{C}^{2j_\ell}$.

**Lemma 42.**

where $x_{kj} = j_k + j - j_\ell$ and $x_{\ell j} = j_\ell + j - j_k$.

*Proof.* This also holds by (PA), although it is less trivial.

Write the $n$-th component of the RHS $(1/a)$-box as a sum via the Chu-Vandermonde identity;

$$\left(\overrightarrow{\frac{1}{a_{2j}}}\right)_n = \binom{2j}{n} = \sum_{i=0}^{n}\binom{m}{i}\binom{2j-m}{n-i} ,$$

for some $m$. Letting $m = x_{kj}$, then we have,

$$\binom{m}{i} = \binom{x_{kj}}{i} = \left(\overrightarrow{\frac{1}{a_{x_{kj}}}}\right)_i \quad \text{and} \quad \binom{2j-m}{n-i} = \binom{2j-x_{kj}}{n-i} = \binom{x_{\ell j}}{n-i} = \left(\overrightarrow{\frac{1}{a_{x_{\ell j}}}}\right)_{n-i} ,$$

using $2j - x_{kj} = 2j - (j + j_k - j_\ell) = j - j_k + j_\ell = x_{\ell j}$. Also, note that,

$$r_0 := \sum_{i=0}^{2j} \left(\overrightarrow{\frac{1}{a_{x_{kj}}}}\right)_i \left(\overrightarrow{\frac{1}{a_{x_{\ell j}}}}\right)_{n-i} = \sum_{i=0}^{2j}\binom{x_{kj}}{i}\binom{2j-x_{\ell j}}{2j-i} = \binom{2j}{2j} = 1 ,$$

again by the Chu-Vandermonde identity. Putting this together,

$$\left(\overrightarrow{\frac{1}{a_{2j}}}\right)_n = \frac{1}{r_0}\sum_{i=0}^{n}\binom{m}{i}\binom{2j-m}{n-i} = \sum_{i=0}^{n}\left(\overrightarrow{\frac{1}{a_{x_{kj}}}}\right)_i \left(\overrightarrow{\frac{1}{a_{x_{\ell j}}}}\right)_{n-i} ,$$

thus, by (PA), the above diagram holds.                                                   □

**Proposition 43** (Qudit Isometry)**.** *Specifically for the setting of the $3j$-symbol (with spins $j, j_k, j_\ell$), the isometry can equivalently be given as a direct dichotomisation of groupings of wires with respect to the spins of the other links; graphically,*

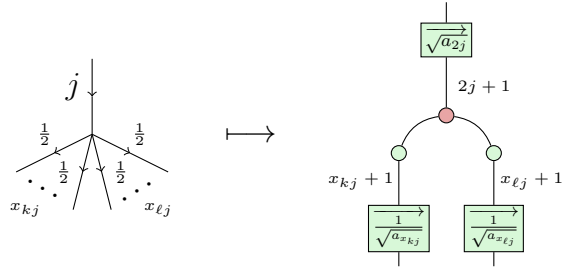

*where $x_{kj} = j + j_k - j_\ell$ and $x_{\ell j} = j + j_\ell - j_k$, hence $x_{kj} + x_{\ell j} = 2j$.*

*Proof.* Use lemma 36 to break into qubits, then prove as in Prop. 35.                       □

**Proposition 44.** *The map of Prop. 43 is an isometry.*

*Proof.* The condition for the map to be an isometry is graphically written as,

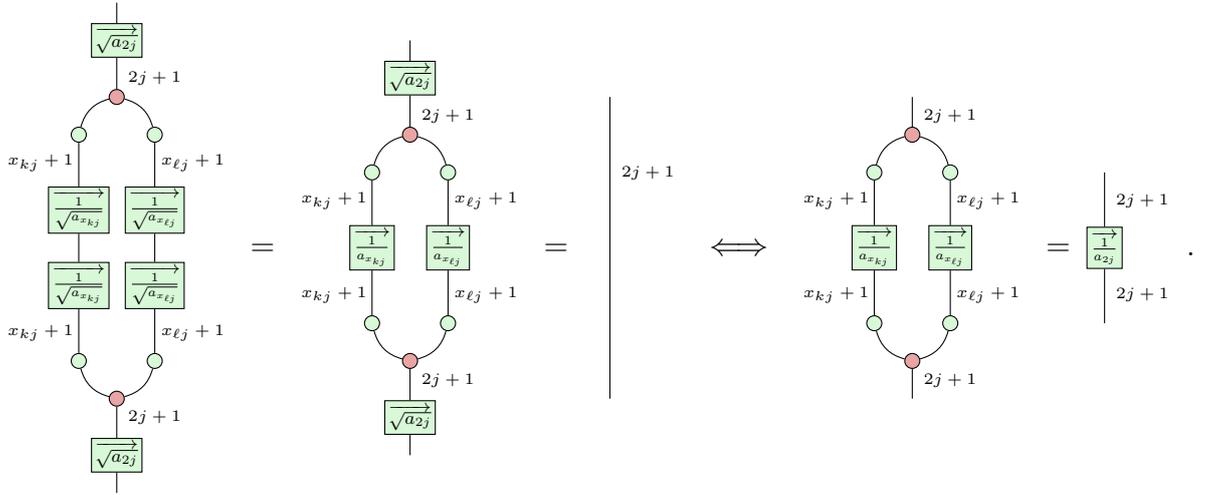

Using lemma 42, we can show that this RHS equality always holds (which is to specialise lemma 37):

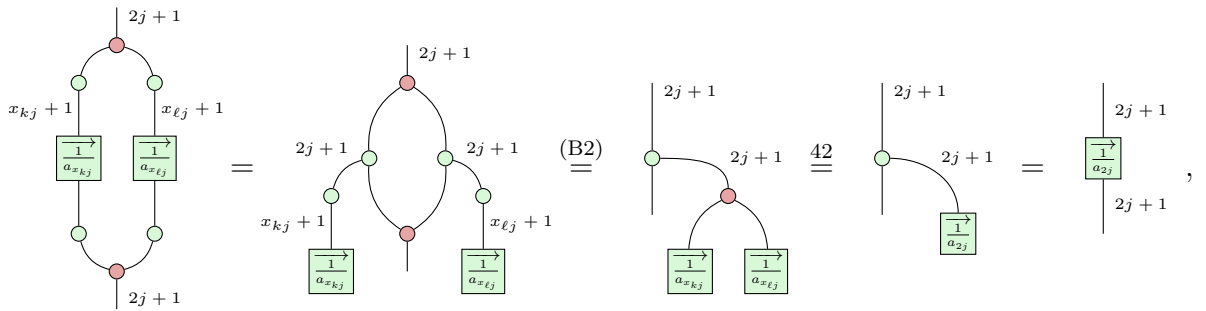

thus, the if and only if condition is satisfied; the map is a well-defined isometry.  □

As you might expect, the symmetriser $\mathcal{S}_n$ specialises in the same way.

**Definition 45** (Qudit Symmetriser).

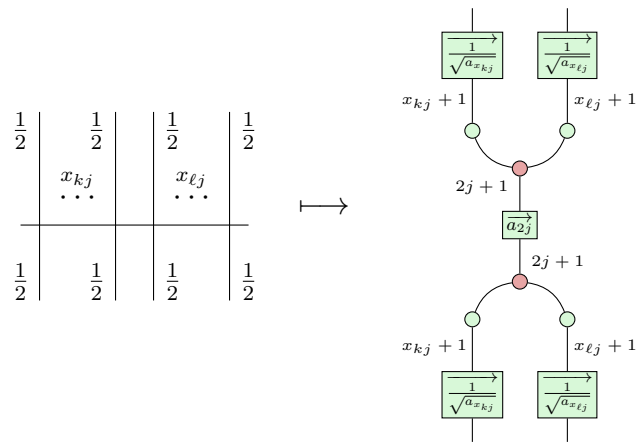

### 3.2.3   The Qu*b*it $3j$-Symbol and Wigner Matrix

Now recall the Wigner $3jm$-symbol of Def. 4, which comes from the desire for a symmetric description of the Clebsch-Gordan coefficients. We started by taking three Hilbert spaces $\mathcal{H}_{j_1}, \mathcal{H}_{j_2}, \mathcal{H}_{j_3}$ without privileging any one space. Then, supposing the Clebsch-Gordan coefficients were satisfied, we had,

$$\dim\left(\mathrm{Inv}_{SU(2)}\left(\mathcal{H}_{j_1} \otimes \mathcal{H}_{j_2} \otimes \mathcal{H}_{j_3}\right)\right) = 1 \quad.$$

Hence, there is a unique unit vector in $\mathrm{Inv}_{SU(2)}(\mathcal{H}_{j_1} \otimes \mathcal{H}_{j_2} \otimes \mathcal{H}_{j_3})$, up to phase. We will now find this vector via ZX-diagrammatic reasoning to present the 3-valent vertex of Yutsis diagrams as a mixed-dimensional ZX-diagram.

First, as we did previously, observe that the following cup diagram is invariant under the action of $SU(2)$: for any $u \in SU(2)$, we have,

Second, let us establish a diagrammatic representation for Wigner matrices.

**Proposition 46** (Wigner Matrix). *For some $u \in SU(2)$, the Wigner matrix $D_u^j$ can be diagrammatically represented using the isometry of Prop. 35,*

*Proof.* To show that this is a Wigner matrix, we can compose the open legs with a basis to obtain the expected matrix element. As a diagram, we can write this as,

$$D^j_{mn}(u) \quad = \quad \text{[diagram]} \quad = \quad E \quad ,$$

where

$$u = \begin{pmatrix} a & b \\ c & d \end{pmatrix}, \quad E = \sum_k \frac{\sqrt{(j-m)!(j+m)!(j-n)!(j+n)!}}{k!(j-m-k)!(j+n-k)!(m-n+k)!} a^{j+n-k} b^{m-n+k} c^k d^{j-m-k} ,$$

with the sum running over all values of $k$ for which the argument of every factorial is non-negative [32]. This can then be proved by pushing the X-spiders inward, where each of the $2j$ closed sub-diagrams are some element of $\{a, b, c, d\}$. Then, by the logical argument presented in Prop 10. of Wang et al. [4], this diagram is equivalent to,

$$\text{[diagram]} = \frac{1}{\sqrt{\binom{2j}{j-n}}} \frac{1}{\sqrt{\binom{2j}{j-m}}} \sum_k \frac{(2j)! a^{j+n-k} b^{m-n+k} c^k d^{j-m-k}}{k!(j-m-k)!(j+n-k)!(m-n+k)!}$$

$$= \sum_k \frac{\sqrt{(j-m)!(j+m)!(j-n)!(j+n)!}}{k!(j-m-k)!(j+n-k)!(m-n+k)!} a^{j+n-k} b^{m-n+k} c^k d^{j-m-k} \quad .$$

$\square$

And lastly, note the following lemma.

**Lemma 47.** *Any $2 \times 2$ unitary matrix $U$ commutes with the symmetriser,*

*Proof.* Any $2 \times 2$ unitary can be decomposed as a product of Z-spiders and Hadamard gates, so this property follows directly from commutation of qubit Z-spiders and Hadamard gates with the symmetriser [proven in 4, lemmas 9 and 10]. □

Following Prop. 46 and this lemma 47, we can find the unique vector in $\mathrm{Inv}_{SU(2)}(\mathcal{H}_{j_1} \otimes \mathcal{H}_{j_2} \otimes \mathcal{H}_{j_3})$ corresponding to the state $|j_1, j_2, j_3\rangle$, up to phase.

**Theorem 48** (Qubit 3-Valent Node). *The diagram,*

*depicts a vector in $\mathrm{Inv}_{SU(2)}(\mathcal{H}_{j_1} \otimes \mathcal{H}_{j_2} \otimes \mathcal{H}_{j_3})$. Since the dimension of this subspace is unit, this vector is necessarily unique, up to phase.*

*Proof.* We can observe the diagram is left invariant after composition of the Wigner matrices $D_u^{j_i}$ for $u$ on each open leg,

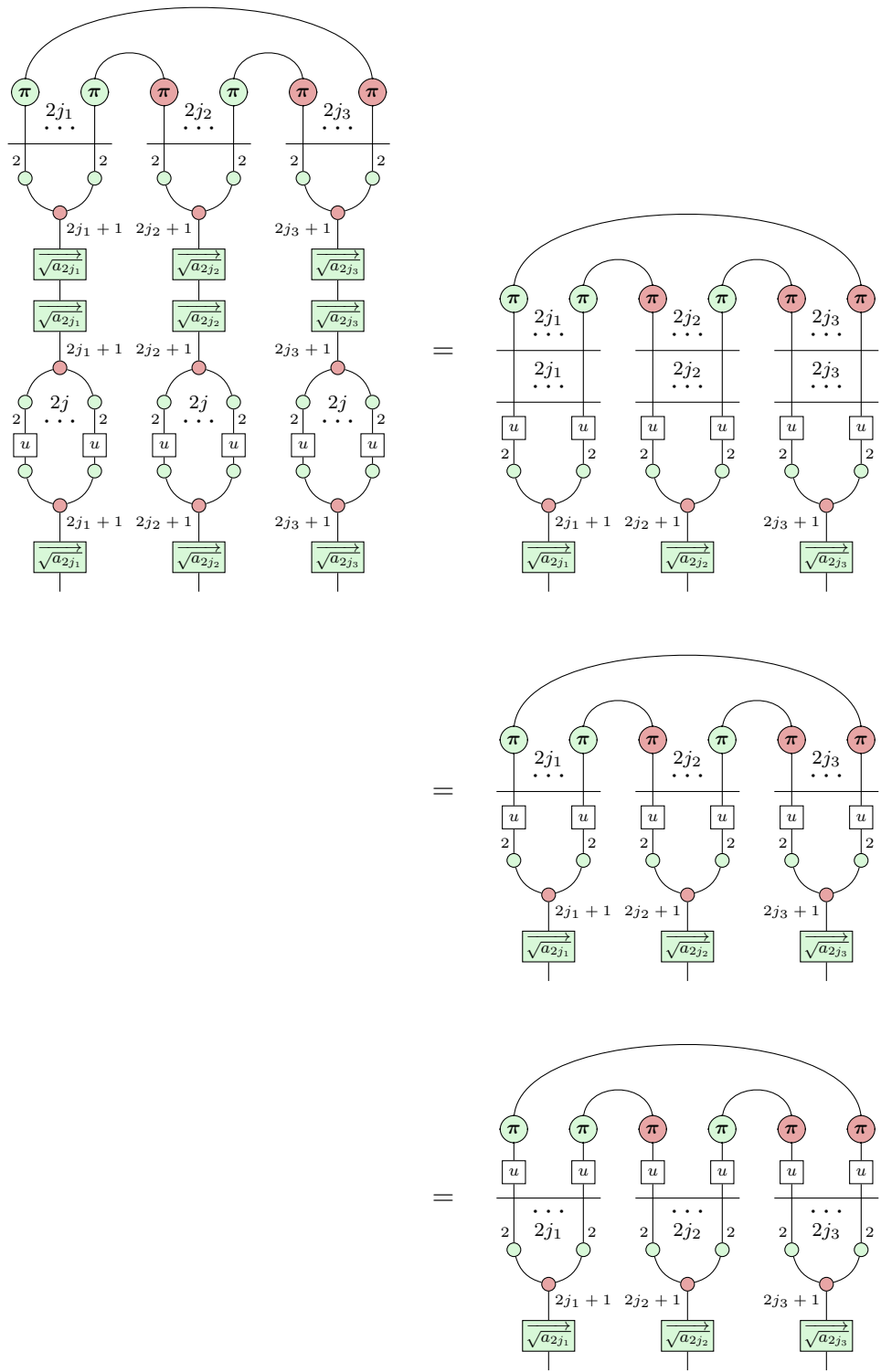

$$= \quad \text{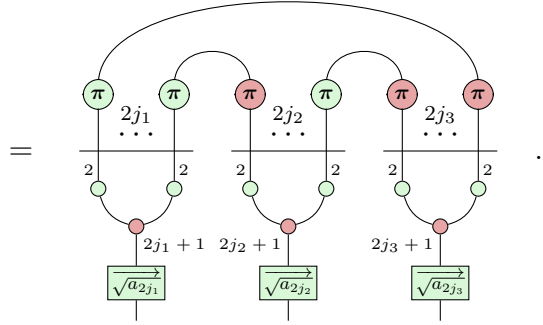} \quad .$$

$\square$

### 3.2.4 The Qu*d*it 3*j*-Symbol

Once again, it would be useful to simplify the 3*j*-symbol of Thm. 48 by drawing the groupings of qubit wires as singular qudit wires. In doing this, we require to also express the cup/cap generator in the qudit form, so this section will finally yield the generators as they appear in Fig. 3.1.

To begin with, we can generalise the cup to arbitrary spin by moving from groupings of qubit wires to singular qudit wires, recovering the form of Fig. 3.1. This gives us the full generality of the change of direction (input to output, and vice versa) by composing with the generalised cup, in an analogue of the *singlet effect*. Similarly for the cap.

**Lemma 49.** *For all pairs* $(k, \ell) \in \{(1, 2), (1, 3), (2, 3)\}$, *we have,*

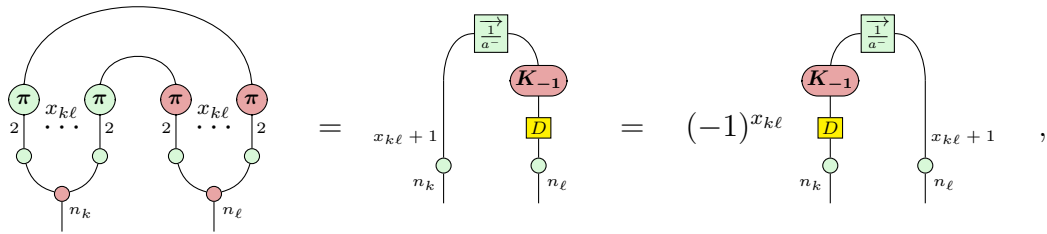

*where* $x_{12} = j_1 + j_2 - j_3$, $x_{13} = j_1 - j_2 + j_3$, $x_{23} = -j_1 + j_2 + j_3$, *and* $n_k = 2k + 1$.

*Proof.*

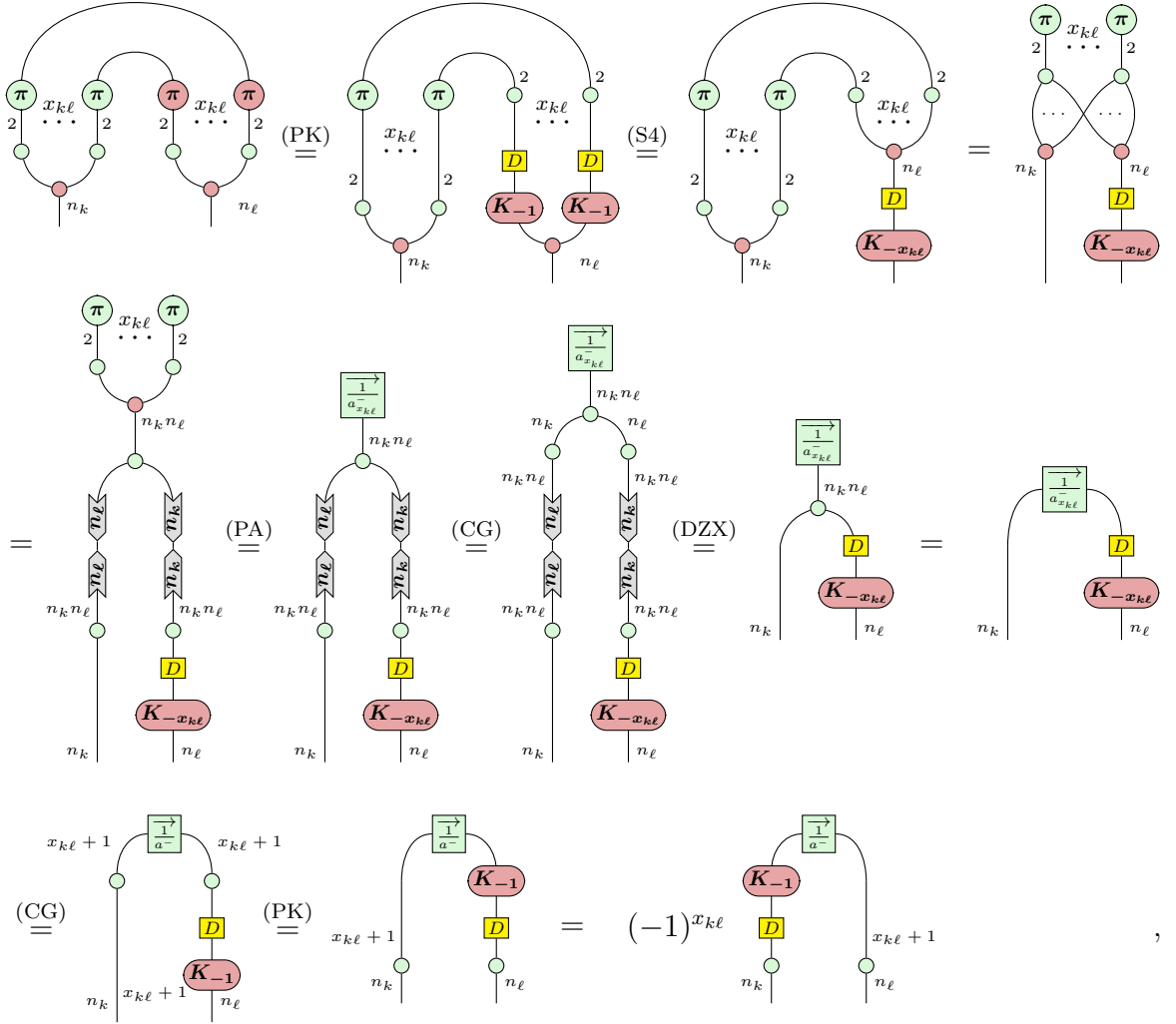

where (CG) is simply an application of the Clebsch-Gordan being satisfied, and where the last equality holds by lemma 12 of Wang et al. [4]. □

Next, we can further take the diagrammatic representation of the state $|j_1, j_2, j_3\rangle$, corresponding to the basic 3-valent vertex generator of the Yutsis calculus, into the specialised qudit form making use of higher dimensions. Together with the forms of section 3.2.2, the usefulness of this rewrite will be immediately clear, as it allows us to simplify our generator into a diagram with a much smaller collection of graphical elements, recovering the form of Fig. 3.1.

**Theorem 50** (Qudit 3-Valent Node)**.**

where $x_{12} = j_1 + j_2 - j_3$, $x_{13} = j_1 - j_2 + j_3$, $x_{23} = -j_1 + j_2 + j_3$, $\overrightarrow{1/a} = \left((-1)^1\binom{d}{1}, \ldots, (-1)^d\binom{d}{d}\right)$ for $d+1$ the dimension of the wire, and the normalisation factor $N(j_1, j_2, j_3)$ is given by,

$$N(j_1, j_2, j_3) = \sqrt{\frac{(j_1 + j_2 + j_3 + 1)!\,(-j_1 + j_2 + j_3)!\,(j_1 - j_2 + j_3)!\,(j_1 + j_2 - j_3)!}{(2j_1)!\,(2j_2)!\,(2j_3)!}} \ .$$

*Proof.* Starting from the qubit 3-valent node of Thm. 48,

$$\overset{49}{=} \quad \frac{1}{N(j_1, j_2, j_3)} \quad$$

$$= \quad \frac{1}{N(j_1, j_2, j_3)} \quad$$

$$= \quad \frac{1}{N(j_1, j_2, j_3)} \quad .$$

$\square$

Now that we have seen where the 3-valent node comes from, we can think about its asymmetric cousin. By the following proposition, we can arrive at the injection $\mathcal{H}_{j_3} \to \mathcal{H}_{j_1} \otimes \mathcal{H}_{j_2}$ [as given in 4, Prop. 12] through composition of the open $j_3$ leg with a cup.

**Proposition 51.**

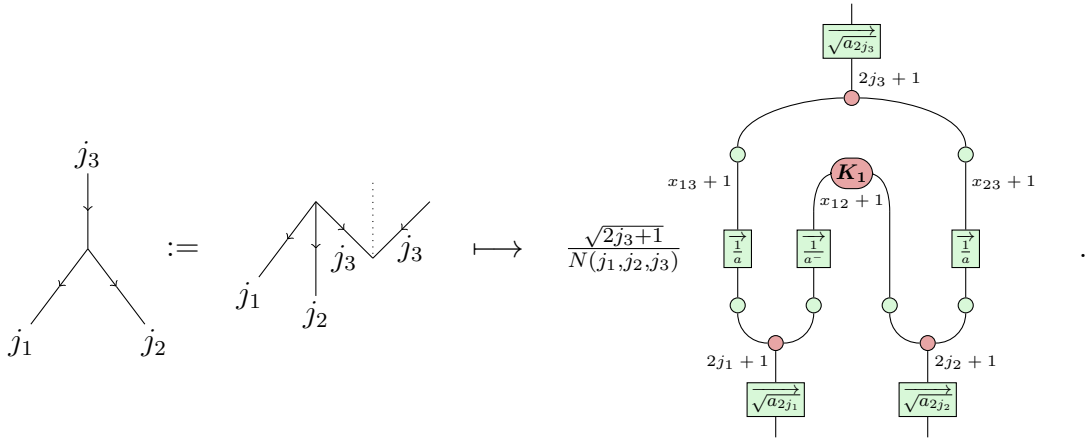

*Proof.* See Prop 12. of Wang et al. [4] for a direct proof in terms of the Clebsch-Gordan coefficients. Otherwise, this is seen immediately through a composition of the open $j_3$ leg with the cup and some planar deformation. $\qquad\square$

### 3.2.5 Dissecting the 3-Valent Node Generator

The diagram of Thm. 50 can feel abstract in the simplified form. Some more explicit intuition for how this object is structured will greatly improve understanding for the subsequent structure of larger spin networks. So, let us break up the qudit 3-valent node into its constituent parts (already introduced throughout this chapter).

In fact, the $3j$-symbol is the connection of three isometries by three caps (or cups, depending on orientation), seen in Fig. 3.1. This is actually quite clear in the qudit case, and to see it, we need only to break apart the $(1/a^-)$-box into a pair of $(1/\sqrt{a})$-boxes and a $(1^-)$ box via the following lemma.

**Lemma 52.** *For any wire dimension $d + 1$, the cap can be decomposed as,*

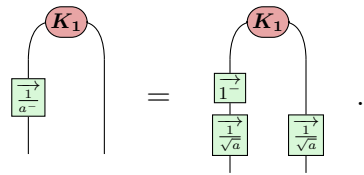

*Proof.*

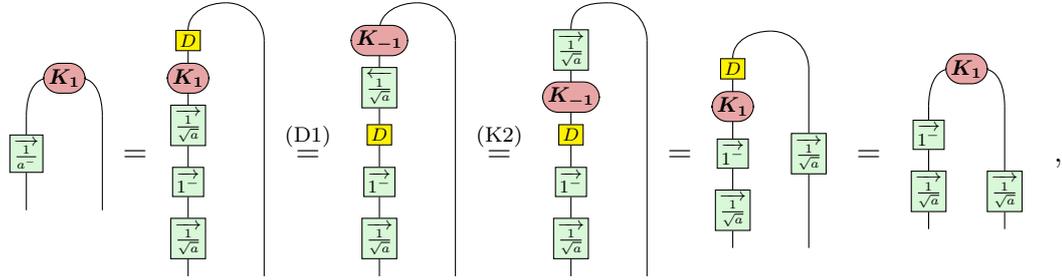

where, after (D1), we yield a box of $\overleftarrow{1/\sqrt{a}} = \left( \binom{d}{d}, \binom{d}{d-1}, \dots, \binom{d}{1} \right)$, and the application of (K2) above follows as,

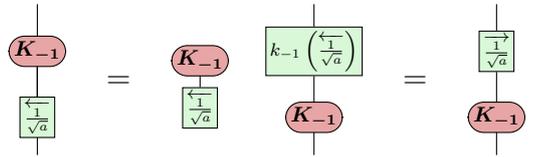

with (writing as a function of $N$, for $1 \leq N < d+1$),

$$k_{-1}\left( \overleftarrow{\frac{1}{\sqrt{a}}} \right) = k_{-1}\left( \binom{d}{d+1-N} \right) = \frac{\binom{d}{d+1-(N+1)}}{\binom{d}{d+1-(d+1)}} = \binom{d}{d-N} = \binom{d}{N} = \overrightarrow{\frac{1}{\sqrt{a}}} \quad,$$

and,

$$\begin{array}{ccc} \boxed{K_{-1} \atop \frac{1}{\sqrt{a}}} & = & \boxed{K_d \atop \frac{1}{\sqrt{a}}} \end{array} \overset{[\![\cdot]\!]}{\longmapsto} \sum_{i=0}^{d} \binom{d}{d+1-i} \langle i|i\rangle \langle i|d+1-d\rangle = \binom{d}{d} = 1 \quad.$$

$\square$

The structure can now be seen plainly as isometries post-composed with caps (recovering the form we see from the $3j$-decomposition in Fig. 3.1),

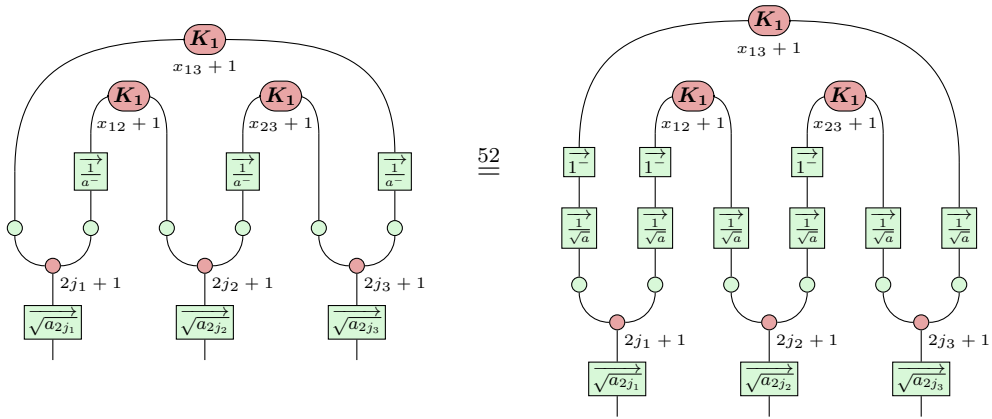

### PSC Generators: Summary

- Our fundamental qu*bit* objects are the isometry $V_j$ (Prop. 35) and cap/cup (Def. 32). The identity generator follows $V_j^\dagger V_j = I$ (Prop. 38), and the reverse gives the symmetriser (Def. 39). These are sufficient to prove the PSC rules (Projector) and (Capping/Cupping).

- Composing three isometries with three caps (all-to-all connected) builds the qu*bit* $3j$-symbol generator (Thm. 48), which can be seen via its invariance under composition with Wigner matrices (Prop. 46).

- Both the isometry and cap/cup can be simplified by merging groupings of qubit wires into qu*dit* wires (Prop. 43 and lemma 49), and thus so can the $3j$-symbol (Thm. 50).

- By decomposing the cap/cup (lemma 52), the compact qu*dit* $3j$-symbol can be dissected to reveal the same structure of isometries composed with caps/cups, intuitively proving the PSC rule ($3j$-Decomposition).

## 3.3   The W-Node Perspective

The objects discussed in section 3.2 could equivalently be expressed using W-nodes [2]. The motivation for this is to give us a framework for more direct rewrite rules once we build up spin networks out of multiple 3-valent nodes and encounter 'barriers of symmetriers' in the notation of Wang et al. [4] that can be immediately resolved with W-nodes.

Another motivation for the W-node perspective is to expose *gadgets*[2] in our spin networks that can carry information about the spin and connectivity directly. Our normal form of chapter 5 explores this idea thoroughly. However, to get there, we will need the following lemma to see how gadgets can be pulled out of diagrams.

**Lemma 53.**

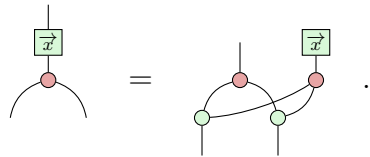

*Proof.*

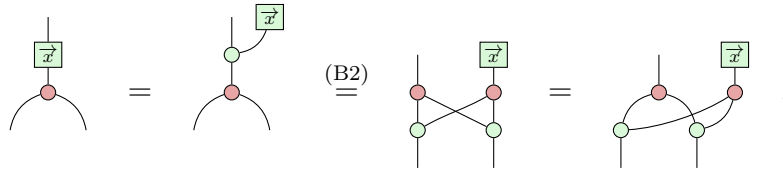

$\square$

Now we can translate the objects of section 3.2 into the W-node perspective. Note that, when discussing W-nodes, it is common to move away from writing the coefficients of our Z-boxes as vectors, and towards writing them as functions of $N$. We start, as before, with the isometry.

---

[2]In the context of this work, we loosely use the familiar term "gadget" to refer to 'dangly bits' of our diagrams. In particular, the W-node dangly bits we reveal bare a strong resemblance and carry a similar role in our rewrites to gadgets commonly seen in ZX literature [e.g. 49, 51].

**Theorem 54** (W-Node Isometry)**.** *The qudit isometry of Prop. 43 can be written equivalently in terms of the W-nodes of Def. 28,*

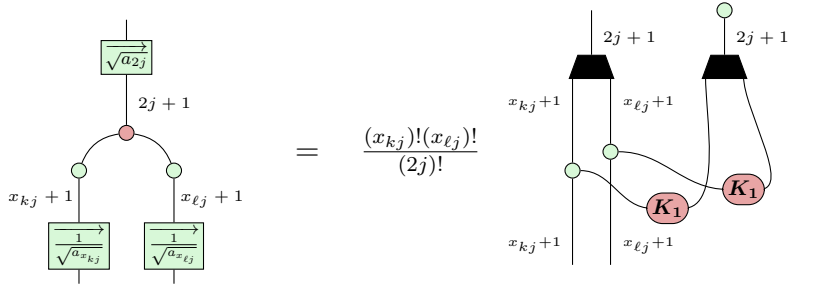

*where the dimensions coming in to, and out of, the gadget-like W-node (right) are the same as the other W-node's (left); i.e. the Z-spiders in the RHS diagram do not involve dimension changes. The Z-spider input to the gadget-like W-node is not strictly necessary (since it will just fuse with the upper Z-box in the W-node), but we draw it anyway to align with the established notation.*

*Proof.* The coefficients of the $a$-box can be written as the following function of $N$,

$$(\overrightarrow{a_d})_N \;=\; \frac{1}{\binom{d}{N}} \;=\; \frac{N!(d-N)!}{d!} \;=\; \frac{1}{d!} \cdot N! \cdot (d-N)! \quad,$$

and similarly for the reciprocal and square-root variants. Writing the isometry of 43 in this functional notation (rather than the vector notation), we have,

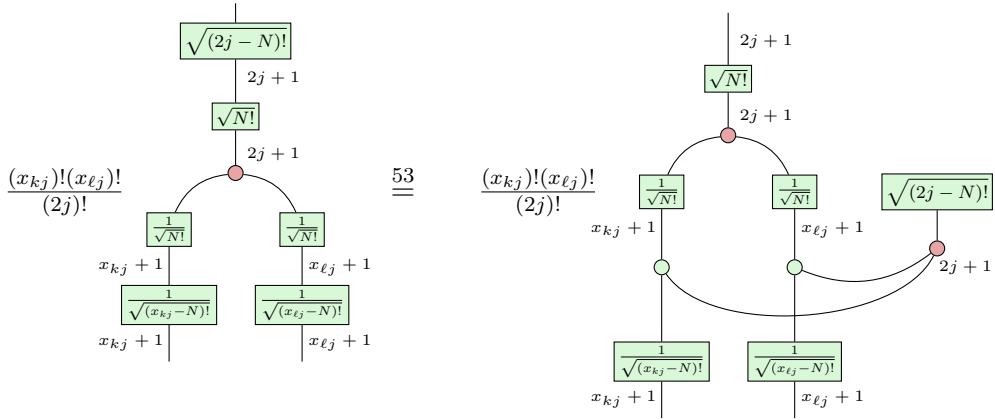

$$\overset{x_{kj} \leq 2j}{\underset{x_{\ell j} \leq 2j}{\equiv}} \frac{(x_{kj})!(x_{\ell j})!}{(2j)!}$$

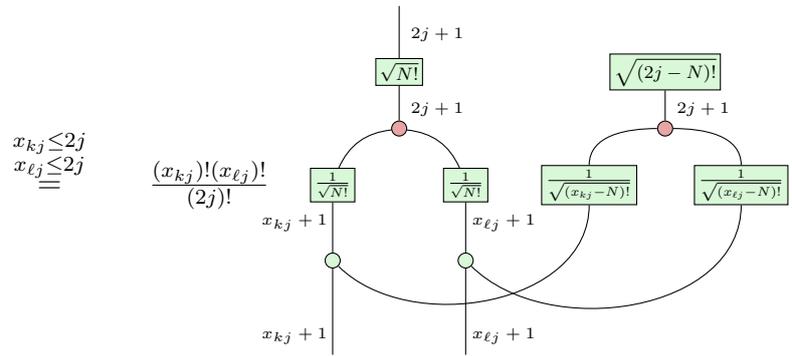

$$= \frac{(x_{kj})!(x_{\ell j})!}{(2j)!}$$

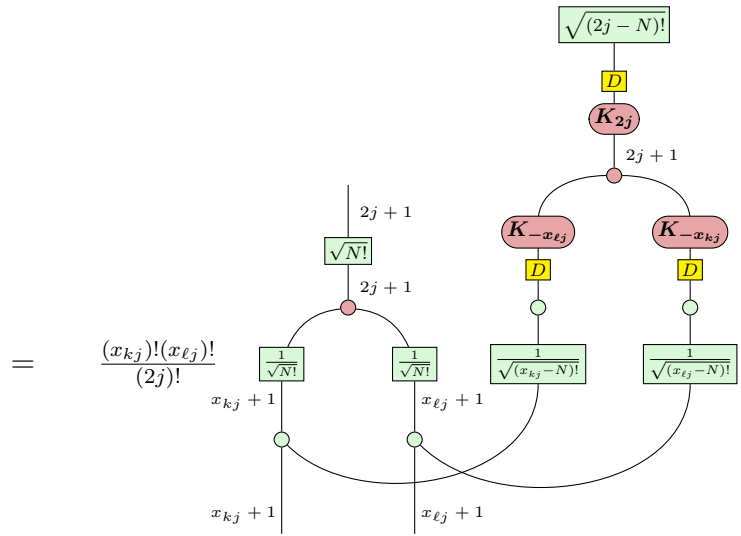

$$\overset{(\text{PK})}{=} \frac{(x_{kj})!(x_{\ell j})!}{(2j)!}$$

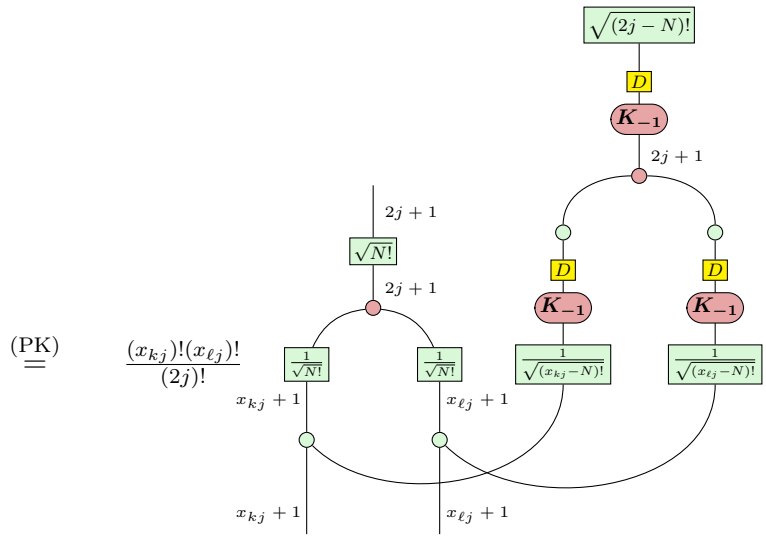

$$\stackrel{(\mathrm{D1})}{=} \frac{(x_{kj})!(x_{\ell j})!}{(2j)!}$$

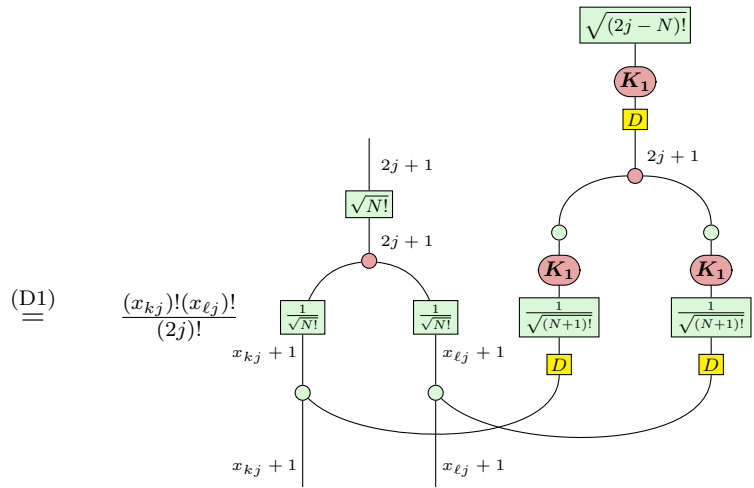

$$\stackrel{(\mathrm{K2})}{=} \frac{(x_{kj})!(x_{\ell j})!}{(2j)!}$$

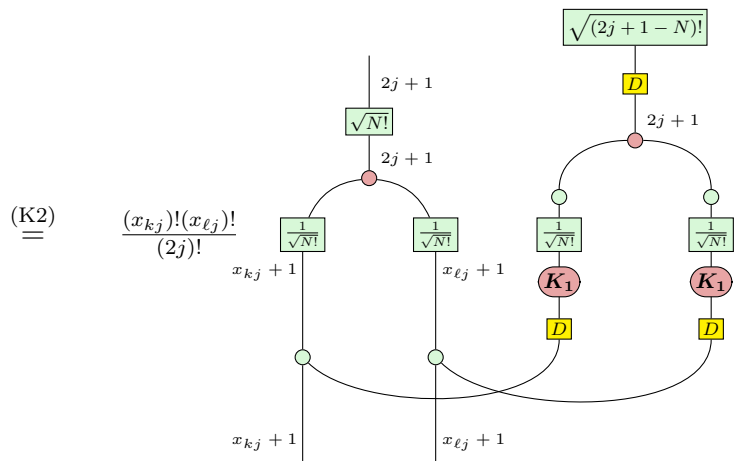

$$\stackrel{(\mathrm{D1})}{=} \frac{(x_{kj})!(x_{\ell j})!}{(2j)!}$$

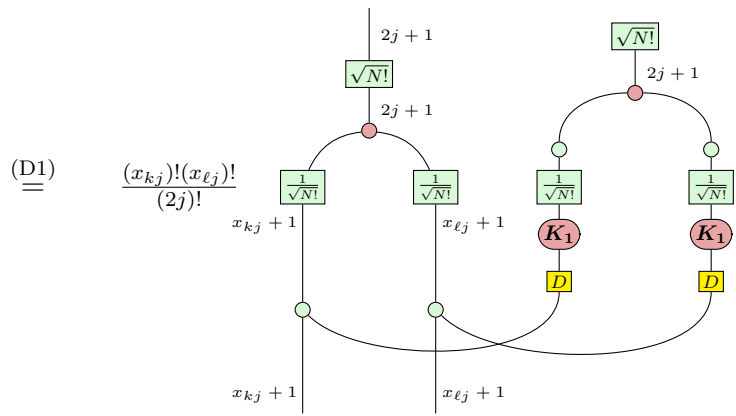

$$= \quad \frac{(x_{kj})!(x_{\ell j})!}{(2j)!} \qquad \text{}$$

$$= \quad \frac{(x_{kj})!(x_{\ell j})!}{(2j)!} \qquad \text{} \quad .$$

We could further notice that,

$$\frac{(x_{jk})!(x_{\ell j})!}{(2j)!} \;=\; \frac{1}{\binom{2j}{x_{kj}}} \;=\; \frac{1}{\binom{2j}{x_{\ell j}}} \quad .$$

$\square$

We will refer to these W-node gadgets, rather plainly, as *W-gadgets*.

The cap (and cup) can also be written in this way. While the translation alone may seen contrived, its new form will permit simpler rewrites due to the interaction of the W-nodes with the cup/cap and those in the isometry.

**Proposition 55** (W-Node Singlet)**.** *For any dimension,*

*Proof.* The $(1^-)$-box in the cap can easily be seen to correspond to a function $-1^N$ of $N$. The core idea to this proof is to create a W-node around the X-spider, then use bi-algebra to reveal a second W-node,

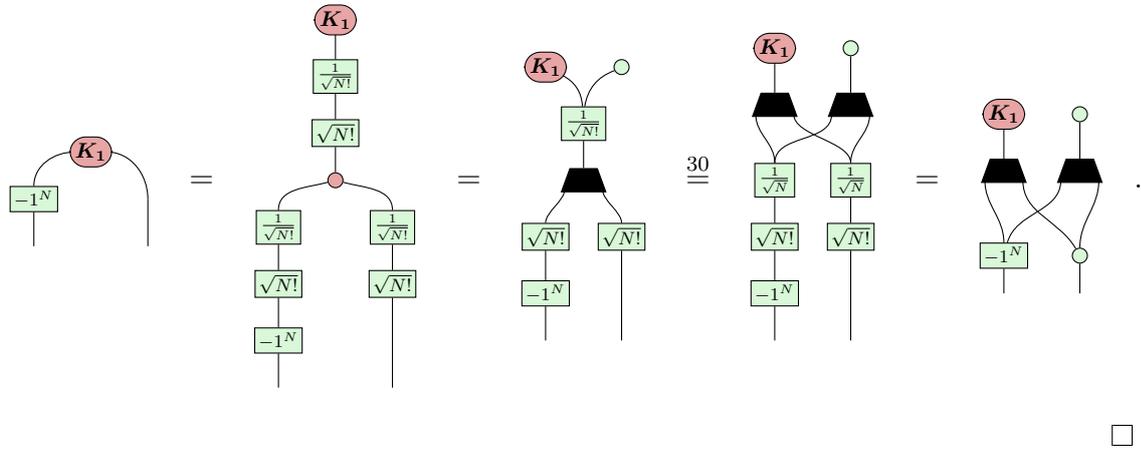

$$\square$$

We can then translate the 3-valent node generator of the PSC, with correctness following immediately from our derivations in section 3.2.

**Proposition 56** (W-Node 3-Valent Injection)**.**

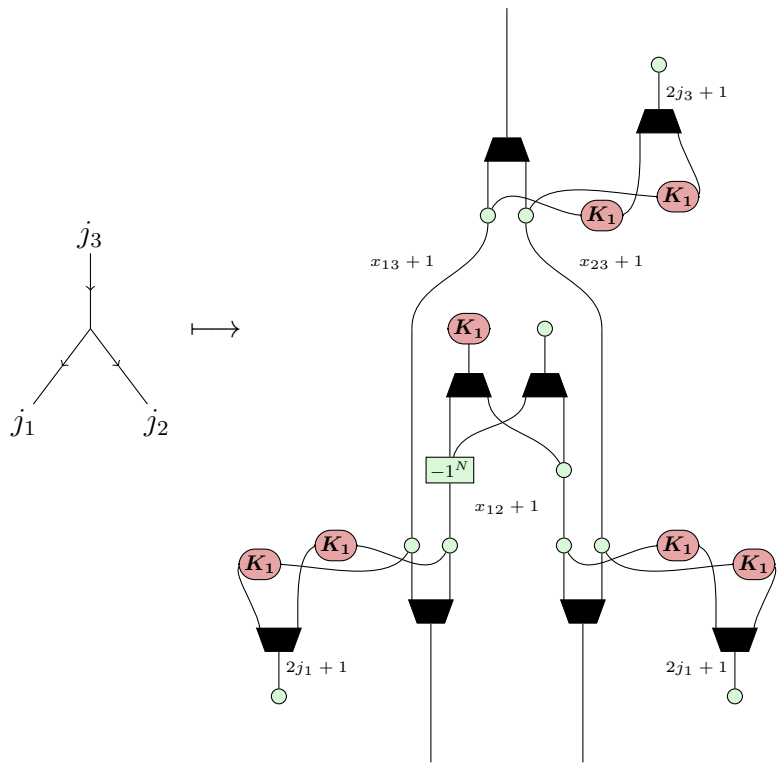

*Proof.* This is a translation of Prop. 51, using Thm. 54 and Prop. 55.          $\square$

We have already caught a glimpse of the power W-nodes give us in the proof of Prop. 55. Let us now give an explicit example of how this notation improves on the standard PSC introduced by Wang et al. [4]. A prominent issue that arises when we try to rewrite spin networks in the standard PSC is the *symmetriser barrier* problem: composing any two 3-valent nodes gives a symmetriser on the connecting links (by definition),

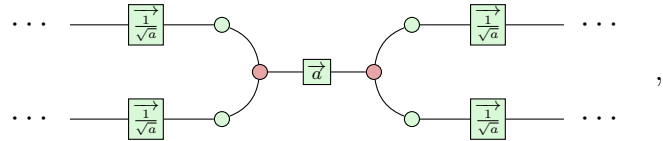

,

which thus leaves an isolated *a*-box that cannot be pushed through either X-spider[3]. Once we resolve to using the W-node notation, however, we can immediately apply bi-algebras to remove the bottleneck,

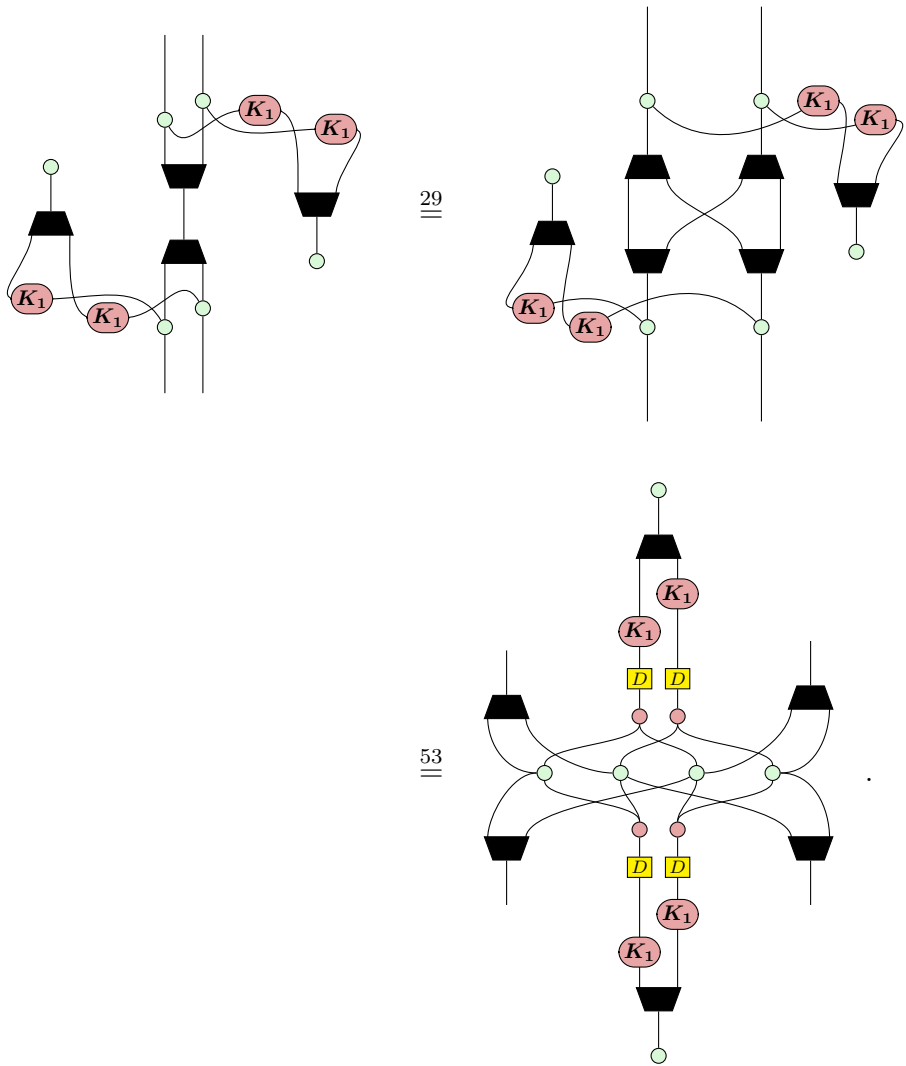

.

---

[3]A convolution rule, like (PC), cannot be used here since the coefficients of the *a*-box contain factorials; there does not exist a discrete function $f$ such that $(x+y)! = f(x) \cdot f(y)$ for all $x, y$.

Indeed, this is just about the *only* thing we can do in such a situation; since the $a$-box cannot be moved (and neither can the $(1/a)$-boxes be pushed into the bottleneck), the only way to rewrite this $a$-box is to resort to gadget-like operations. Of course, this bi-algebra could be performed without using ZXW notation, but the visual intuition would certainly be lacking, whereas it is immediately apparent how to resolve the symmetriser barrier with W-nodes.

The above technique—of bi-algebra-ing the opposing W-nodes in the symmetriser and pulling the W-gadgets into the melee between them—will be the foundation of our spin network normal form in chapter 5. Do not worry about the mess; we will clean it up in the next section.

### 3.3.1   Clean Notation for W-Gadgets

As we move up to larger spin networks, we acquire many more W-gadgets that interact and stretch across many wires, which has the unfortunate consequence that it can quickly become difficult to draw them clearly. To address this concern, we will now introduce some new notation to clean up the interactions and traversals of our W-gadgets; this notation has two features:

- *Colouring* (labels)—a means to identify a W-gadget with a single colour or label to replace the full diagrammatic representation of the gadget; and

- *Orientation*—a means to distinguish the left leg from the right leg, arbitrarily choosing the left leg as the in-direction and the right as out.

For a particular gadget, we assign a label (usually denoting the dimension of its top side, deriving from the spin of the link it belonged to in the original spin network's structure). Having defined this, we can then write the label beside the Z-spiders to which the gadget connects—on the left if it connects to the gadget's left leg, or on the right otherwise,

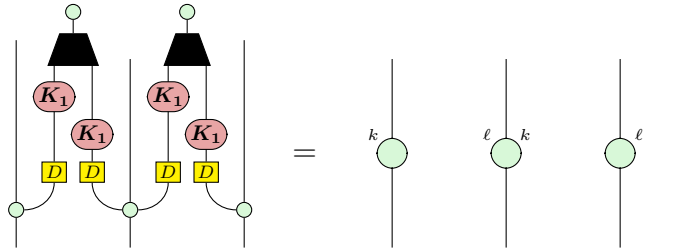

This notation scales up seamlessly to gadgets whose left- or right-side leg branches out to more than one wire; simply write the label beside all Z-spiders to which it connects (again, on the left or right),

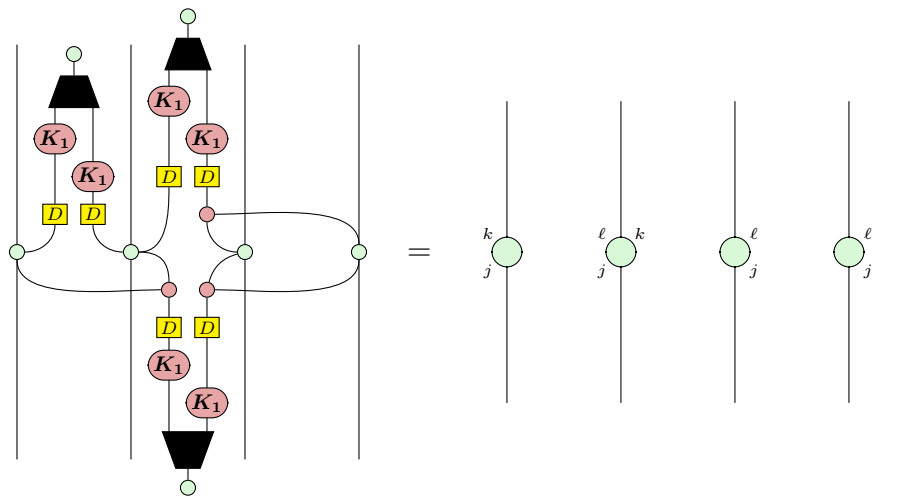

No two gadgets will ever share the same label (even if they are mathematically equivalent; e.g. they comprise the same dimensions). With this promise made, we can always assume that *all* left-sided occurrences of a label in our diagram refer to wires destined for *the same gadget*'s left leg (joined by an X-spider), and likewise for the right-sided occurrences.

As an example, here is the notation in action for the symmetriser barrier of the previous section, dramatically cleaning up the diagram,

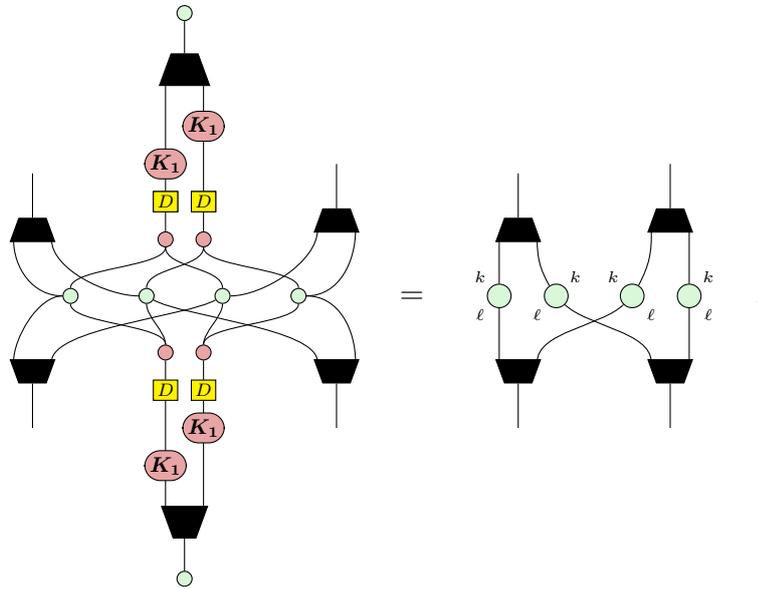

Great, now we have the language to work with many—possibly interacting—W-gadgets without getting lost in the lack of planarity. This also acts as a handy pedagogical tool for abstracting the locale and orientation of our gadgets into (hyper)graph decorations in chapter 5. All in good time.

**The W-Node Perspective: Summary**

- The PSC generators of table 3.1 can be written with W-nodes to abstract the tricky *a*-boxes into easily-handled graphical elements. Hence, we can write the isometry (Thm. 54), cup/cap (Prop. 55), and 3*j*-symbol (Prop. 56) with W-nodes.

- We indicate that the W-node perspective natively handles the "symmetriser barrier" problem by pulling the W-gadgets associated with the 3-valent nodes into a fully-connected 'nest' between the output links of the symmetriser.

- Drawing many gadgets can quickly make a diagram overly busy, so we introduce a cleaner notation: write a W-gadget's label beside the (enlarged) Z-spider to which it connects—left/right distinguished.

*We are stories, contained within the twenty compli-*
*cated centimetres behind our eyes.*

— Carlo Rovelli [94]

# 4

# Removal of $n$-Loop Spin Network Portions

## Contents

Surrounding our discussion of (G27), we have noted the possibility to "surgically remove" spin network portions (something we prefer to call "loop removal"). Indeed, the algorithmic approach taken in Guedes et al. [1] is laced, start to finish, in loop removal (almost exclusively 2- and 3-loop removal) in an effort to handle changing graphs. In this chapter, we will derive the ZX-form for 2- and 3-loop removal (sections 4.1 and 4.3) and prove their correctness in the PSC.

To present exact scalars, we also require to derive the $\Theta$-graph (subsection 4.1.2), and the $6j$-symbol and tetrahedral net symbol (section 4.2). Following this, we present a proof for the associativity of spin network nodes (section 4.4) and a novel argument for $n$-loop removal based on these results (section 4.5). Our proofs here are ZX-diagrammatic and self-contained; namely, they do not depend on any results from Schur's lemma [67], as most works do [e.g. the 2-loop removal and $\Theta$-graph in 66].

For this work, we consider full normalisation embedded directly into the generators of our calculus. There are various different conventions for such normalisation; namely, Guedes et al. [1] do not normalise their generators in the same fashion as they use implicitly contracted graphical elements (i.e. the magnetic indicies $m_1, m_2, m_3$ in the Wigner $3jm$-symbol), whereas we define higher-level linear maps (i.e. an uncontracted $3j$-symbol). As such, our graphical relations will differ from Guedes et al. [1] by a normalisation factor $N(\cdots)$ for each 3-valent node.

## 4.1  2-Loop  Removal

A *2-loop spin network portion* (sometimes referred to as a "*bubble*") is the meeting of two injective 3-valent nodes so as to form an internal 'loop' structure with two (3-valent) nodes. In the special case that these two nodes also happen to have their third links connect, we obtain the closed Θ-graph; otherwise, we have an open bubble.

By the *bubble identity* of (G13), we know that this object is proportional to the identity wire (given $j = j'$; it is the zero map otherwise). Existing proofs for this relation make use of Schur's lemma [e.g. 66], but a diagrammatic proof has not been given without relying on this result. This is perhaps indicative of the difficulties of the graphical languages currently available in the LQG literature.

In the following proposition, we give a remarkably intuitive argument for the proportionality between the 2-loop and the identity (or zero map) by way of permutations, taking Def. 2 as a starting point. For an analysis of the structure of arbitrary spin networks, particularly as we manipulate them into normal forms, this may be a useful result on its own.

**Proposition 57** (2-Loop Removal by Permutations)**.** *Writing (G13) in the PSC,*

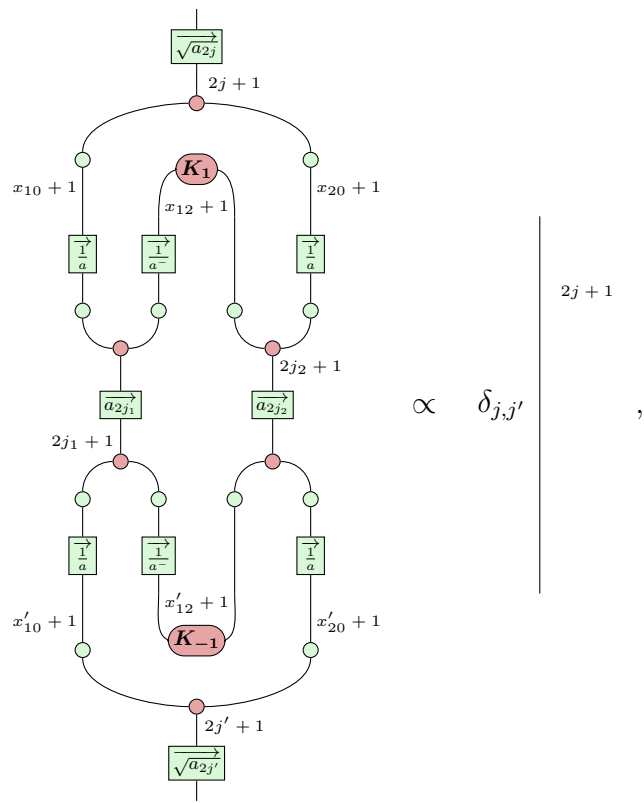

*where we use the subscript* 0 *to denote the spin* $j$ *and add a prime to denote that* $j'$ *is used in place of* $j$ *in the wire dimensions. For example,* $x_{10} = j + j_1 - j_2$ *and* $x'_{10} = j' + j_1 - j_2.$

*Proof.* First, let $j = j'$. Start by breaking open the $1/a$ boxes and cap/cup to reveal a pair of qubit symmetrisers with $x_{12}$ internal legs each involved in shared (mutually non-interacting) loops;

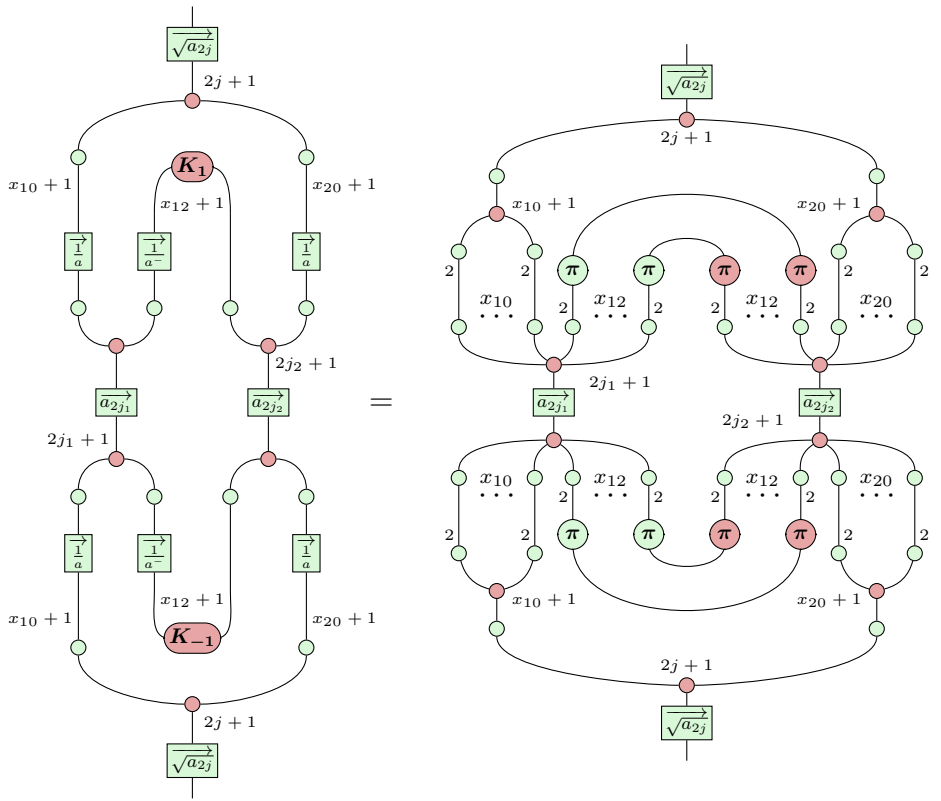

Since these loops do not interact with one another beyond their involvement in the symmetrisers, we can, without loss of generality, look at the simplest setting for only a single set of internal loops (i.e. letting $x_{12} = 1$) to observe how a loop affects the diagram at-large.

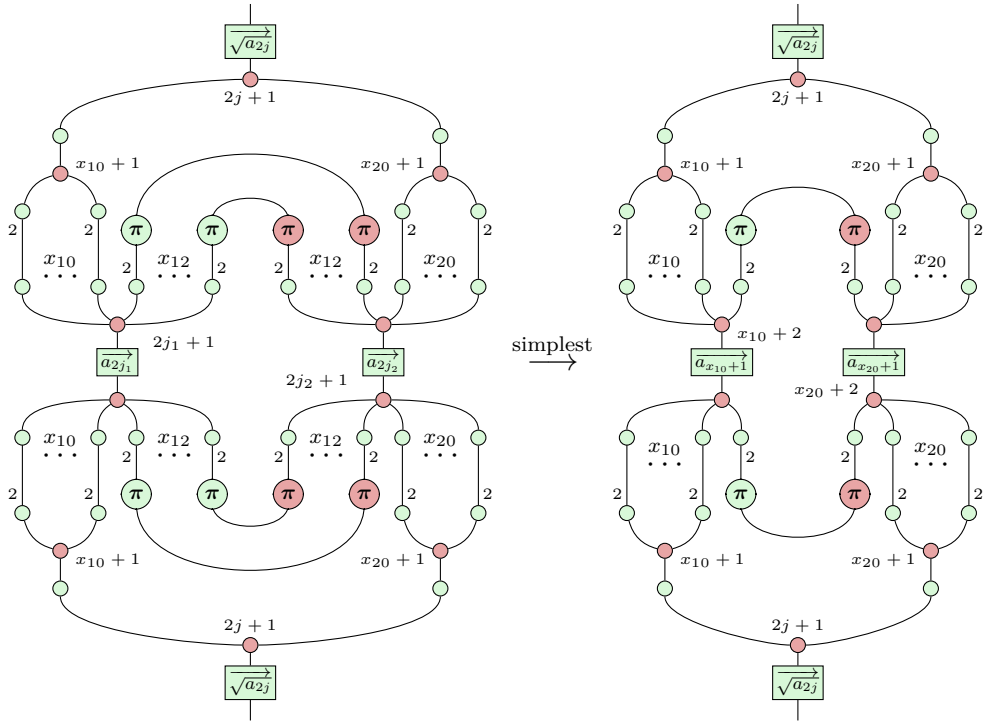

Consider the following sub-diagram, written in the defining view of symmetrisers as sums over permutations $P_{i_x}^{(x)}$ of qubit wires,

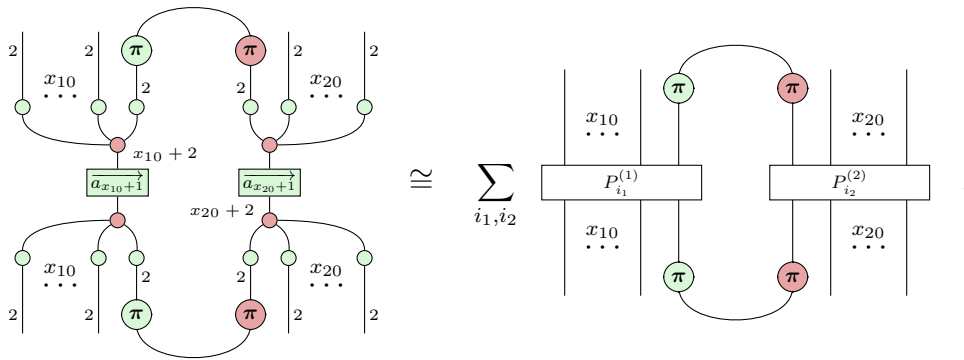

There are three distinct cases for these permutations to handle in each summand:

Case 1. Both $P^{(1)}$ and $P^{(2)}$ leave the loop-wires unswapped.

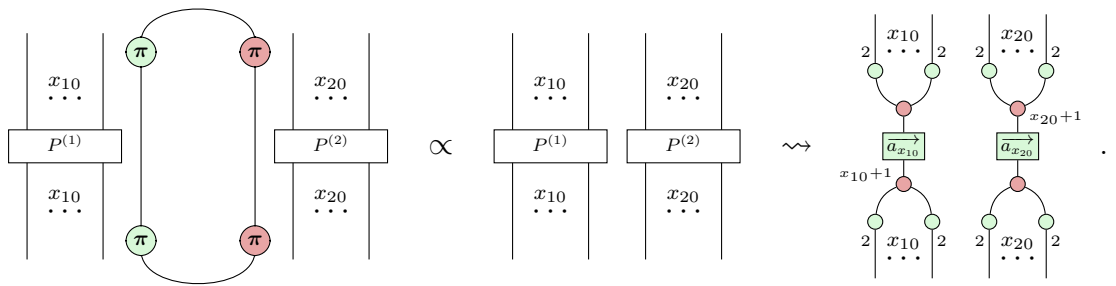

Case 2. Both $P^{(1)}$ and $P^{(2)}$ swap the loop-wires with some other wires.

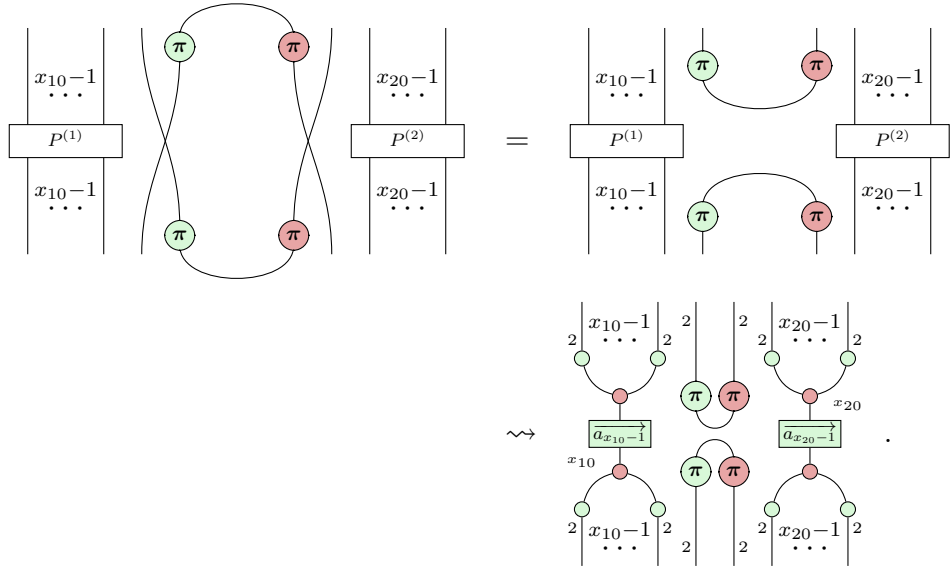

Case 3a. $P^{(1)}$ leaves the left loop-wire unswapped, and $P^{(2)}$ swaps the right loop-wire with some other wire (from the $x_{20} - 1$ right wires).

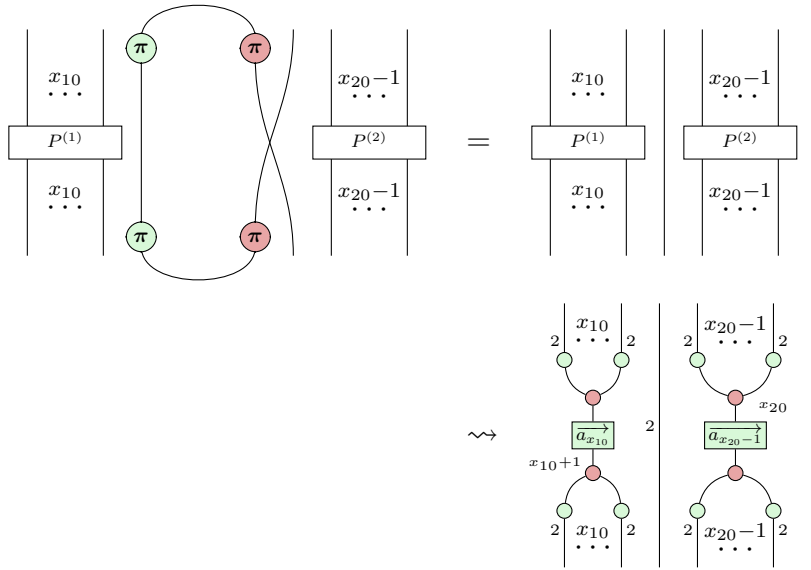

Case 3b. $P^{(1)}$ swaps the left loop-wire with some other wire (from the $x_{10} - 1$ left wires), and $P^{(2)}$ leaves the right loop-wire unswapped.

We can now substitute each of these cases into the context of the full (simplest) diagram to observe that each summand yields the identity wire or zero map. The easiest way to see this is by reforming and fusing the $(1/a)$-boxes by lemma 37.

Case 1. Immediately follows from Prop. 44 once the $(1/a)$-boxes are formed;

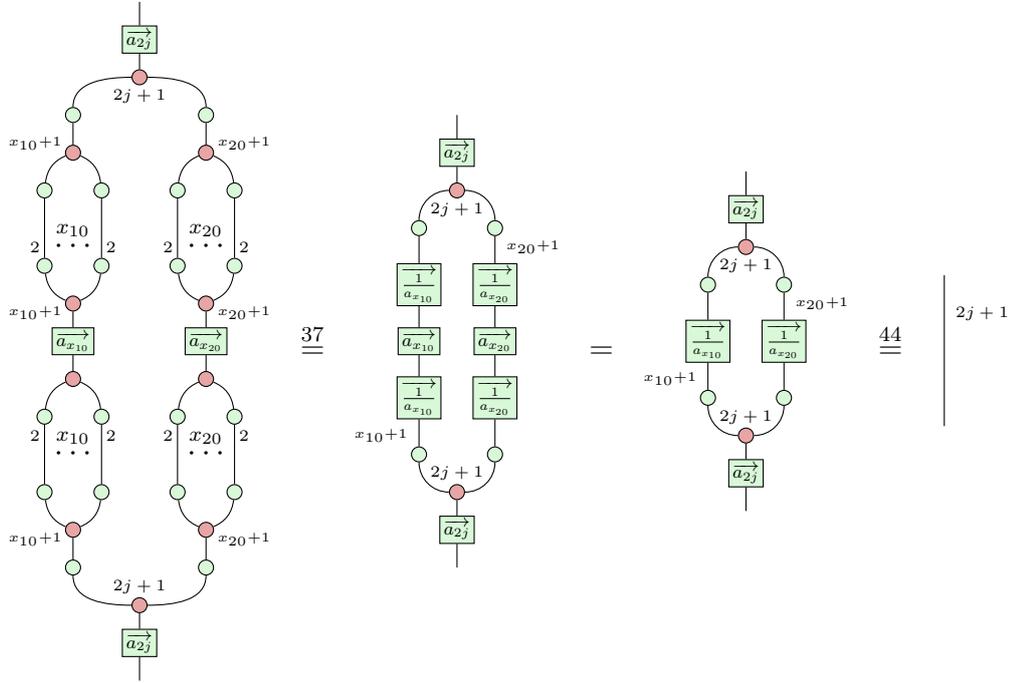

Case 2. This introduces cups/caps that, by 41, yields the zero map;

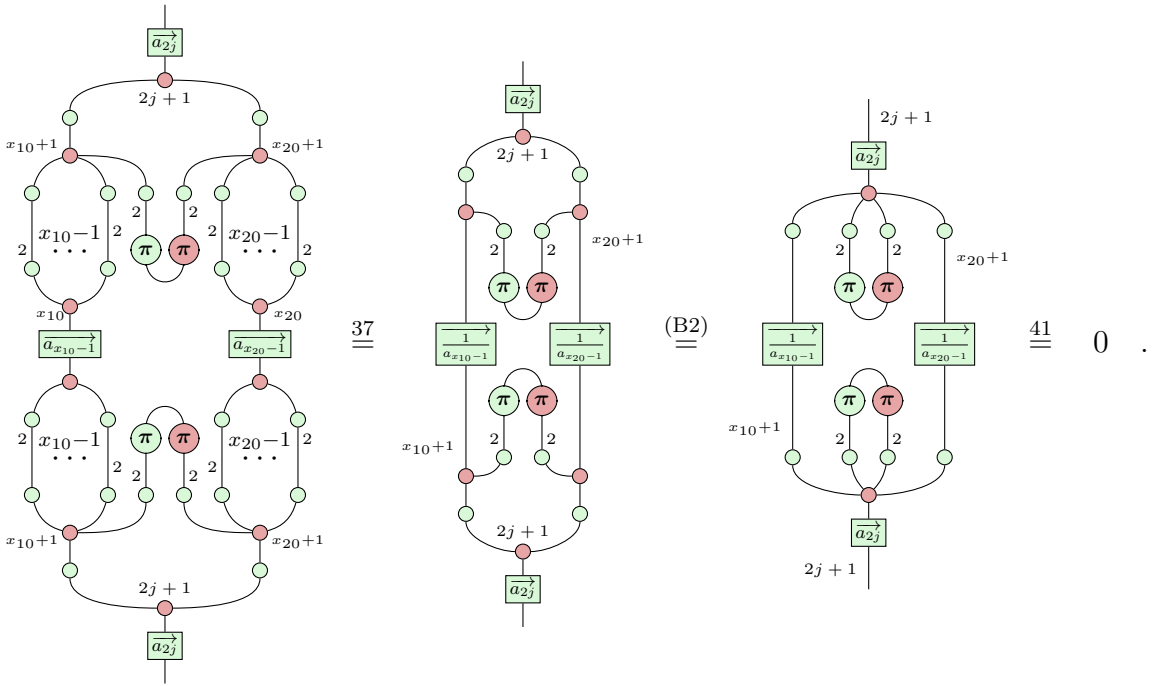

Case 3a. Similar to case 1, but first requires integrating the identity wire with one of the $(1/a)$-boxes;

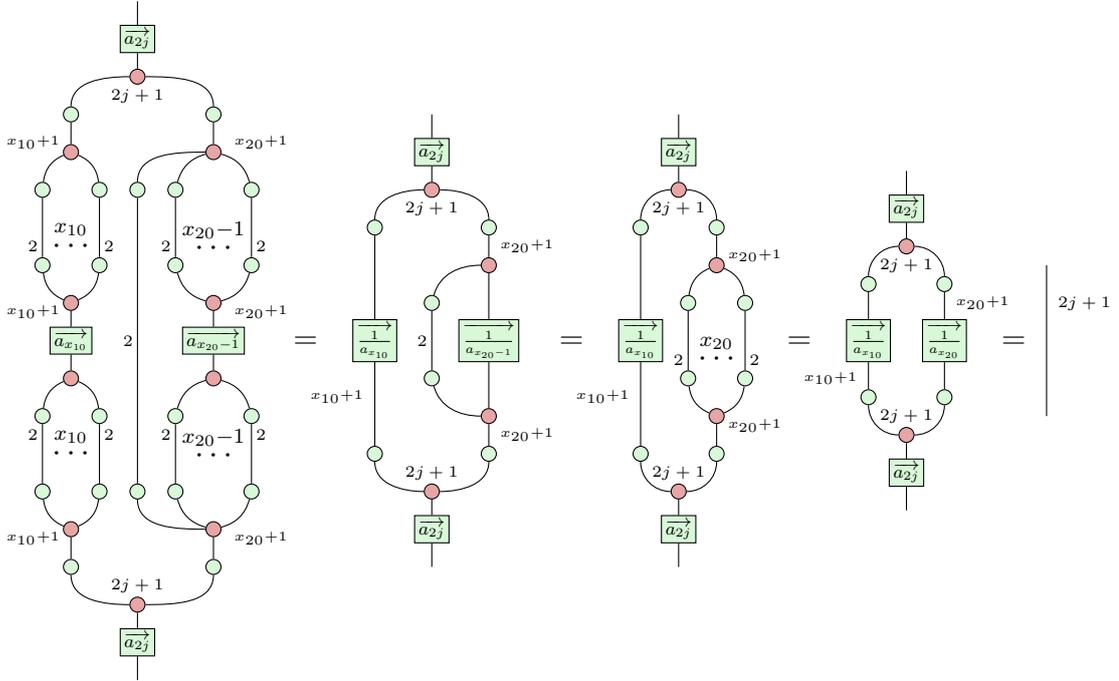

Case 3b. Symmetrically to 3a.

Since every summand either reduces the overall diagram to an identity wire or yields the zero map, we can conclude that, for $j = j'$, the full diagram must be identity, up to scalar.

Now let $j \neq j'$, and without loss of generality, suppose that $j > j'$ (the argument is symmetric for $j < j'$). We are now expecting the zero map, which will result from the imbalance of caps/cups introduced by mismatched in- and out-spins (similar to case 2 above, though asymmetric).

Take the full diagram as before, and look at the setting wherein only a single excessive cap is present on the top side of the diagram (i.e. ignore all caps completing an internal loop, and letting $x_{12} - x'_{12} = 1$), which we can see via $x_{12} = -j + j_1 + j_2 < -j' + j_1 + j_2 = x'_{12}$. Whenever $j \neq j'$, an excess will always be present on one side.

And again, consider the sub-diagram, written in terms of permutations,

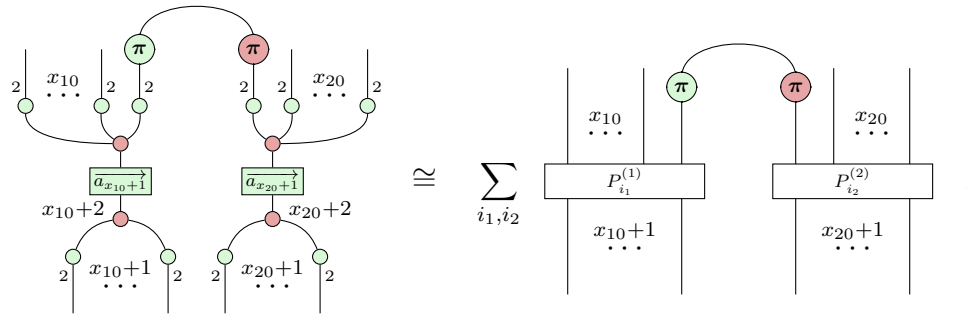

There is only one case to handle here: that of the cap passing to any two output wires, which invariably produces a cap output,

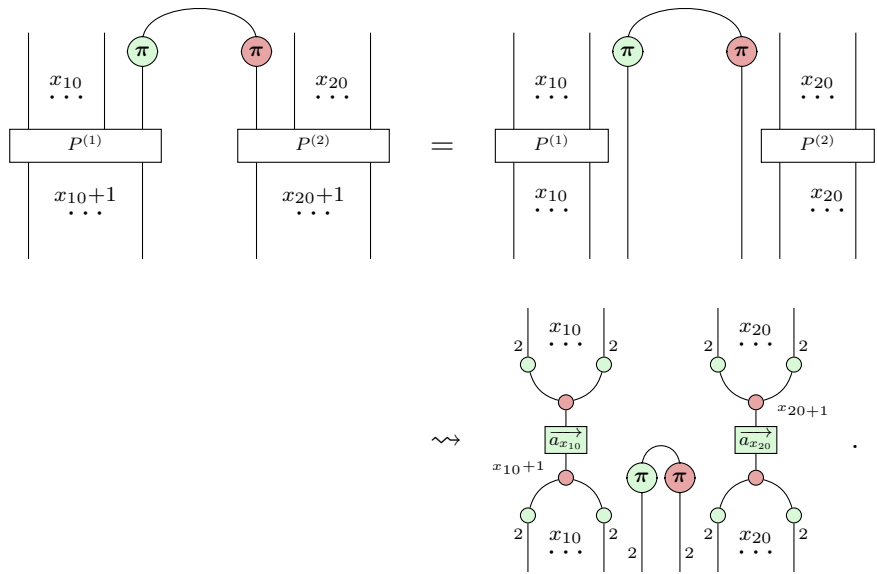

which leads to a zero map in the same argument as with case 2,

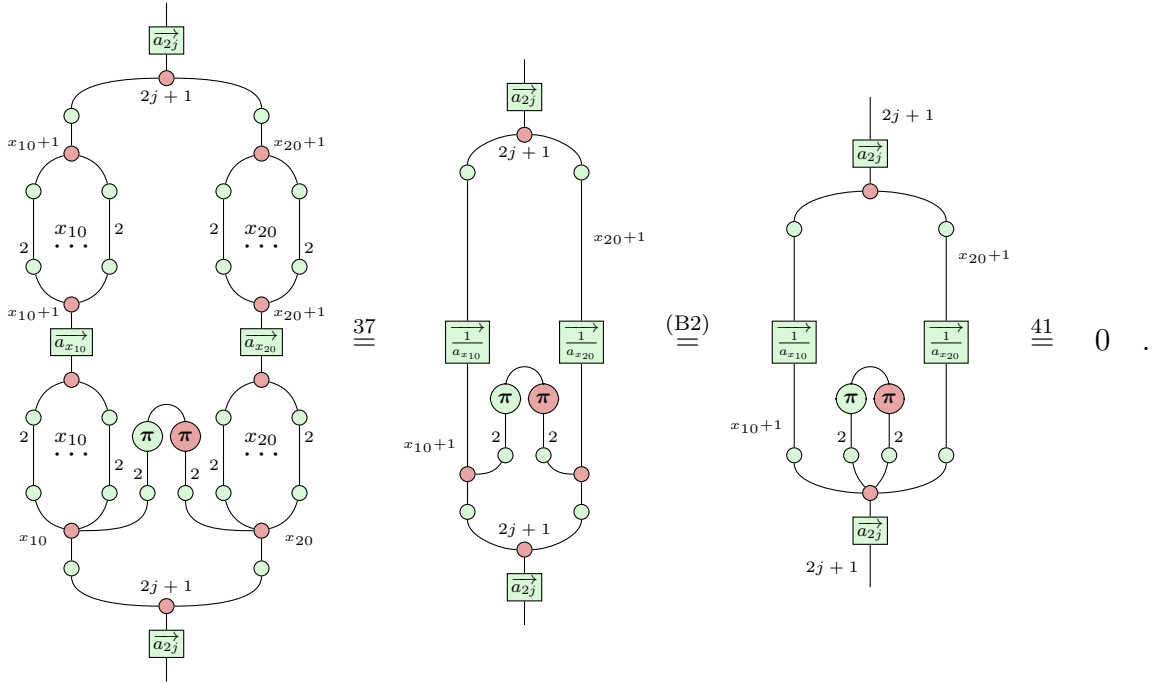

So, since every summand reduces the overall diagram to the zero map, we can conclude that, for $j \neq j'$, the full diagram must be the zero map.

Hence, the translation as given holds for arbitrary $j, j', j_1, j_2 \in \mathbb{N}/2$.

$\square$

Next, we give a direct diagrammatic rewrite of the ZX-diagram by splitting the singlet states and symmetrisers into sums of basis states. Following the above proposition to leverage our knowledge of its proportionality to the identity (or otherwise relying on results from Schur's lemma), we can then give a proof of the equality of (G13) by finding the constant. We will not reprove the $\delta_{j,j'}$, as the above is sufficient for this detail.

**Theorem 58** (Exact 2-Loop Removal). *Again writing (G13) in the PSC,*

$$= \frac{N(j, j_1, j_2)^2}{2j + 1} \; \delta_{j, j'} \quad \Bigg|^{2j+1} \quad ,$$

*Proof.* Start by splitting the singlets, at the $(1/a^-)$-boxes, into sums of basis states,

Then give the same treatment to the symmetriser, at the $\sqrt{a}$-boxes, and push the X-spiders corresponding to the states/costates together,

Now we should look to bound the indices on these X-spiders:

a) *Bound for $s - i$.* Firstly, $i$ is simply bound as $0 \le i \le x_{12}$, thus $-x_{12} \le -i \le 0$. Secondly, $s$ is similarly bound as $0 \le s \le 2j_1$. Hence, taken together,

$$-x_{12} \le s - i \le 2j_1 \quad .$$

We can restrict this even further. Note that if $-x_{12} \le s - i < 0$, then $K_{s-i} = K_{2j_1+1+s-i}$, so we can add $2j_1 + 1$ throughout the inequality,

$$-x_{12} \le s - i < 0 \implies 2j_1 + 1 - x_{12} = x_{10} + 1 \le 2j_1 + 1 + s - i < 2j_1 + 1 \quad ,$$

which gives,

Therefore, we need only consider the case where $0 \le s - i \le x_{10}$.

*b)* *Bound for* $t + i - x_{12}$. Now we have $-i$ bounded as $-x_{12} \leq -i \leq 0$, thus $-x_{12} \leq i - x_{12} \leq 0$, and $t$ bounded as $0 \leq t \leq 2j_2$. Hence, taken together,

$$-x_{12} \leq t + i - x_{12} \leq 2j_2 \quad .$$

And again, if $-x_{12} \leq t + i - x_{12} \leq 0$, then $K_{t+i-x_{12}} = K_{2j_2+1+t+i-x_{12}}$, so we can add $2j_2 + 1$ throughout the inequality,

$$-x_{12} \leq t + i - x_{12} \leq 0 \implies 2j_2 + 1 + 2 - x_{12} = x_{20} + 1 \leq 2j_2 + 1 + t + i - x_{12} < 2j_2 + 1 \quad ,$$

which gives,

therefore, we need only consider the case where $0 \leq t + i - x_{12} \leq x_{20}$.

*c)* Similarly, we only need consider the cases where $0 \leq s - \ell \leq x_{10}$ and where $0 \leq t + \ell - x_{12} \leq x_{20}$ (do not get tripped up by the negation of the phase on the spider as we move between states and costates!).

*d)* Taking these bounds together gives,

$$0 \leq s + t - x_{12} \leq x_{10} + x_{20} = 2j \quad ,$$

which is what we are expecting for our map.

With these bounds noted, we can finish pushing our X-spider states/costates all the way through the diagram, picking up the coefficients on the $(1/a)$- and

$\sqrt{a}$-boxes on their path via (K0), eventually yielding,

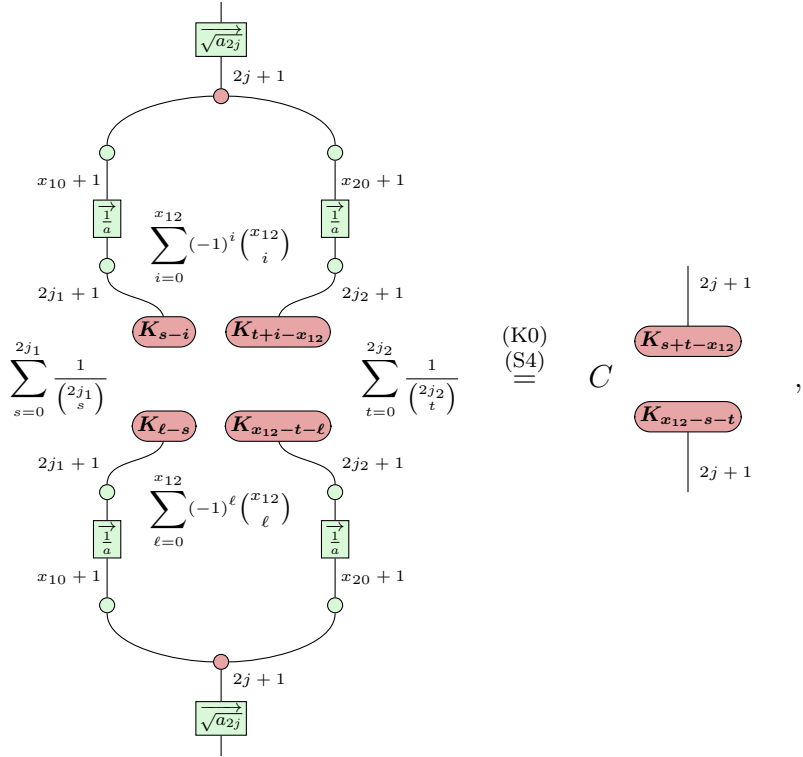

where the coefficient $C$ is given by,

$$C = \sum_{s=0}^{2j_1}\sum_{t=0}^{2j_2}\sum_{i=0}^{x_{12}}\sum_{\ell=0}^{x_{12}} \frac{(-1)^i(-1)^\ell\binom{x_{12}}{i}\binom{x_{12}}{\ell}\binom{x_{10}}{s-i}\binom{x_{20}}{t-x_{12}+i}\binom{x_{10}}{s-\ell}\binom{x_{20}}{t-x_{12}+\ell}}{\binom{2j_1}{s}\binom{2j_2}{t}\binom{2j}{s+t-x_{12}}} .$$

Note that, given $s + t$, the coefficient $C$ must be constant. Let $s + t = x_{12}$, and denote the corresponding value of $C$ by $C_0$. Then,

$$t - x_{12} + i = i - s \qquad \text{and} \qquad t - x_{12} + \ell = \ell - s .$$

For the binomials to be meaningful, we should have $s \geq i$ and $i \geq s$, hence $s = i$ (similarly, $s = \ell$). Hence, the coefficient simplifies to,

$$C_0 = \sum_{s=0}^{x_{12}} \binom{x_{12}}{s}\binom{x_{12}}{s} \frac{1}{\binom{2j_1}{s}\binom{2j_2}{x_{12}-s}} .$$

For the remainder of the proof, we will show that $C_0 = N(j, j_1, j_2)^2/(2j + 1)$. First, observe

$$C_0 = \frac{N(j_1, j_2, j_3)^2}{2j + 1} \iff \sum_{s=0}^{x_{12}} \frac{x_{12}!}{s!(x_{12}-s)!} \frac{x_{12}!}{s!(x_{12}-s)!} \frac{s!(2j_1-s)!}{(2j_1)!} \frac{(x_{12}-s)!(x_{20}+s)!}{(2j_2)!}$$

$$= \frac{(j_1 + j_2 + j + 1)!(-j_1 + j_2 + j)!(j_1 - j_2 + j)!(j_1 + j_2 - j)!}{(2j_1)!(2j_2)!(2j + 1)!}$$

$$= \frac{(j_1 + j_2 + j + 1)!(x_{20})!(x_{10})!(x_{12})!}{(2j_1)!(2j_2)!(2j + 1)!}$$

$$\iff \sum_{s=0}^{x_{12}} \frac{x_{12}!(2j_1 - s)!(x_{20} + s)!}{s!(x_{12} - s)!} = \frac{(j_1 + j_2 + j + 1)!(x_{20})!(x_{10})!}{(2j + 1)!}$$

$$\iff \sum_{s=0}^{x_{12}} \frac{(2j_1 - s)!(x_{20} + s)!}{s!(x_{12} - s)!} = \frac{(j_1 + j_2 + j + 1)!(x_{20})!(x_{10})!}{(2j + 1)!(x_{12})!} \quad .$$

Next, from Peterson [66, page 50], we have,

$$\sum_{u+v=j} \binom{a + b - j}{b - v}(a+u)!(b+v)! = \binom{a + b + j + 1}{2j + 1}(a+j-b)!(b+j-a)!(a+b-j)! \quad ,$$

hence, we can write,

$$\sum_{u+v=j} \frac{(a + b - j)!}{(b - v)!(a + v - j)!}(a + u)!(b + v)!$$
$$= \frac{(a + b + j + 1)!}{(2j + 1)!(a + b - j)!}(a + j - b)!(b + j - a)!(a + b - j)!$$

$$\iff \sum_{u+v=j} \frac{(b + v)!(a + u)!}{(b - v)!(a - u)!} = \frac{(a + b + j + 1)!(a + j - b)!(b + j - a)!}{(2j + 1)!(a + b - j)!}$$

So, finally, let $a = j_1$ and $b = j_2$, and also let $a - u = s$ and $b - v = x_{12} - s$. Then,

$$u = a - s = j_1 - s, v = b - x_{12} + s = j_2 - x_{12} + s = j - j_1 + s,$$

$$u + v = j_1 - s + j - j_1 + s = j,$$

$$a + u = j_1 + j_1 - s = 2j_1 - s, b + v = j_2 + j - j_1 + s = x_{20} + s \quad ,$$

and hence the if and only if condition is satisfied,

$$\sum_{s=0}^{x_{12}} \frac{(2j_1 - s)!(x_{20} + s)!}{s!(x_{10} - s)!} = \sum_{u+v=j} \frac{(b + v)!(a + u)!}{(b - v)!(a - u)!} = \frac{(j_1 + j_2 + j + 1)!(x_{20})!(x_{10})!}{(2j + 1)!(x_{12})!}$$

We can further observe, for completeness sake, that when we include the $\sqrt{2j + 1}/N(j, j_1, j_2)$ normalisation terms for both injections, they multiply together to cancel the $N(j, j_1, j_2)^2/(2j + 1)$, leaving strict identity (together with the Kronecker delta). $\qquad\square$

### 4.1.1   Rewriting 2-Loops with Qubits

Part of the beauty of a diagrammatic language like the finite-dimensional ZX-calculus is its freedom. Having established, with Prop. 57 and Thm. 58, that the bubble identity holds true in our ZX formalism, we can redraw 2-loops in other useful ways. An especially helpful back-and-forth is the relation between the *qubit* and *qudit* 3-valent node (the former exposing an intuitive view written with symmetrises are groupings of wires; the latter giving an efficient and concise description for direct proofs). We have already seen both used, but let us now give a direct relation between them. The following lemma will ease the proof.

**Lemma 59.**

where $\overrightarrow{1/a^-} = \left( (-1)^1 \binom{d-1}{1}, \ldots, (-1)^{d-1} \binom{d-1}{d-1} \right)$. *This also holds for* $\overrightarrow{\sqrt{1/a^-}}$.

*Proof.*

using that,

$$
\begin{aligned}
k_1 \left( \overrightarrow{\frac{1}{a^-}} \right) &= (-1)^{-(d-1)} \left( (-1)^0 \binom{d-1}{0}, \ldots, (-1)^{d-2} \binom{d-1}{d-2} \right) \\
&= \left( (-1)^{-(d-1)} \binom{d-1}{d-1}, \ldots, (-1)^{-1} \binom{d-1}{1} \right) \\
&= \left( (-1)^{d-1} \binom{d-1}{d-1}, \ldots, (-1)^1 \binom{d-1}{1} \right) = \overleftarrow{\frac{1}{a^-}} \quad .
\end{aligned}
$$

□

**Proposition 60** (Qudit vs Qubit Injection)**.**

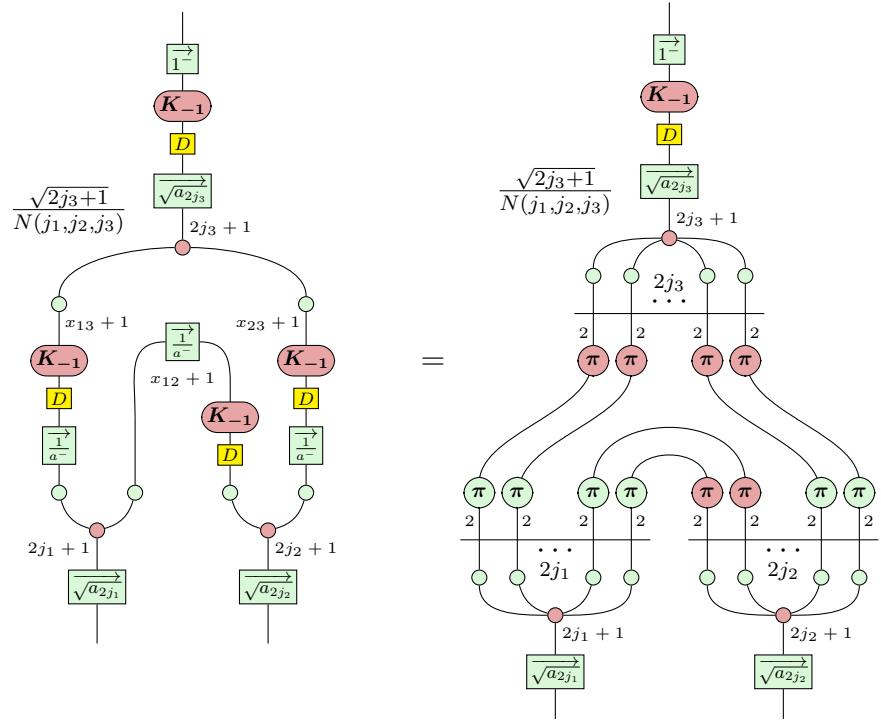

*Proof.*

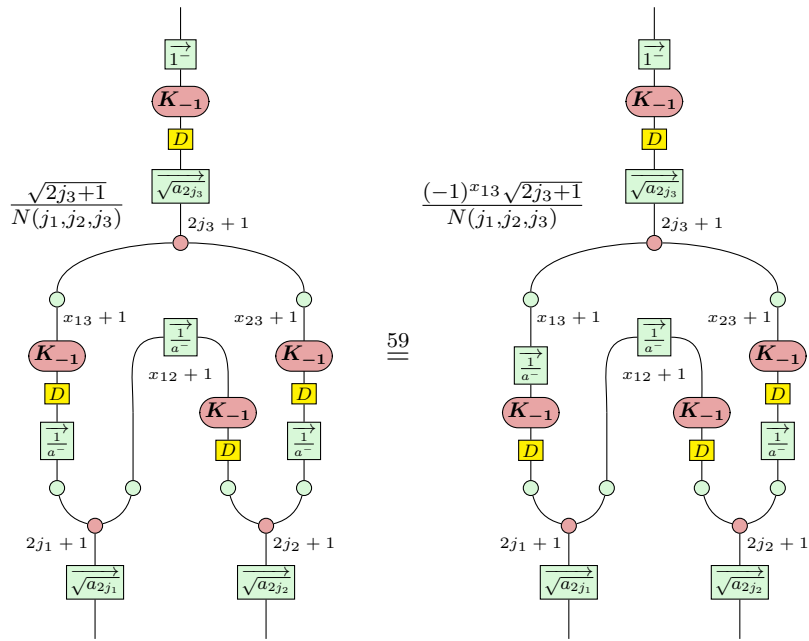

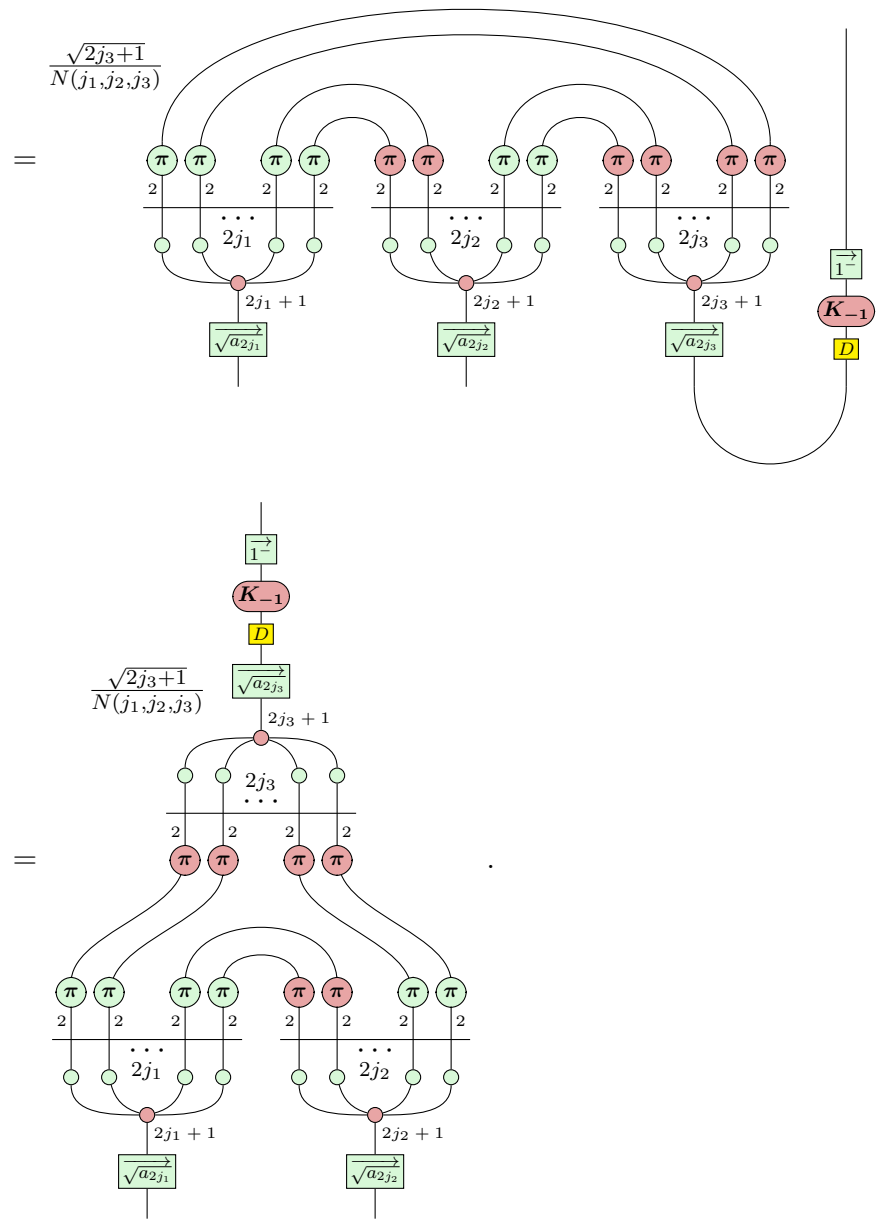

Using the result of Prop. 60, we can rewrite the bubble with this intuitive qubit perspective (again, setting aside the proven Kronecker delta),

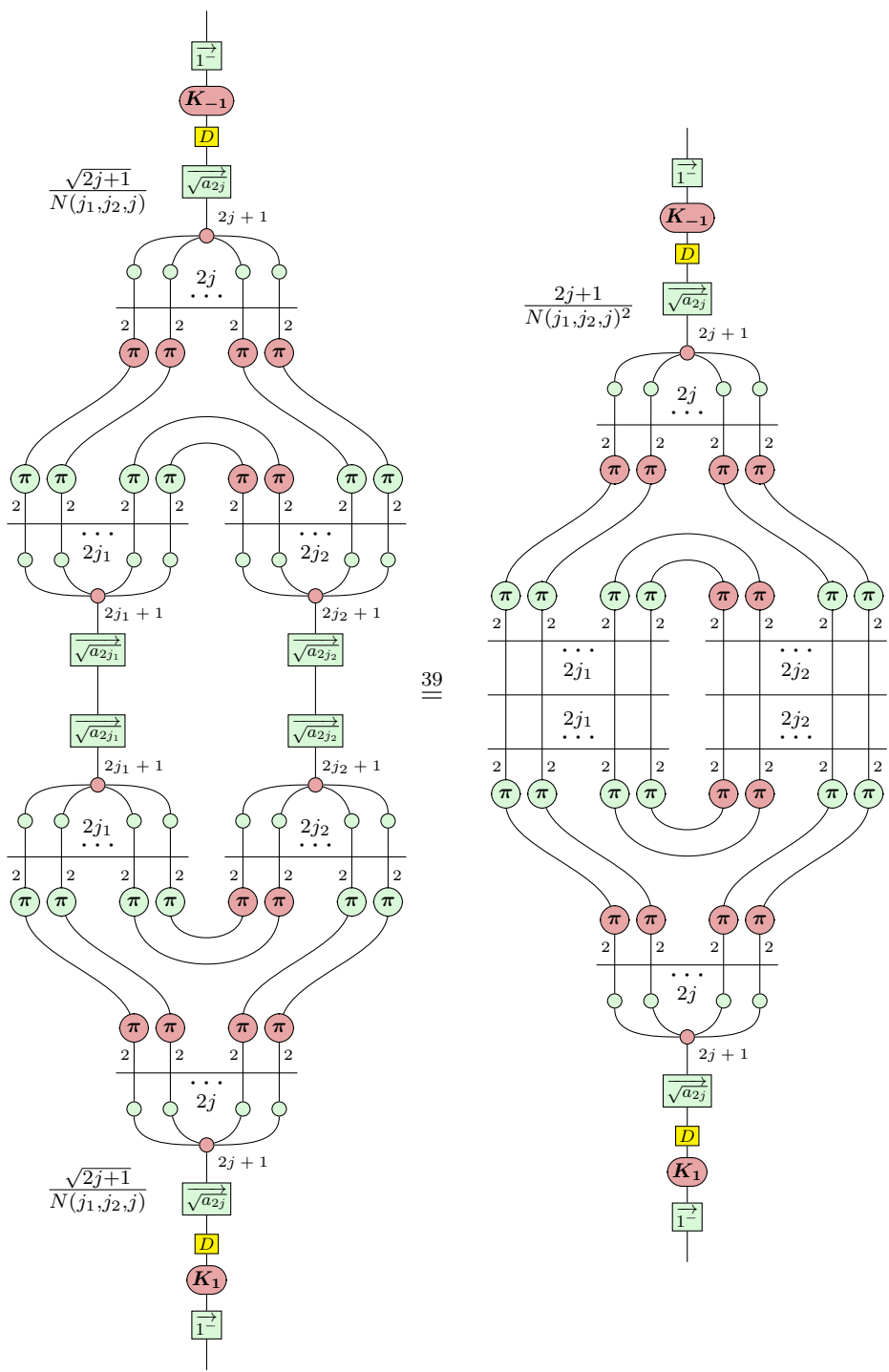

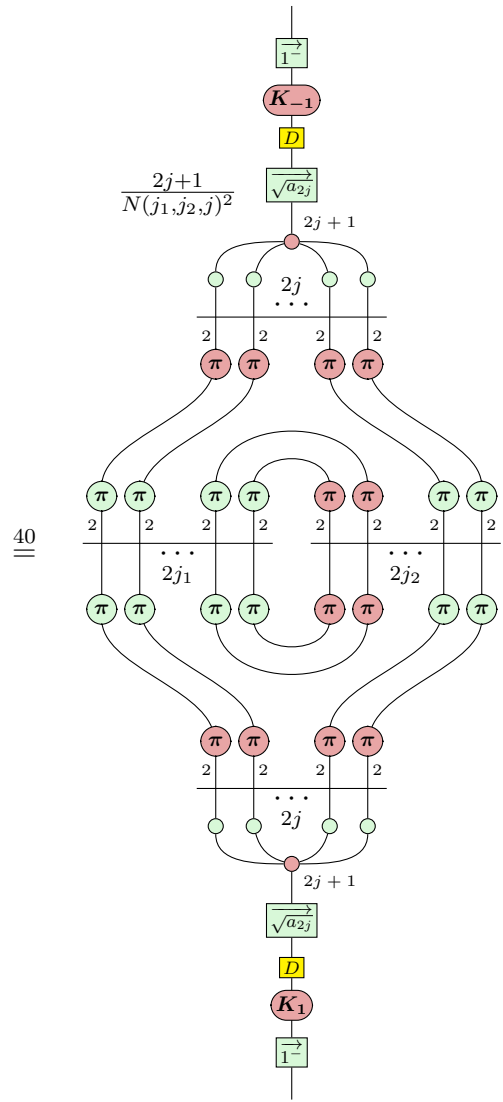

which in turn, by Thm. 58, implies that,

$$\frac{2j+1}{N(j_1,j_2,j)^2} \qquad = \qquad . \qquad (4.1)$$

### 4.1.2 Θ-Graphs as Special Case Bubbles

Being able to write the symmetriser in the form of Eq. (4.1) is extremely useful to us. For example, it leads immediately to a diagrammatic representation for the Θ-graph of (G12) in the PSC that holds without reliance on Schur's lemma; using a similar idea as with Peterson [66] (proposition 3.25) to represent the Θ-graph as the trace of a bubble identity, we can write an analogue to (G12) as,

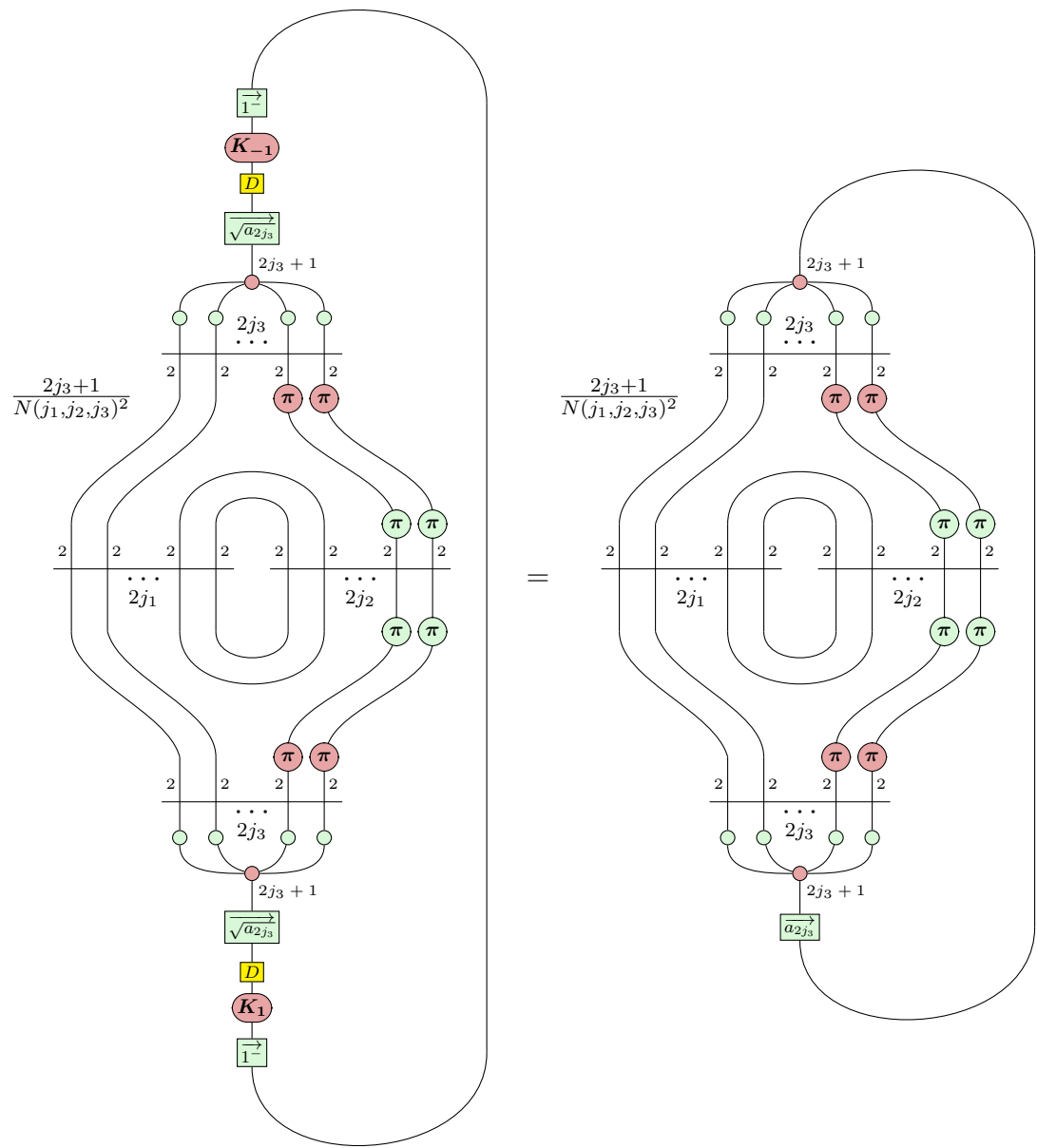

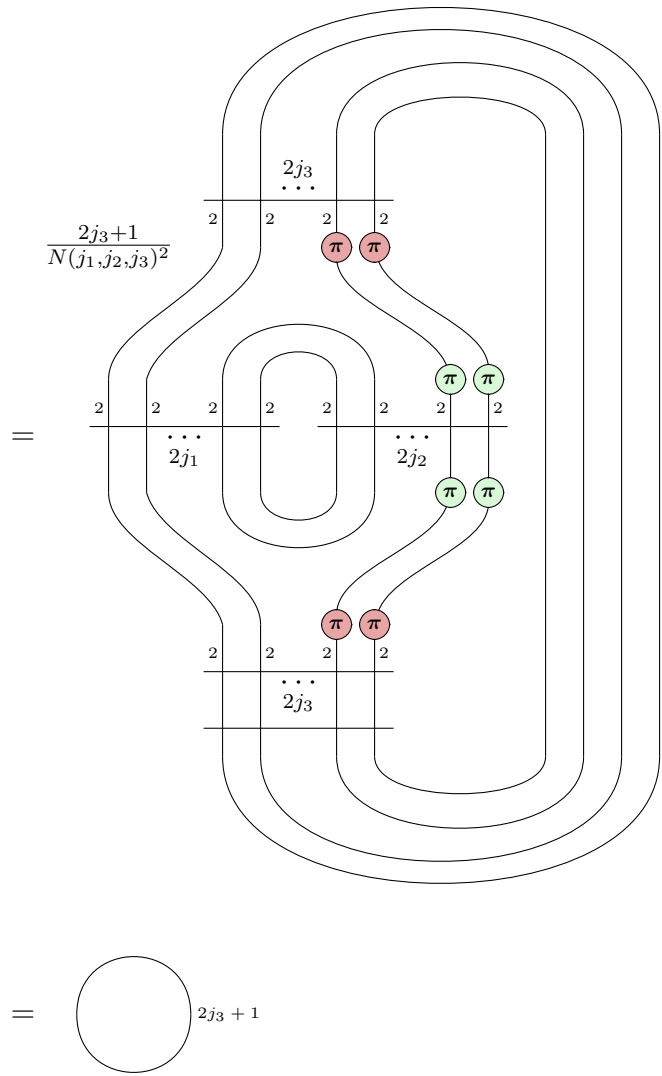

$$= \quad 2j_3 + 1 \quad ,$$

where the penultimate equality is using the implication (of Thm. 58) above, and hence we have that,

$$\Theta(j_1, j_2, j_3) \quad = \quad$$  $$\quad = \quad N(j_1, j_2, j_3)^2 \quad .$$

Drawing this in the stripped-back Yutsi-style clearly presents the relationship between the bubble and the $\Theta$-graph,

$$\Theta(j_1, j_2, j_3) \quad := \quad$$  $$,$$

ignoring scalars (which are not needed for our discussion).

**2-Loop Removal and the Θ-Graph: Summary**

- Using an intuitive permutation-based argument (Prop. 57), we can show that a 2-loop ("bubble") is proportional an identity (or a zero map if the input and output spins differ).

- By breaking $a$-boxes into sums over basis states/costates, we can determine the exact scalar obtained by removing the bubble (Thm. 58). Up to normalisation, we find the correct scalar, as in (G13).

- Taking the trace over a bubble presents as the Θ-graph (subsection 4.1.2). Using 2-loop removal, we can determine the exact value of the Θ-graph as a scalar (a.k.a. the "Θ-coefficient"), which concurs, up to normalisation, with (G12).

## 4.2  Tetrahedral Net Symbols and $6j$-Symbols

We begin our discussion of the $6j$-symbol by extending $\Theta$-graphs (or "$\Theta$-coefficients", in this context) to *tetrahedral coefficients*. In fact, we can appreciate that the tetrahedral coefficient is to the triple bubble (G28) as the $\Theta$-coefficient is to the bubble (G13), which becomes clear in the Yutsis-style,

$$\text{Tet}(j_1, j_2, j_3, j_4, j_5, j_6) \quad := \quad \text{} \quad ,$$

ignoring scalars (which are again not needed for our discussion). When drawn diagrammatically like this, we may refer to the "*tetrahedral net symbol*", in the same way we have been referring to the diagrammatic $\Theta$-coefficient as the $\Theta$-graph.

By Prop. 3.29 of Peterson [66], we can express this tetrahedral coefficient with respect to a $6j$-symbol,

$$\text{Tet}(j_1, j_2, j_3, j_4, j_5, j_6) \;=\; \left( \frac{\Theta(j_1, j_3, j_4)\Theta(j_4, j_5, j_6)}{2j_4 + 1} \right) \begin{Bmatrix} j_1 & j_3 & j_4 \\ j_6 & j_5 & j_2 \end{Bmatrix} \quad ,$$

where the $6j$-symbol itself is defined as a sum over $3j$-symbols,

$$\begin{Bmatrix} j_1 & j_2 & j_3 \\ j_4 & j_5 & j_6 \end{Bmatrix} := \sum_{m_1, \cdots, m_6} (-1)^{\sum_{i=1}^{6}(j_i - m_i)} \begin{pmatrix} j_1 & j_2 & j_3 \\ -m_1 & -m_2 & -m_3 \end{pmatrix} \begin{pmatrix} j_1 & j_5 & j_6 \\ m_1 & -m_5 & -m_6 \end{pmatrix}$$
$$\times \begin{pmatrix} j_4 & j_2 & j_6 \\ m_4 & m_2 & -m_6 \end{pmatrix} \begin{pmatrix} j_3 & j_4 & j_5 \\ m_3 & -m_4 & m_5 \end{pmatrix} \quad .$$

We will now give a novel representation of the $6j$-symbol in the PSC by using the above expression to rewrite the diagram for the $6j$-symbol as a tetrahedral net symbol. The proof will make use of the following lemma for commuting singlet states through 3-valent injections.

**Lemma 61.**

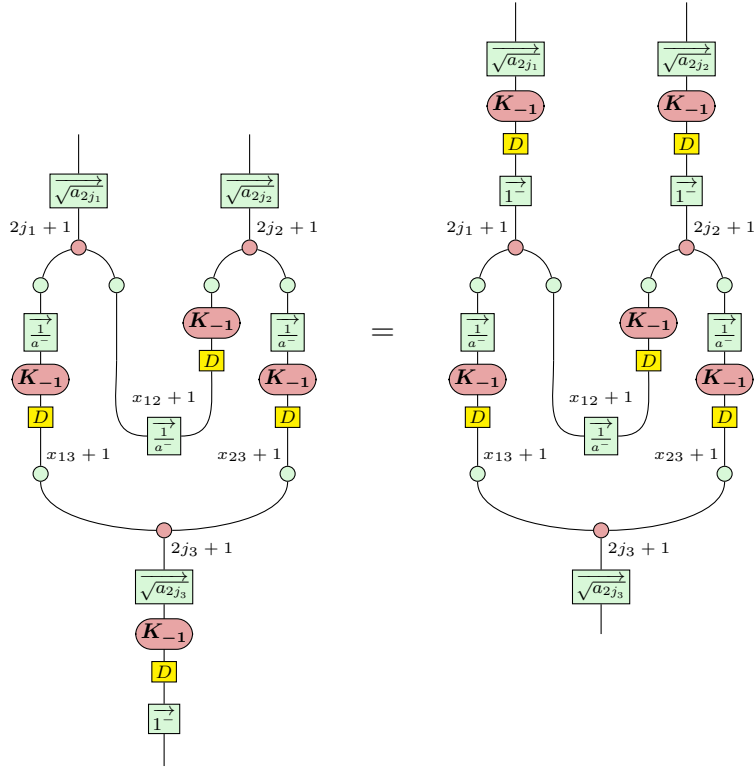

*Proof.*

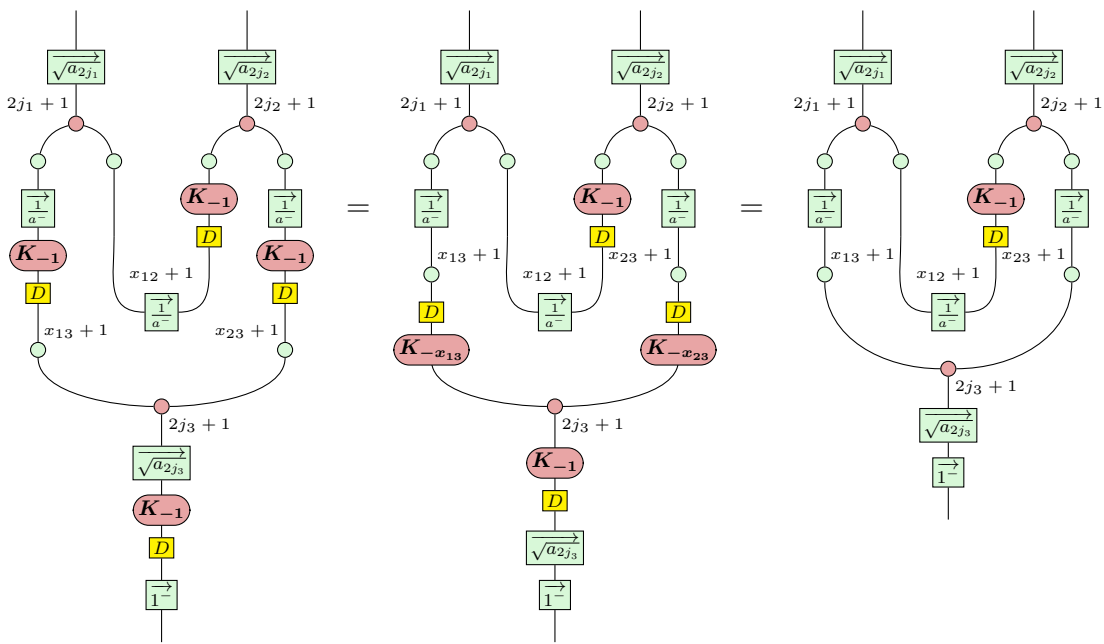

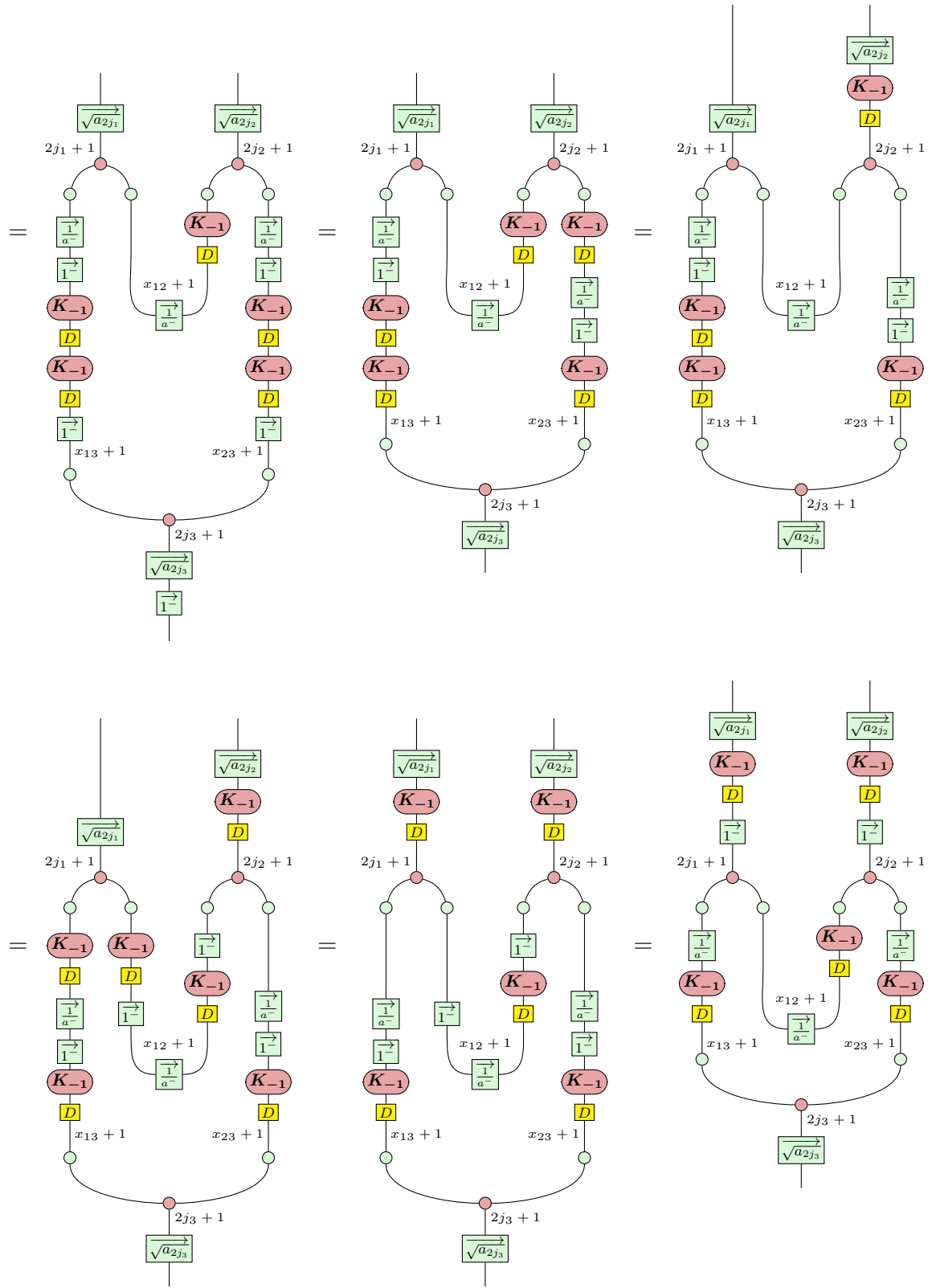

**Theorem 62** (6*j*-Symbol in the PSC)**.** *The Wigner* 6*j-symbol, for spins* $j_1, \ldots, j_6$,

*can be diagrammatically expressed in the PSC as,*

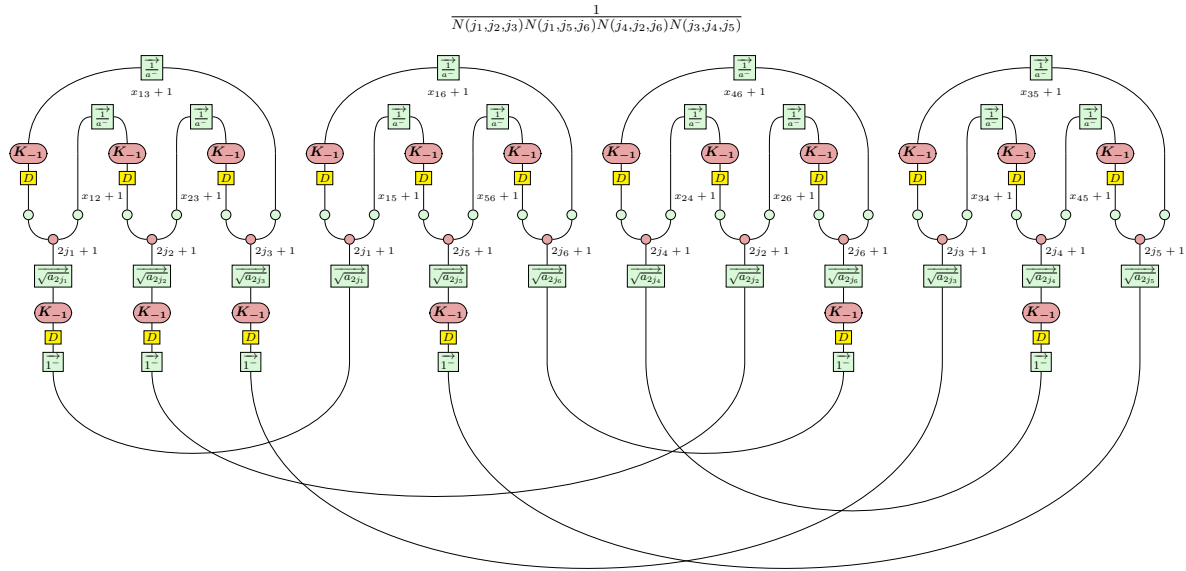

.

*Proof.* To prove that this is a valid representation for the $6j$-symbol, we can show that this diagram rewrites into a tetrahedral net symbol (with the appropriate normalisation), and thus, by Prop 3.29 of Peterson [66], is valid.

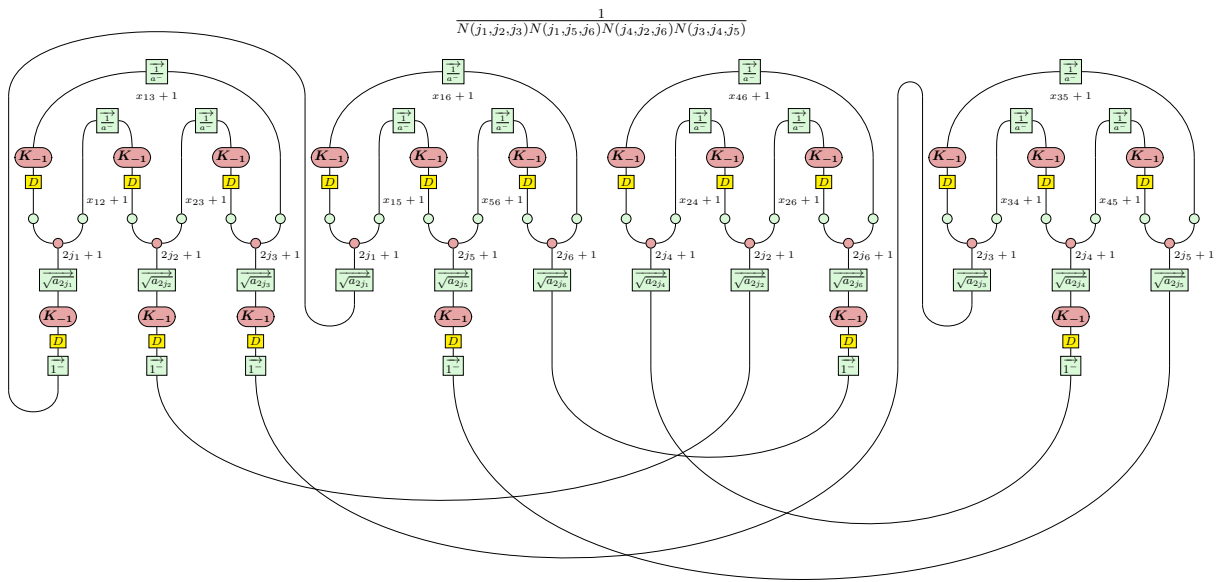

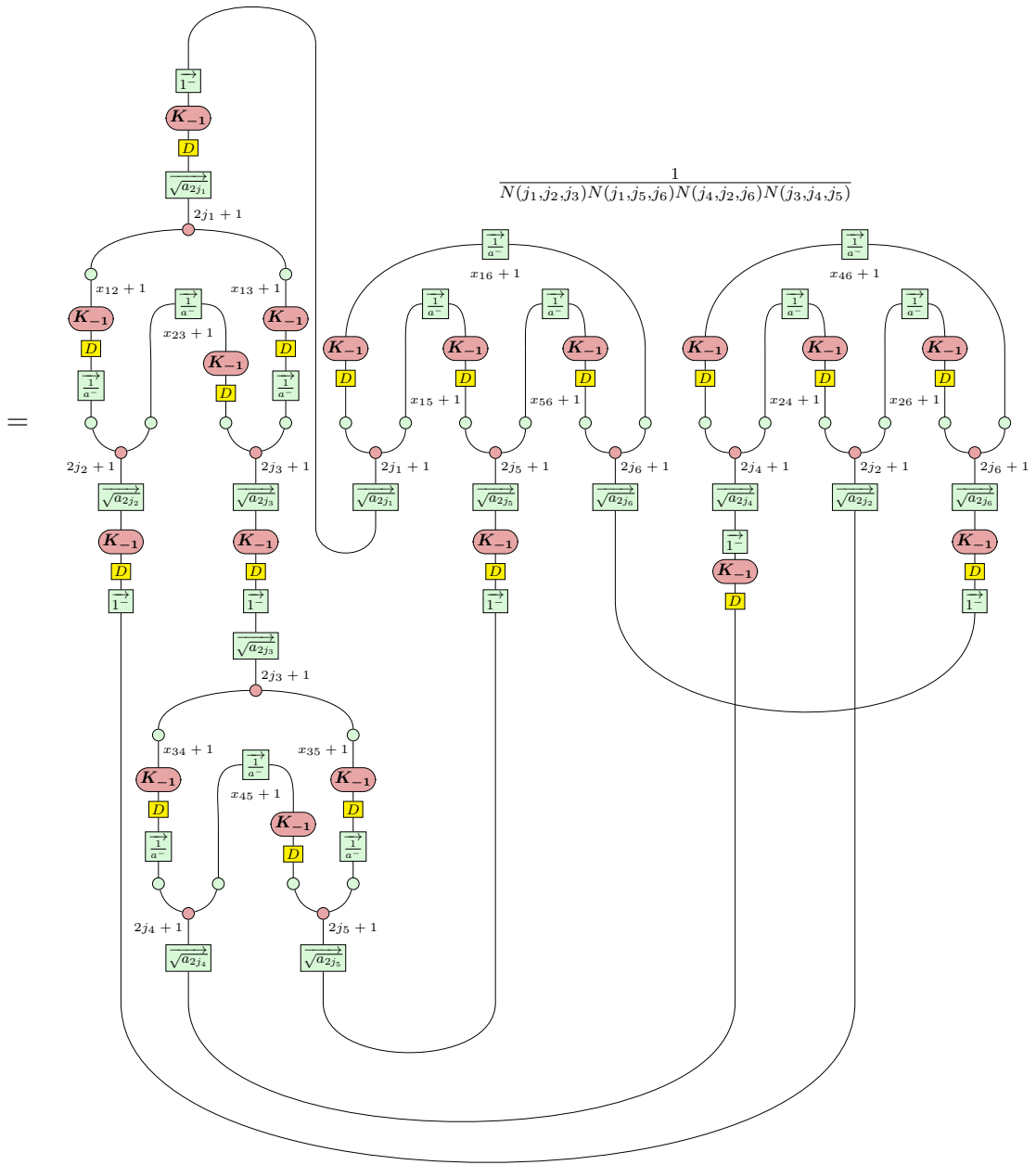

$$= \frac{1}{N(j_1,j_2,j_3)N(j_1,j_5,j_6)N(j_4,j_2,j_6)N(j_3,j_4,j_5)}$$

$$= \frac{1}{N(j_1,j_2,j_3)N(j_1,j_5,j_6)N(j_4,j_2,j_6)N(j_3,j_4,j_5)}$$

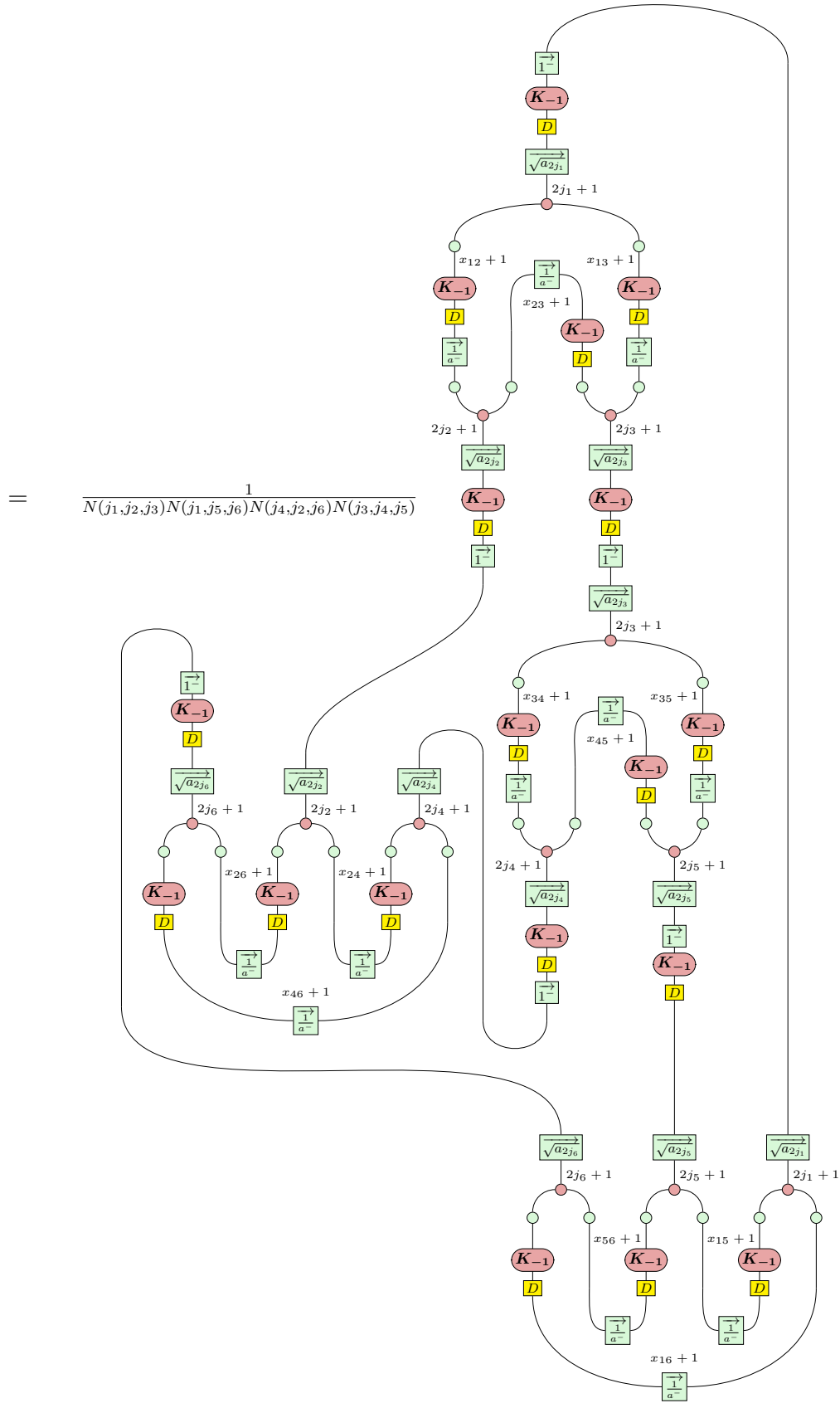

$$= \frac{1}{N(j_1,j_2,j_3)N(j_1,j_5,j_6)N(j_4,j_2,j_6)N(j_3,j_4,j_5)}$$

$$\overset{61}{=} \frac{1}{N(j_1,j_2,j_3)N(j_1,j_5,j_6)N(j_4,j_2,j_6)N(j_3,j_4,j_5)}$$ 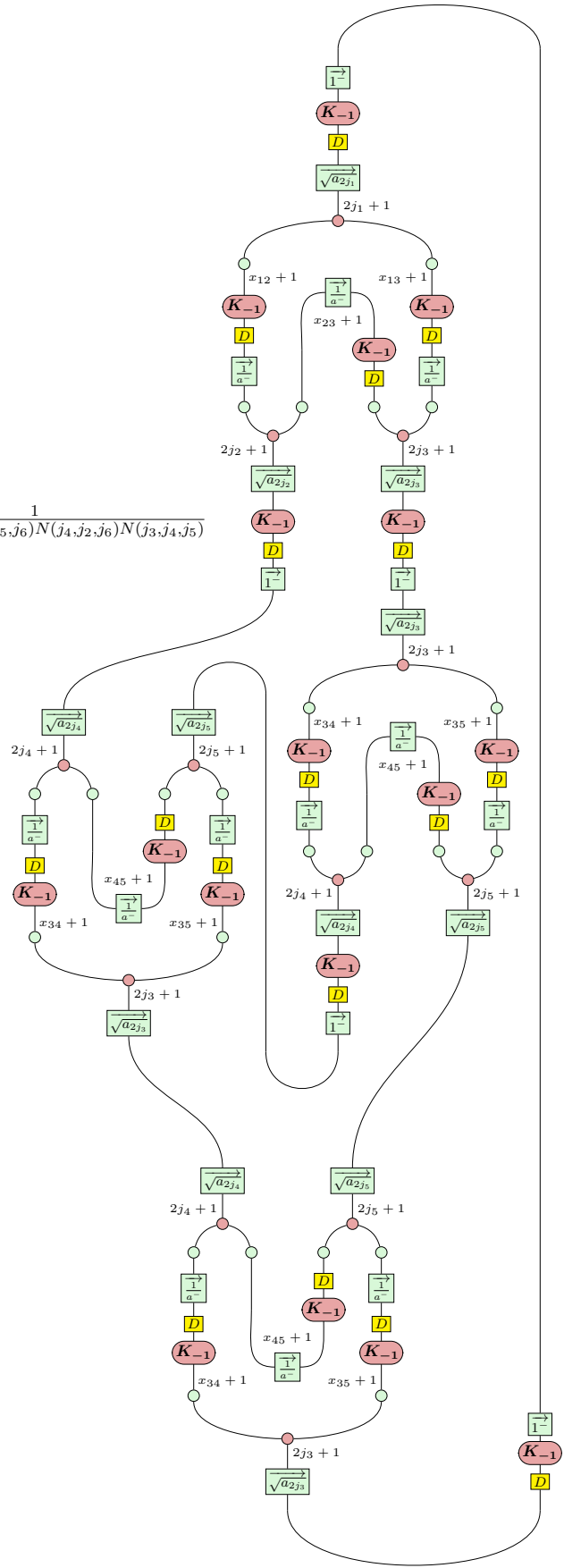

$$\stackrel{61}{=} \frac{(-1)^{2j_4}}{N(j_1,j_2,j_3)N(j_1,j_5,j_6)N(j_4,j_2,j_6)N(j_3,j_4,j_5)}$$

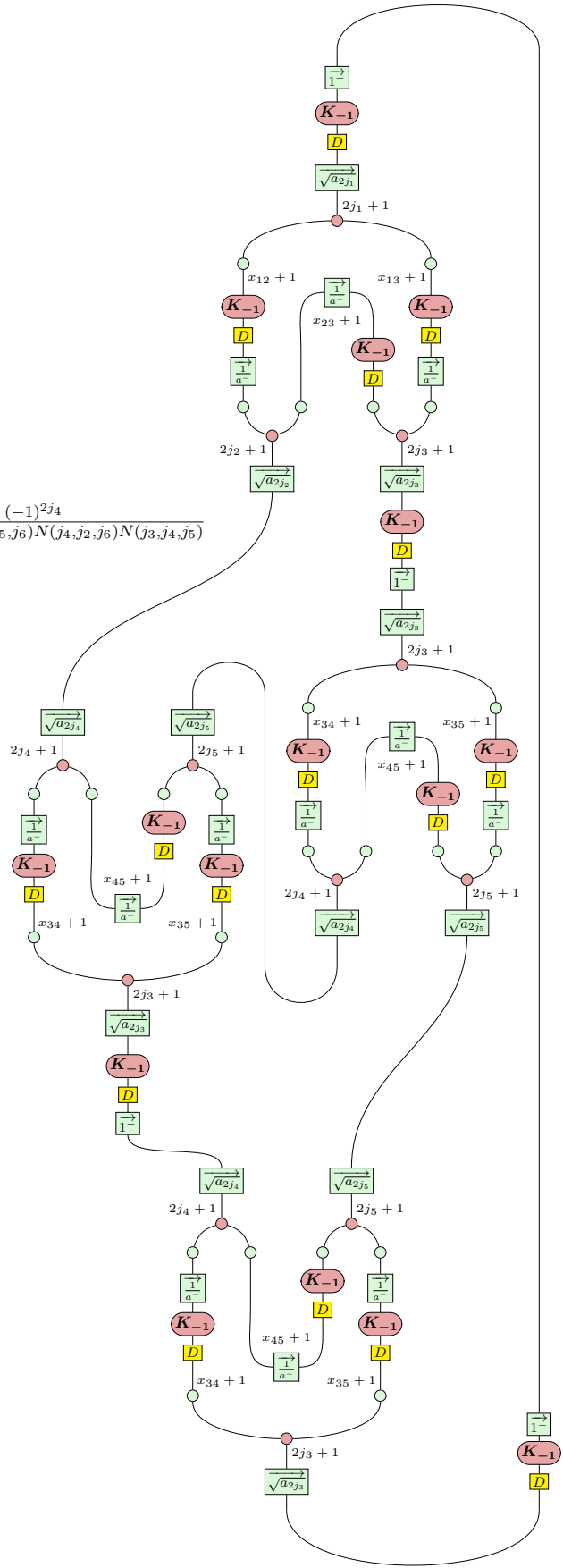

$$= \frac{(-1)^{j_1+j_2+j_4+j_5}}{N(j_1,j_2,j_3)N(j_1,j_5,j_6)N(j_4,j_2,j_6)N(j_3,j_4,j_5)}$$

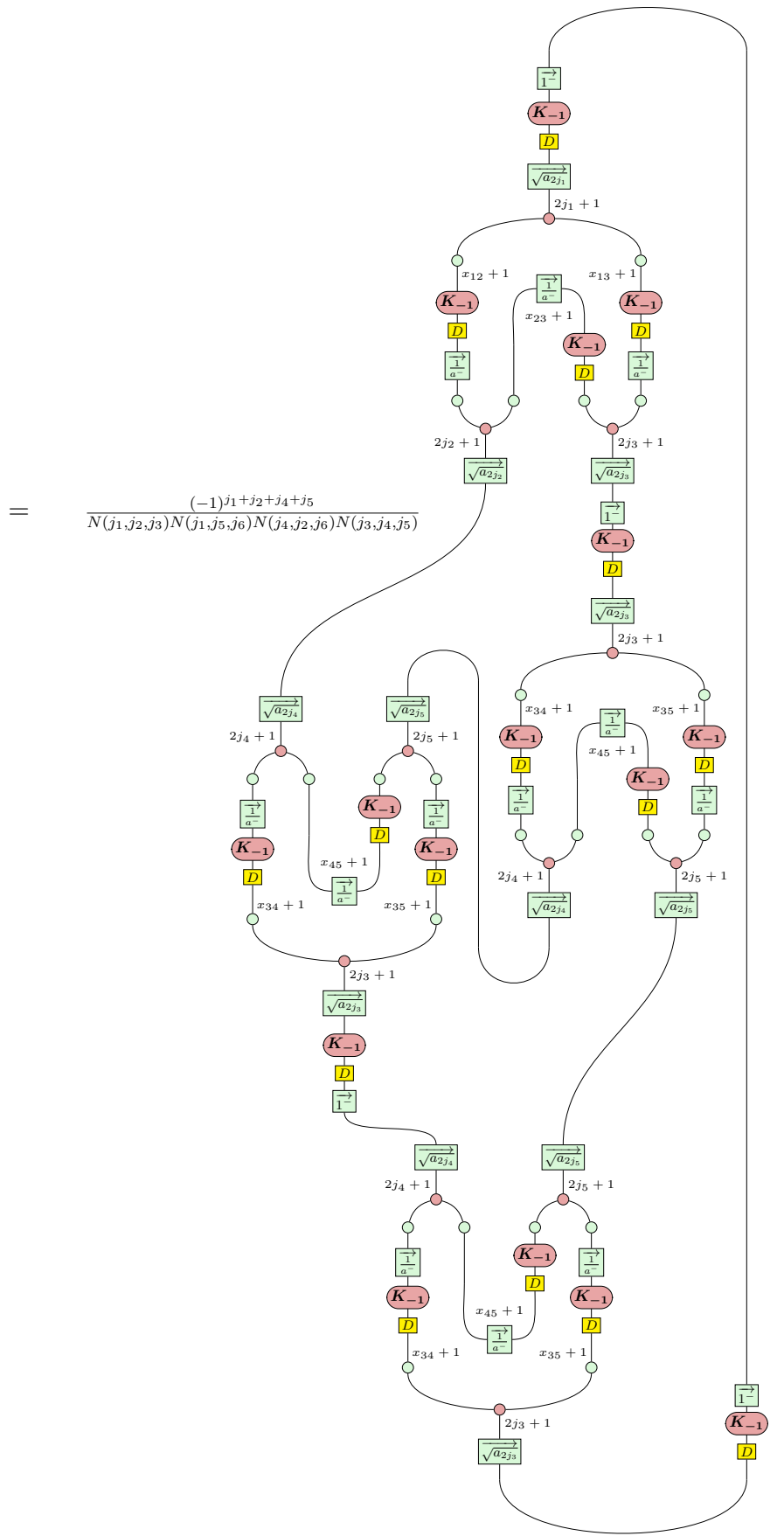

,

which exactly translates as the tetrahedral net symbol built from the injections of table 3.1. Since we have shown that $\Theta(j_1, j_2, j_3) = N(j_1, j_2, j_3)^2$, then by the definition of the $6j$-symbol, we have,

$$\begin{Bmatrix} j_1 & j_2 & j_3 \\ j_4 & j_5 & j_6 \end{Bmatrix} = (-1)^{j_1+j_2+j_4+j_5} \frac{Tet(j_1, j_2, j_3, j_4, j_5, j_6)}{\sqrt{\Theta(j_1, j_2, j_3)\Theta(j_1, j_5, j_6)\Theta(j_4, j_2, j_6)\Theta(j_3, j_4, j_5)}} \,,$$

corresponding exactly to the above diagram. $\qquad\square$

---

**Tetrahedral Net Symbols and $6j$-Symbols: Summary**

- The tetrahedral net symbol is to the triple bubble as the $\Theta$-graph is to the bubble. Following the definition in Peterson [66], this object is proportional to the $6j$-symbol.

- Using this known proportionality, we can give a diagrammatic form of the $6j$-symbol in the PSC (Thm. 62)) by showing that it can be rewritten as the trace over a triple bubble. Since we have embedded proper normalisation, the proof is exact.

## 4.3   3-Loop Removal

The next special kind of loop we can encounter in a spin network is a *3-loop spin network portion* (sometimes referred to as a "*triple bubble*"). This extends on the standard bubble by interacting three injective 3-valent nodes so as to form an internal loop with three (3-valent) nodes. The special cases wherein a pair of these nodes also connect on their third links exposes a pair of standard bubbles that can be removed ('popped', if you are feeling whimsical) by Thm. 58 of the previous section, so we expound only on the general case in this section.

By the *triple bubble identity* of (G28), we know that this object is proportional to a single injective 3-valent node, with the scalar being given by a 6j-symbol determined by the spins involved in the triple bubble. Extending Thm. 58 to 3-loop removal, we can give a diagrammatic argument in the PSC, including the computation of the scalar. The following lemma will ease the proof.

**Lemma 63.**

*Proof.*

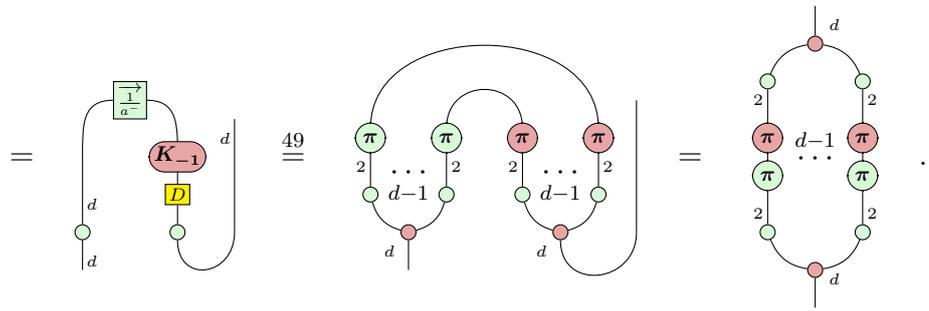

**Theorem 64** (3-Loop Removal). *Writing (G28) in the PSC,*

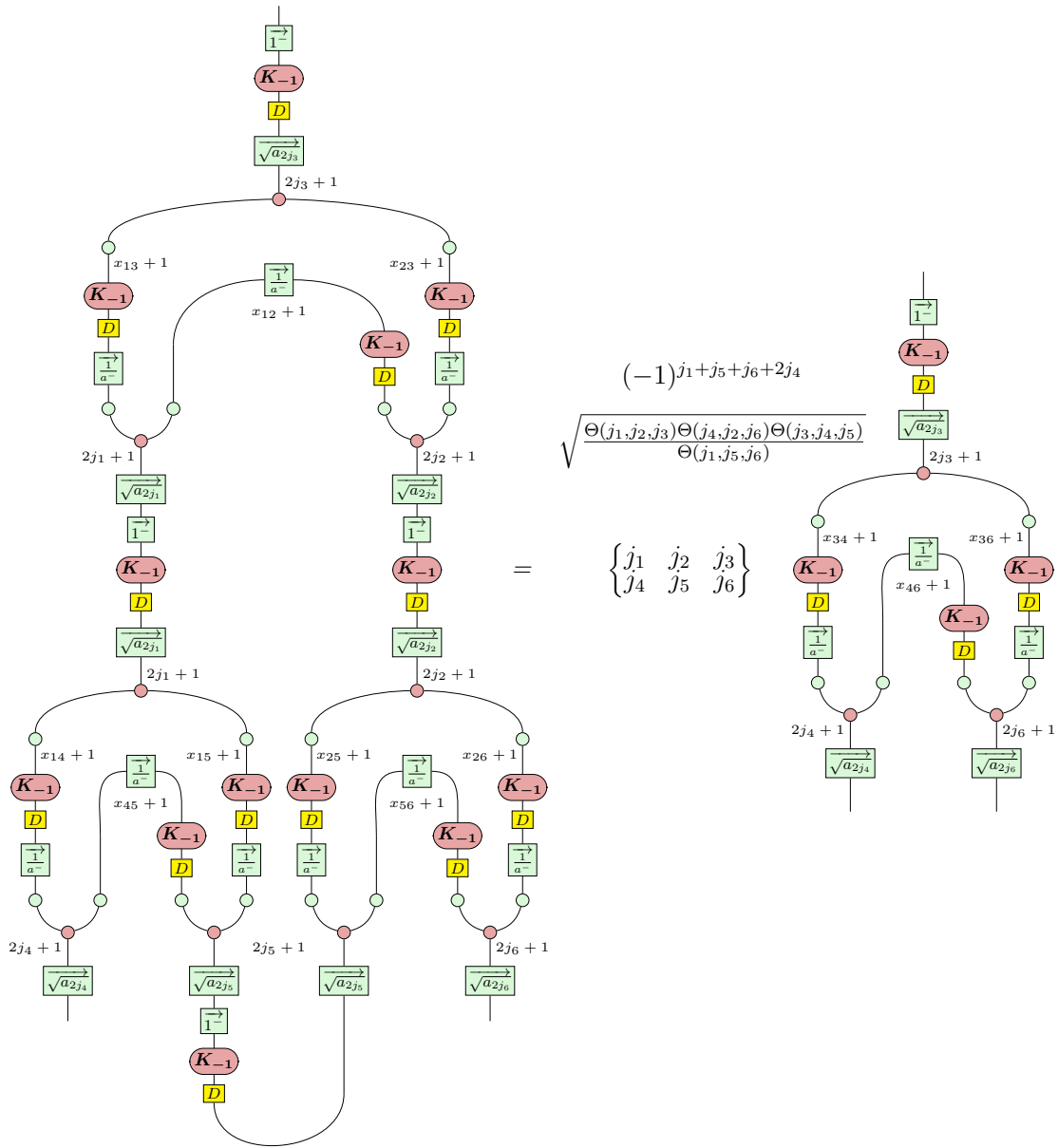

*Proof.* Begin with the qubit-style triple bubble by interacting three (qubit) injections and reducing the diagram to a simpler form,

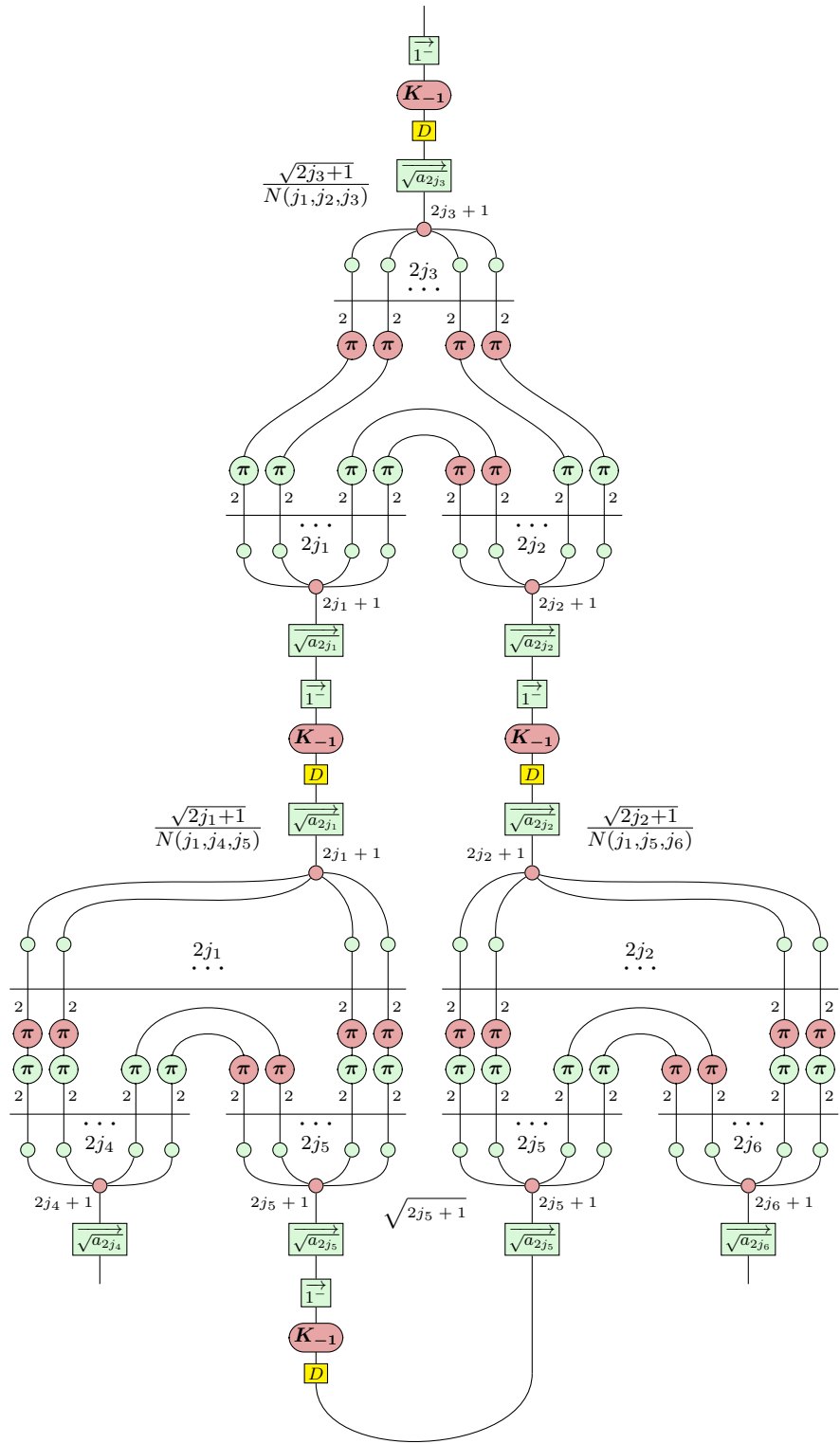

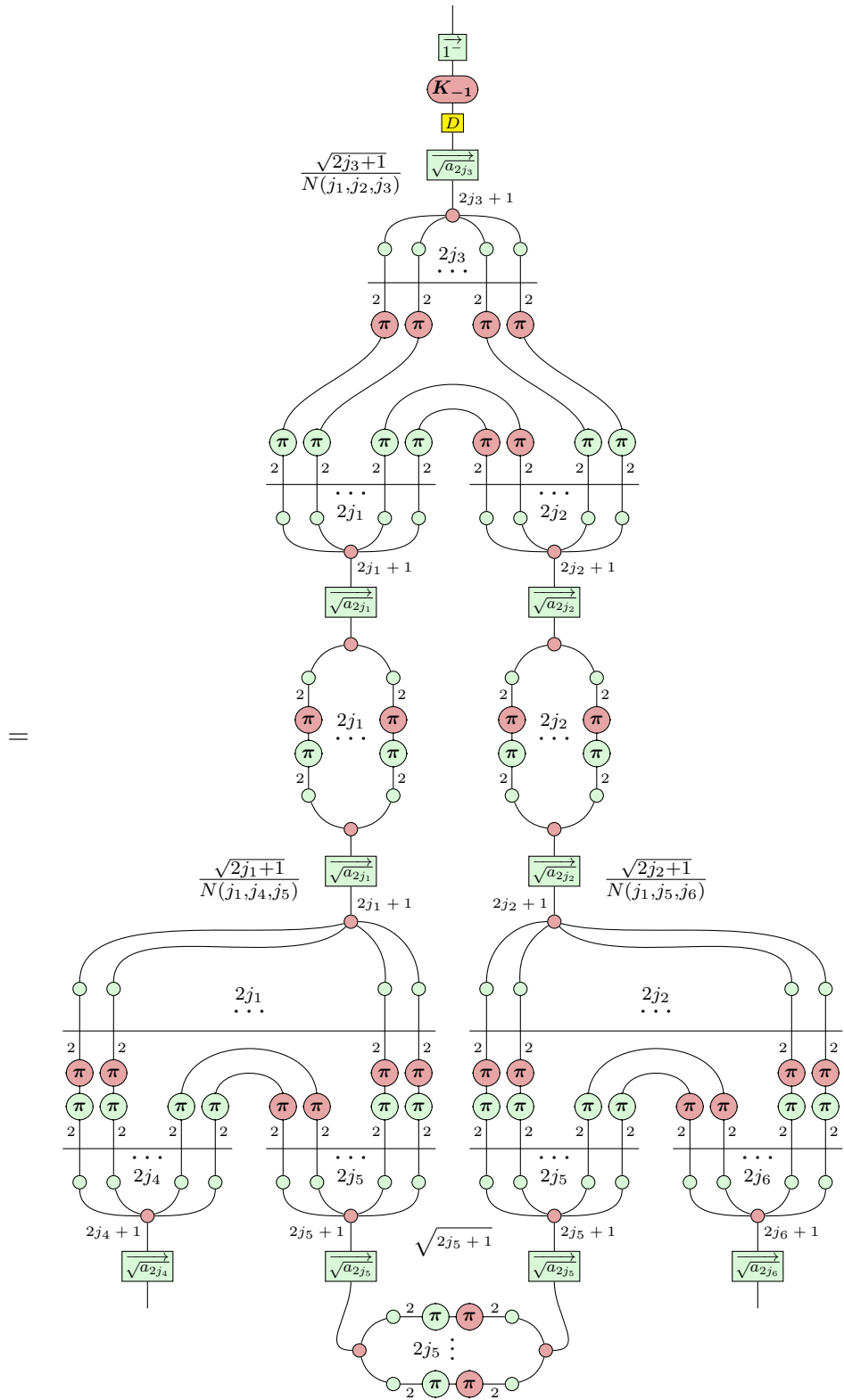

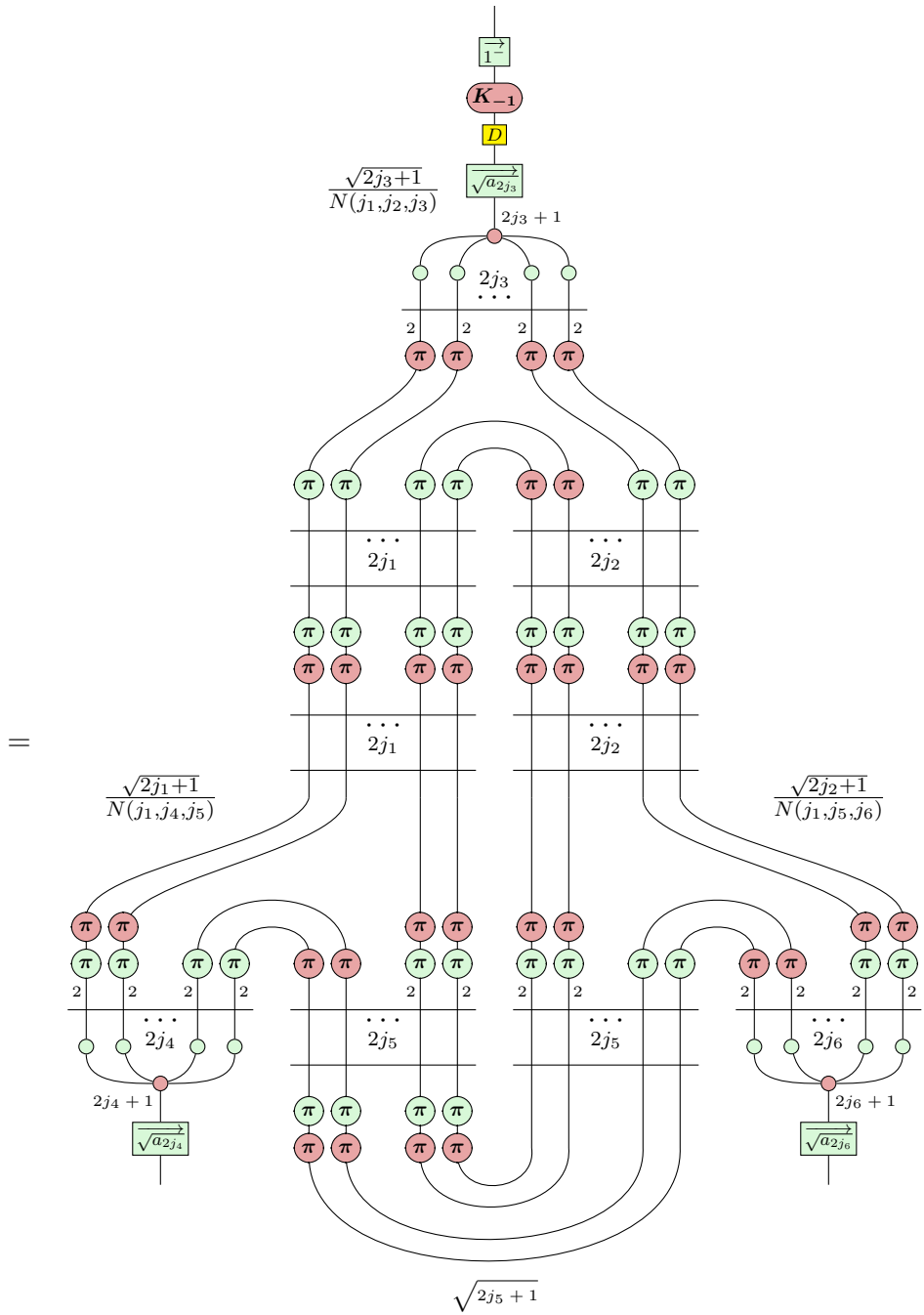

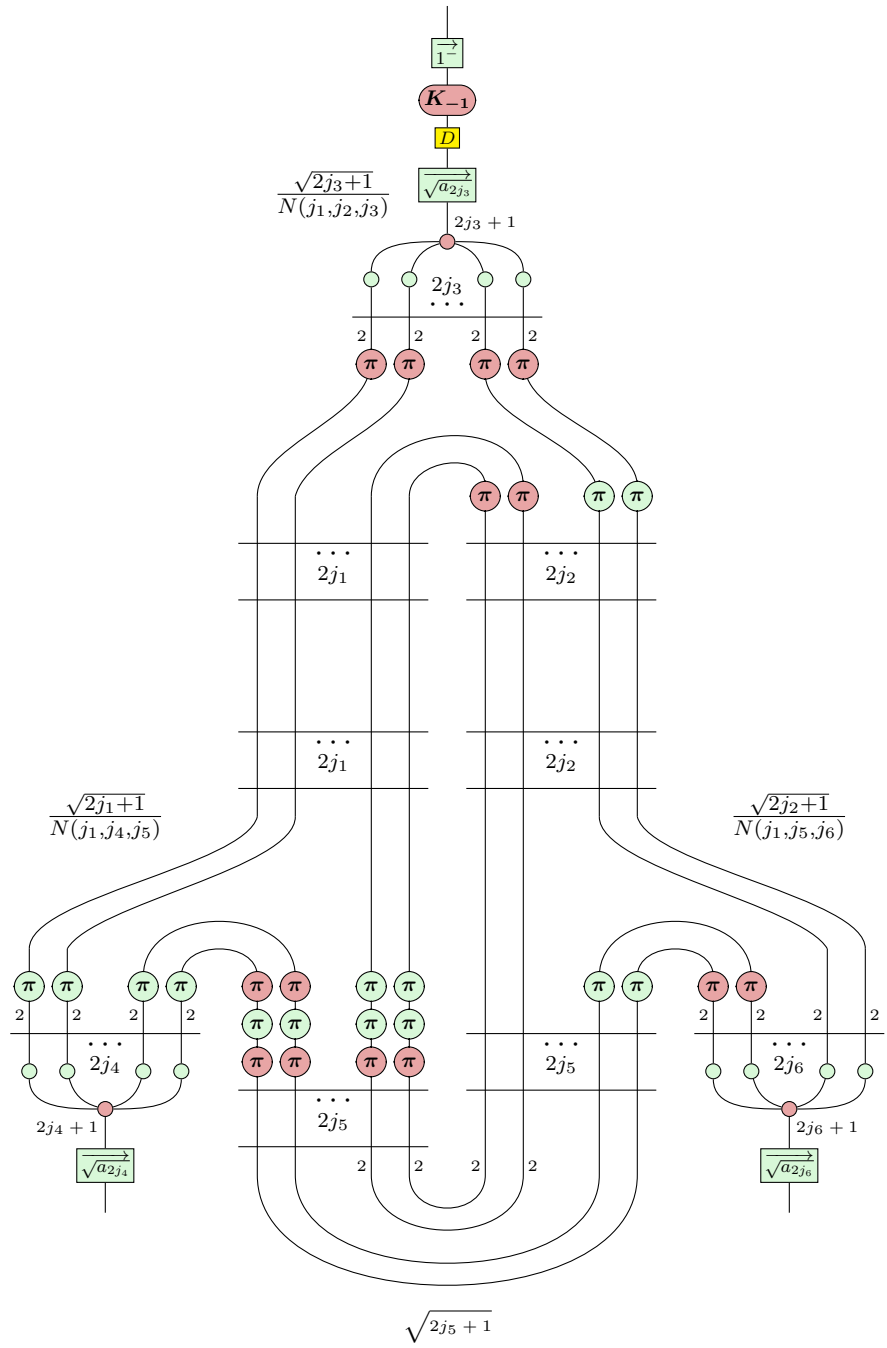

$$\overset{47}{=}$$

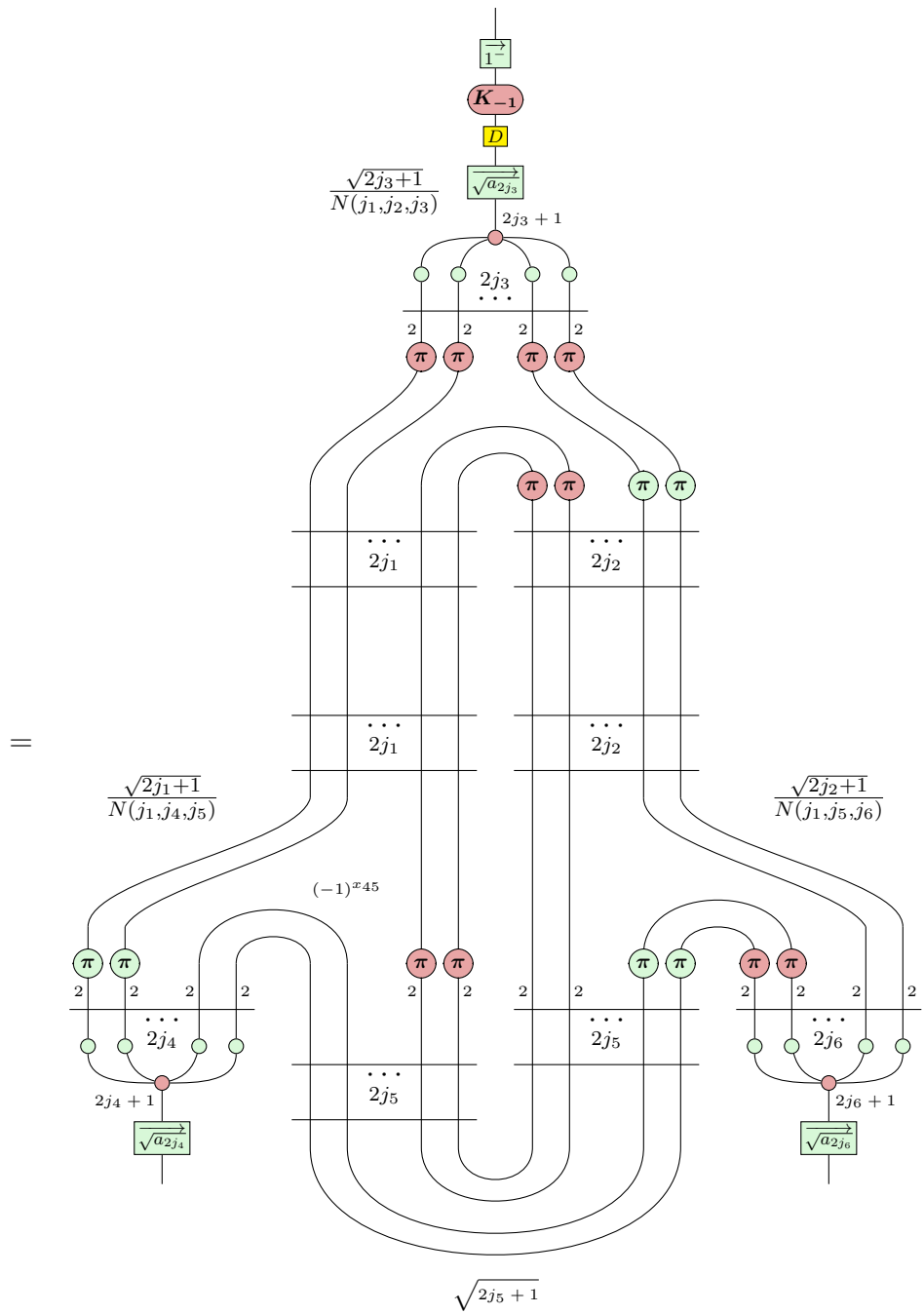

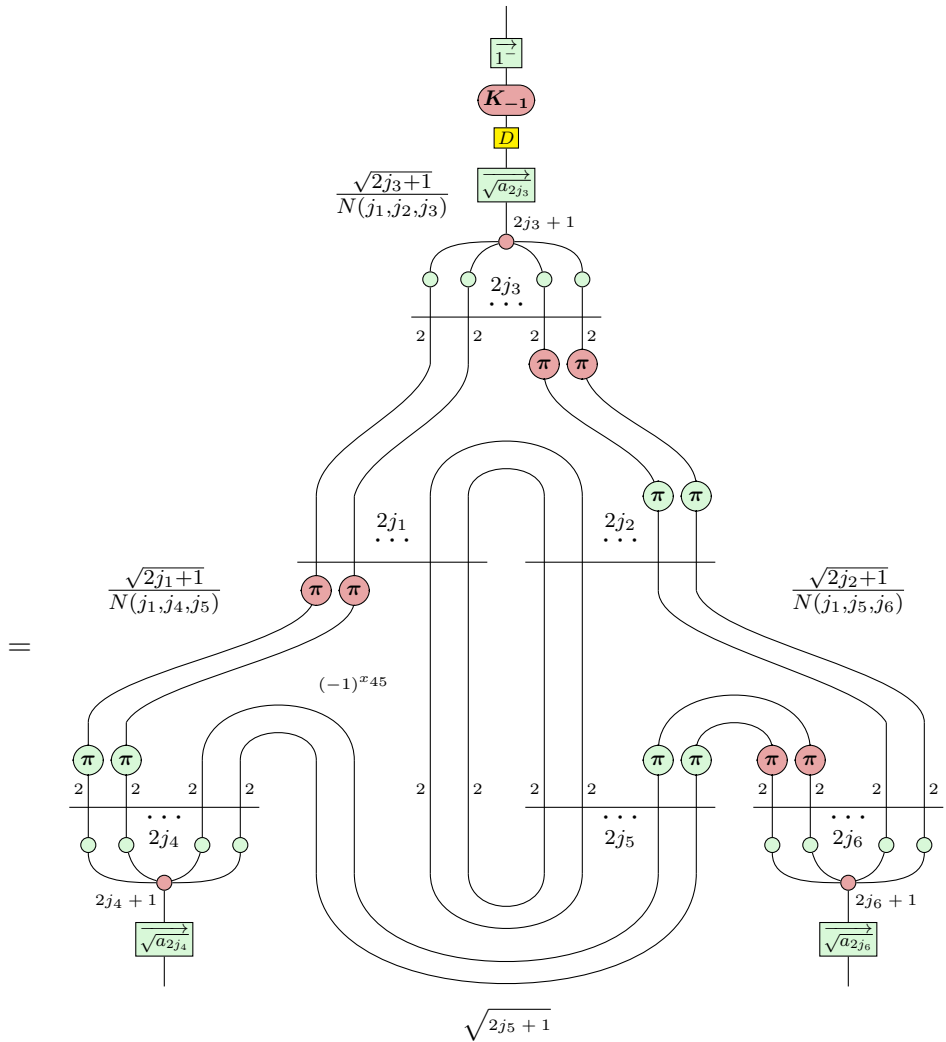

We can now readily analyse this diagrammatic form of the triple bubble, decomposing each symmetriser as a sum of permutations in a similar style as with the bubble identity of Prop. 57. Notice that any wire leaving one of the symmetrisers at each of the open links (those being $j_3$, $j_4$, and $j_6$) necessarily 'arrives' at one of the other open-link symmetrisers (assuming that they do not go back on themselves—this would yield the zero map, by capping). And also notice that there are a fixed number of connections between these open-link symmetrisers:

$x_{34}$, $x_{36}$, and $x_{46}$. Then we may write the above as, for some constant $K$,

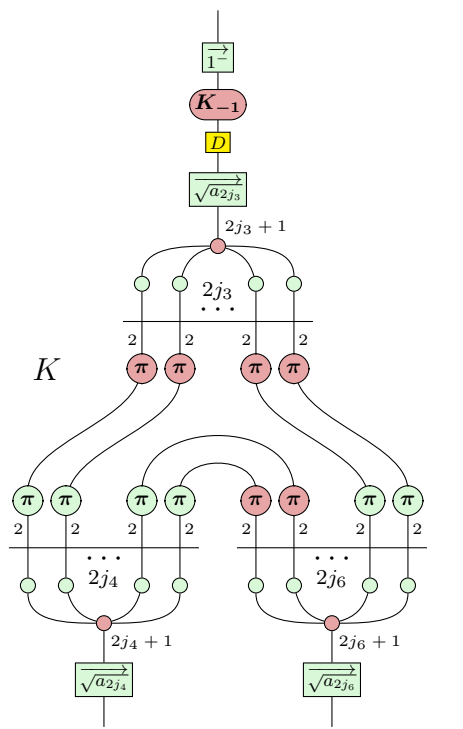

,

or, equivalently, in the qudit-style,

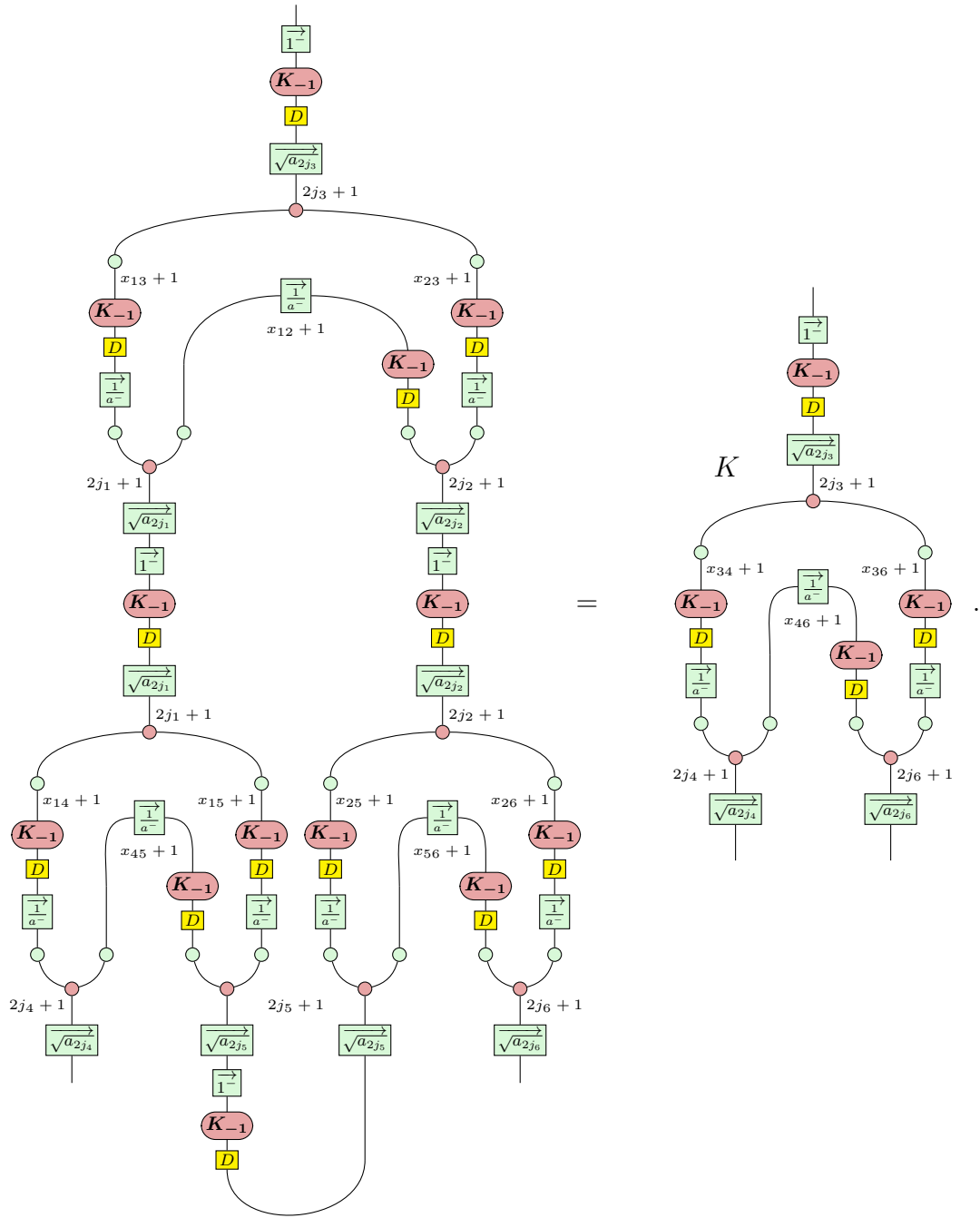

To determine the value of $K$, we can compose another 3-valent node to the bottom of these diagrams and take the trace. This will reveal the $\Theta$-graph on the RHS, scaled by $K$, and the *tetrahedral net* on the LHS—an object whose definition is closely tied to the $6j$-symbol (discussed in section 4.2),

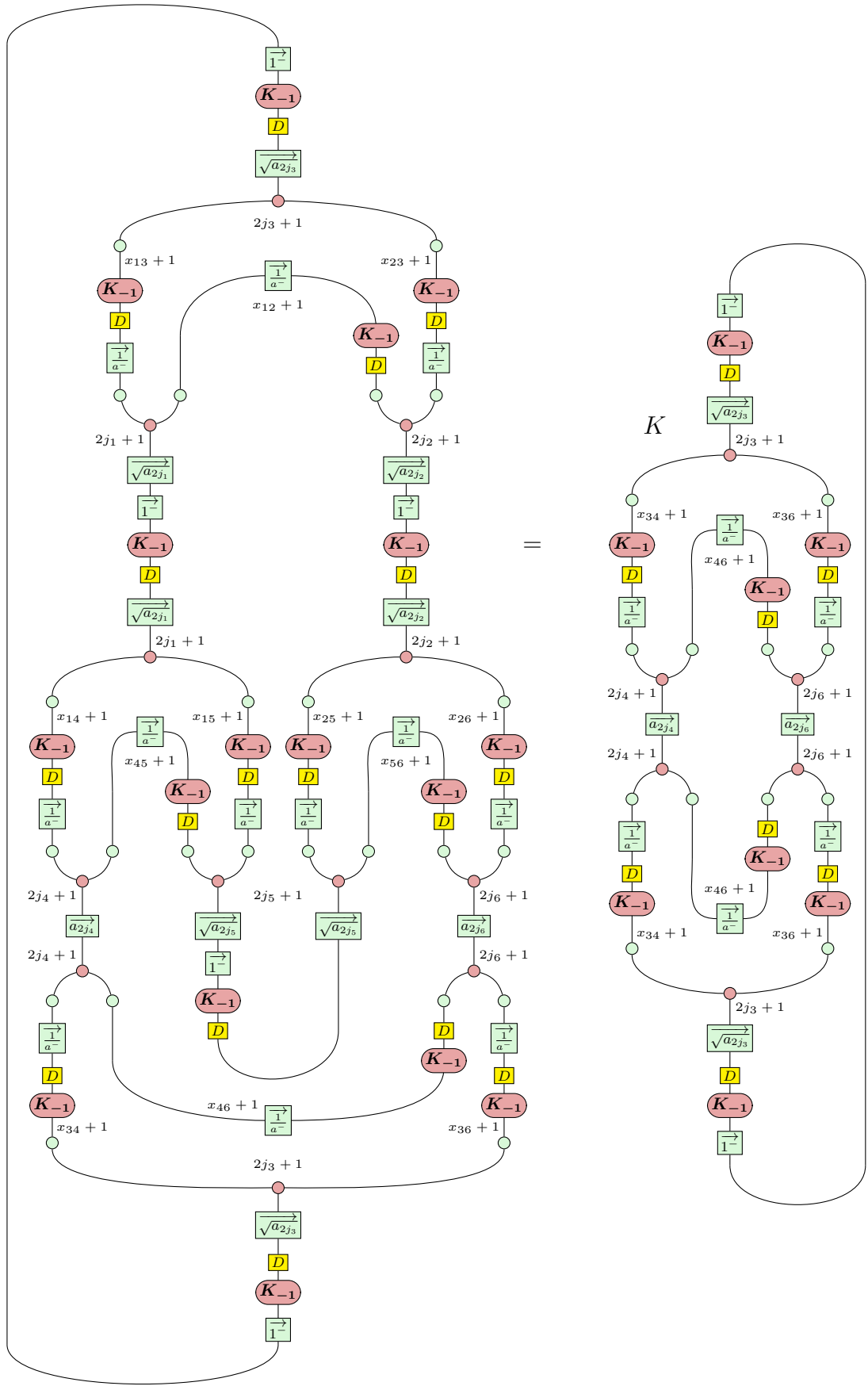

Taking the LHS to be equal to $(-1)^{x_{46}}\text{Tet}(j_1, \dots, j_6)$, and recalling that $\Theta(\cdots) = N(\cdots)^2$ from section 4.1.2 for the RHS, we thus we have,

$$K = (-1)^{j_1+j_5+j_6+2j_4} \sqrt{\frac{\Theta(j_1, j_2, j_3)\Theta(j_4, j_2, j_6)\Theta(j_3, j_4, j_5)}{\Theta(j_1, j_5, j_6)}} \begin{Bmatrix} j_1 & j_2 & j_3 \\ j_4 & j_5 & j_6 \end{Bmatrix} \quad,$$

where we have expanded the definition of $\text{Tet}(j_1, \dots, j_6)$. $\qquad\square$

> **3-Loop Removal: Summary**
>
> - By reducing the 3-loop to the form of three isometries connected via symmetrisers and singlets, we can make a permutation argument to immediately see that a triple bubble is proportional to a $3j$-symbol (first half of Thm. 64).
>
> - To determine the exact scalar, we can compose another $3j$-symbol to form a bubble, then take the trace to form a relation between the tetrahedral net symbol and $\Theta$-graph (second half of Thm. 64). This relation resolves via Thm. 62 and subsection 4.1.2 to obtain the scalar (up to normalisation) of (G28).

# 4.4 Associativity

Since the Clebsch-Gordan decomposition is an isomorphism,

$$C : \mathcal{H}_{j_1} \otimes \mathcal{H}_{j_2} \cong \bigoplus_{j=|j_1-j_2|}^{j_1+j_2} \mathcal{H}_j$$

and the diagram of Prop. 51 is an injection map from $\mathcal{H}_j$ to $\mathcal{H}_{j_1} \otimes \mathcal{H}_{j_2}$, then to check a map equality based on a basis of the space $\mathcal{H}_{j_1} \otimes \mathcal{H}_{j_2}$, it suffices to check the bases of $\mathcal{H}_j$ for all possible $j$ via this injection map. That is to say that, instead of plugging a product basis state into the open links, we need only plug a single basis state into the injection map to obtain a product state—and equality can be ascertained by checking the LHS agrees with the RHS for each $j$ we plug into the injection map. Making use of this trick, we can prove the following proposition.

**Proposition 65.**

*for some coefficients $x_k$.*

*Proof.* Using the injection trick discussed above, we can give a proof by showing that, for all $j$,

Immediately, we can apply the bubble identity of Thm. 58 to the RHS,

The LHS can also be seen to be a triple bubble identity of Thm. 64; to see this, we can observe that therefore, by Thm. 64, we have,

$$(-1)^{3j_2+j_4+\ell}K = x_j\frac{N(j,j_3,j_4)^2}{2j+1} \implies x_j = \frac{(2j+1)(-1)^{3j_2+j_4+\ell}K}{N(j,j_3,j_4)^2},$$

with

$$K = (-1)^{j+j_3+j_4+2\ell}\sqrt{\frac{\Theta(j,j_1,j_2)\Theta(j_1,j_3,\ell)\Theta(j_2,\ell,j_4)}{\Theta(j,j_3,j_4)}}\begin{Bmatrix}j & j_1 & j_2\\ \ell & j_4 & j_3\end{Bmatrix}.$$

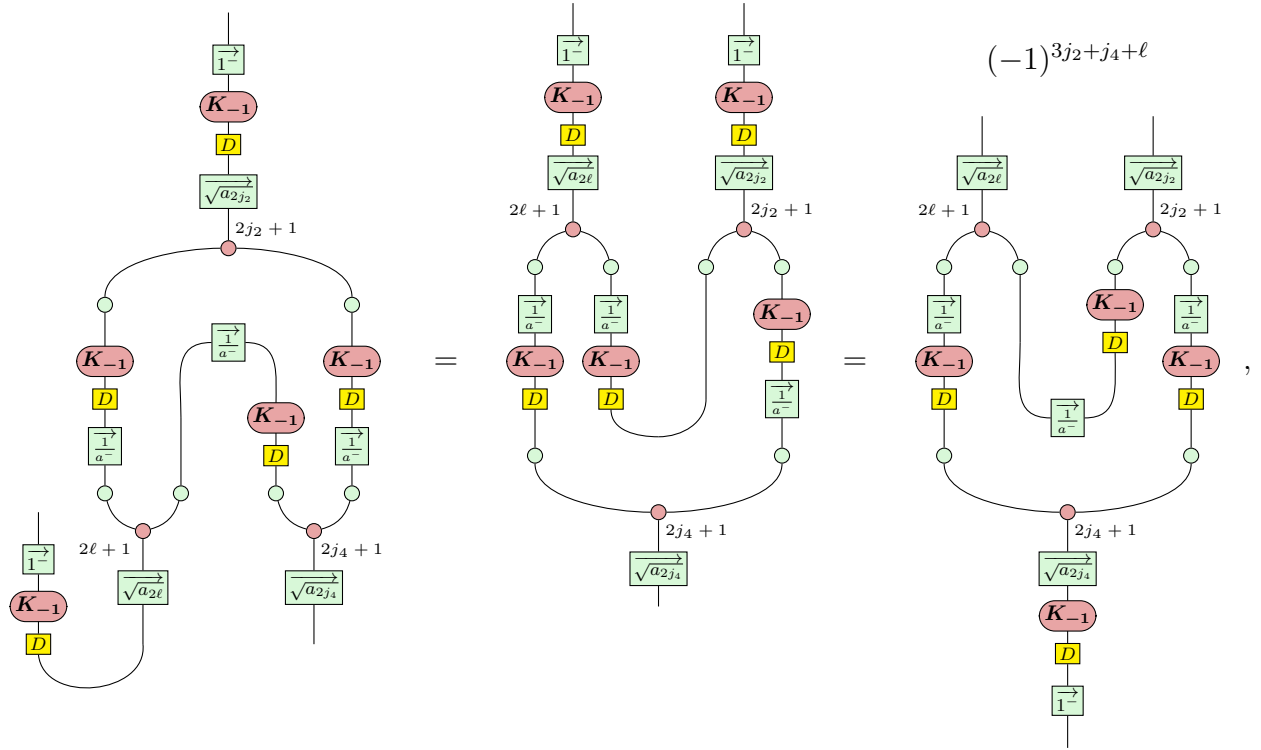

$\square$

Now we want to derive the following associativity of 3-valent spin network nodes.

**Proposition 66** (Associativity)**.**

$$\cdots = \sum_k x_k(-1)^{-j_1+2j_2+j_4+k-\ell}\cdots.$$

*Proof.* First, we can show that,

$$\begin{array}{ccc} j_1 & & j_1 \\ j_3 \quad \ell & = & j_3 \\ j_4 \quad j_2 & & j_4 \quad (-1)^{x_{24}} \quad j_2 \end{array} \quad ,$$

for which it suffices to prove that,

$$\begin{array}{ccc} \ell & & \ell \quad (-1)^{x_{24}} \\ j_4 \quad j_2 & = & j_4 \quad j_2 \end{array} \quad .$$

In the PSC, this can be seen via,

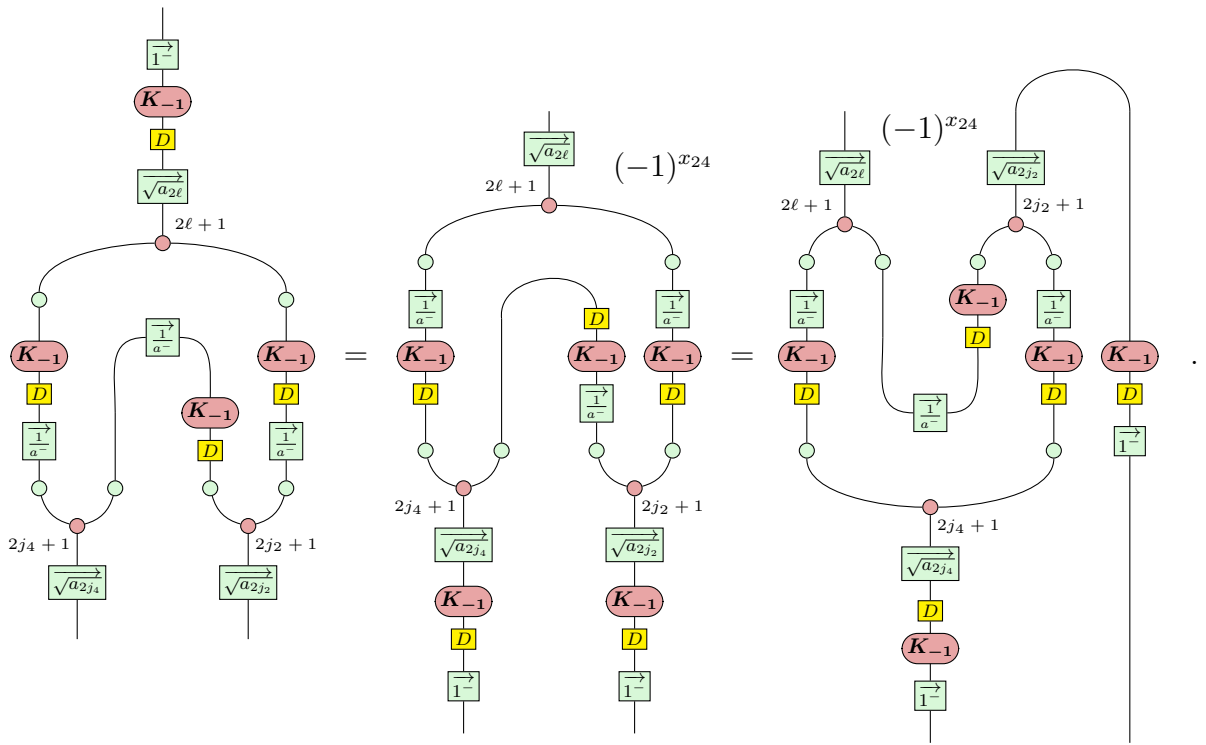

The property of associativity then follows as,

$$\begin{array}{ccccc} j_1 & & j_1 & & j_1 \quad (-1)^{x_{24}} \\ j_3 \quad \ell & = & j_3 \quad \ell \quad (-1)^{x_{24}} & \overset{65}{=} & \sum_k x_k \quad k \quad j_2 \\ j_4 \quad j_2 & & j_4 \quad j_2 & & j_3 \quad j_4 \end{array}$$

$$= \sum_k \; x_k (-1)^{-j_1 + 2j_2 + j_4 + k - \ell} \;\; \vcenter{\hbox{}} \quad .$$

$\square$

## 4.5   *n*-**Loop Removal**

So far, we have established 2- and 3-loop removal in the PSC (Thm. 58 and 64), and shown that 3-valent spin network nodes have the property of associativity (Prop. 66). In this section, we will use these results to give an inductive argument for the removal of any *n*-loop spin network portion, leading to a loop-free spin network normal form.

The bubble identity is a somewhat special case, but 3-loop (triple bubble) removal forms the foundation for removing higher-arity loops. Importantly, we can combine Thm. 64 with Prop. 65 to show that 4-loop removal is expressible as a sum of 3-loop removals (ignoring the finer details—the $x_k$ coefficients and $6j$-symbols should be given for a rigorous proof),

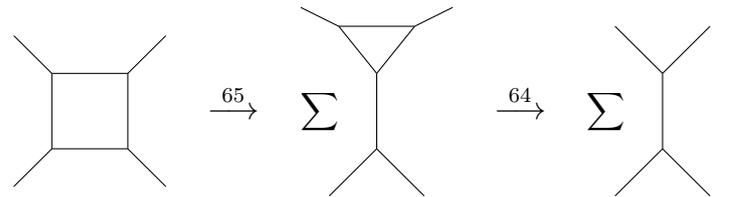

For the sake of this argument, take 4-loops to be the base case (we could take 3-loops as our base case, but 4-loops present a clearer argument, and we know from the above that 4-loop removal follows from 3-loop removal). Let us now see how we can then inductively handle 5- and 6-loops, as an illustration of the proof. The core idea is to introduce a second internal loop through a reverse triple bubble removal, then to use associativity (Prop. 66) to manoeuvre it through the first loop.

Consider the following 5-loop, to which we introduce a second internal loop and manoeuvre to give two roughly equal-sized loops,

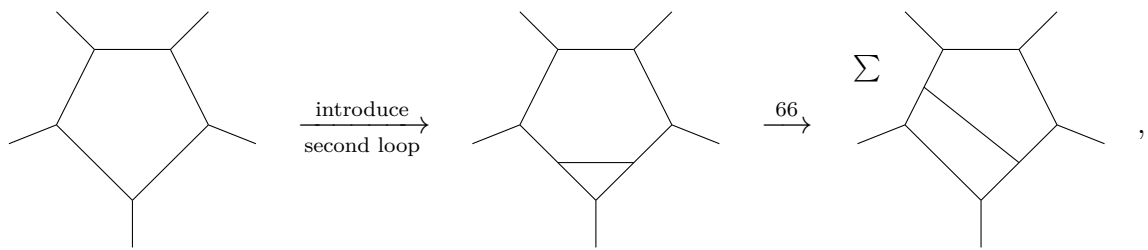

then remove the lower-left 4-loop portion, revealing another 4-loop to be removed,

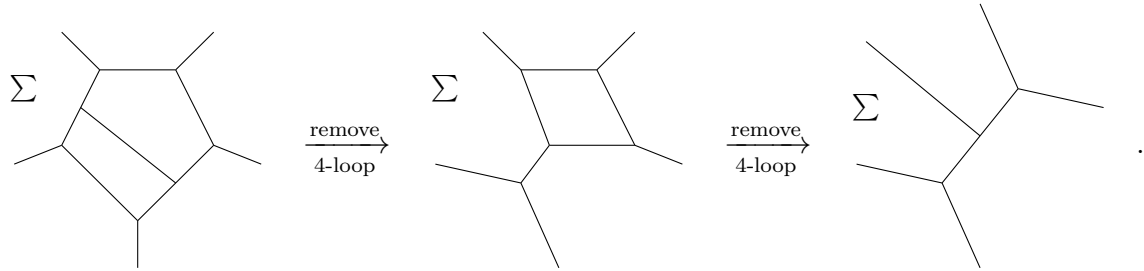

Furthermore, consider the following 6-loop, to which we again introduce and manoeuvre a second internal loop,

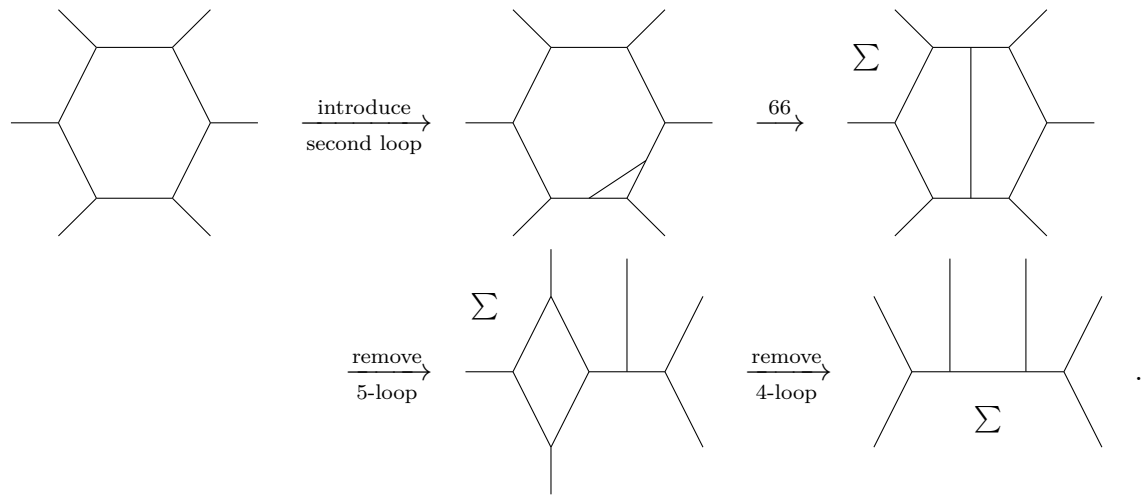

We can observe the following:

- For *odd*-arity $(n+1)$-loops, we can introduce and manoeuvre a second internal loop to yield two loops of $(n+1)$- and $n$-nodes. Removing the $n$-loop (by induction) reduces the arity of the former loop to become a leftover $n$-loop, which can again be removed.

- For *even*-arity $(n+1)$-loops, the introduction and manoeuvring of a second internal loop yields two loops each of $n$-nodes. Again, removing one reduces the other to an $(n-1)$-loop, which can also be removed.

Hence, assuming that an $n$-loop can be removed (taking 2-, 3-, and 4-loops to be removable, by our results in this chapter and the implication of (G27) as noted by Guedes et al. [1]), then any $(n+1)$-loop can always be removed using our above 'introduce-and-manoeuvre' method. More generally, given any spin network, we can always remove *any* such loops to reduce its form to that of an acyclic graph.

**Removal of $n$-Loop Spin Network Portions: Summary**

- First, we showed the correctness of 2-loop ("bubble") removal (G13) in the PSC, with correct normalisation. We did this by showing an intuitive permutation argument explicitly, then proved exactness with a rigorous derivation of the scalar.

- Then, we diagrammatically proved the $\Theta$-graph (G12) as it relates to the bubble. We also showed how this idea extends to give the tetrahedral net symbol in relation to the triple bubble, and used Peterson [66] to define the $6j$-symbol via a diagrammatic rewrite of the tetrahedral net.

- Second, we showed the correctness of 3-loop ("triple bubble") removal (G28) in the PSC, with correct normalisation. This proof used an intuitive rewrite to reduce the form into a diagram that revealed an implicit permutation argument. The scalar was obtained making use of the tetrahedral net symbol, revealing the $6j$-symbol.

- Then, using arguments from the above, we proved the associativity of spin network nodes, and used this to show that 4-loop removal follows from 3-loop removal (introducing a sum in the process).

- Third and finally, we demonstrated an inductive argument for $n$-loop removal, resting on 4-loop removal (and hence 3-loop removal and associativity) as the base case. 2-loop removal is a special case.

*Normal is nothing more than a cycle on a washing machine.*

— Whoopi Goldberg

# 5

# Towards a Hypergraph Normal Form for Spin Networks

## Contents

By chapter 4, we can always algorithmically remove loops to reduce an arbitrary spin network to an acyclic normal form. From here, we can move into the W-node perspective (section 3.3) to further explore how we might reduce this to a fully-connected graph between the open links of the spin network, accompanied by a hypergraph decorated with W-gadget hyperedges. Our motivations for this rewrite are to indicate future research with spin networks as *computable objects*, representable as dense matrices. The forms we derive in this chapter are intended to be the first steps towards a new description for LQG suitable for (quantum) computation via matrix calculations.

Due to the nature of the diagrams in this section, we will sometimes present left-to-right rather than up-to-down for readability.

# 5.1 Deriving the Structure of the Normal Form

For the discussion in this section, we substitute the singlet state (bottom row of table 3.1) with the Bell state $\frac{1}{\sqrt{2}}(|00\rangle + |11\rangle)$, which is to remove the $\pi$-phase spiders (in the qubit representation; Def. 32) from the cap/cup and thus ignore the directionality that singlet states enforce to yield the characteristic invariance of the intertwiner. This will make deriving the normal form more convenient and pedagogically easier to intuit, and is not uncommon as a first step in the literature. For example, Fan, Korepin, and Roychowdhury [95] also make this substitution in the context of acyclic Cayley trees of spin networks, and note that this does not change the properties of the entanglement of the ground states of the AKLT model that they consider. We leave modifications to this form that resolve singlet states to future work.

## 5.1.1 The Fully-Connected Foundations

There are two primary mechanisms involved in the rewrite from a loop-free spin network to a fully-connected graph with a 'nest' of gadgets:

- *Expansion* of the symmetriser via the (WW) bi-algebra; and

- *Absorption* of internal W-nodes, eventually into the external W-nodes.

To demonstrate these mechanisms, consider the following simple (loop-free) spin network with only two nodes, and omit wire dimension labels for the time being:

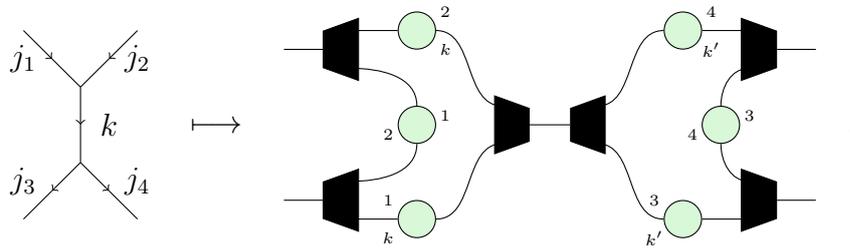

First, we *expand* the $k$-to-$k'$ symmetriser and pass the gadgets by lemma 53,

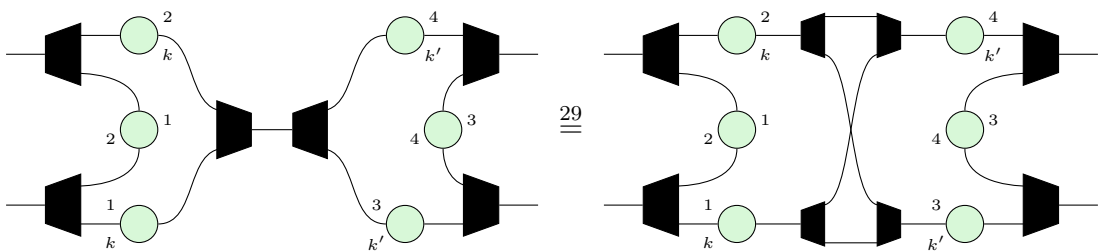

$$\overset{53}{\equiv}$$ 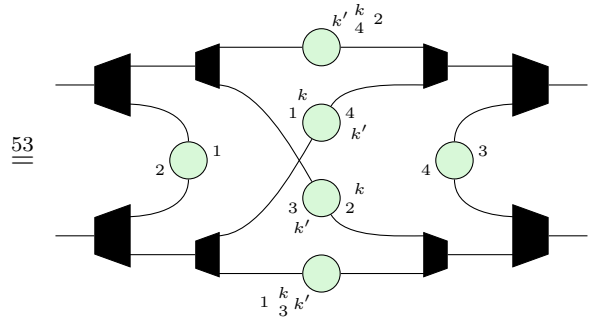  .

Then, we *absorb* the internal W-nodes into the external $j_i$-labelled W-nodes by lemma 30 of Poór, Shaikh, and Wang [2].

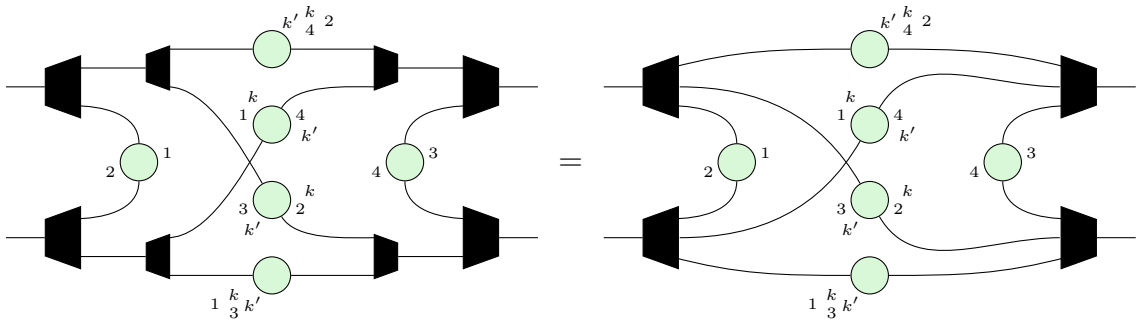  .

## 5.1.2   A W-Gadget Hypergraph

The first thing we can do is write a matrix describing the placement of the W-gadgets. This will be the most informative piece of the normal form, since knowing where each W-gadget connects tells us things as nuanced as *how 'deep' into the spin network its corresponding node is* and *which links have influence on which other links*.

Notice that: (1) multiple gadgets can occupy a given edge; and (2) multiple edges can be occupied by a given gadget. Thus, it will not suffice to simply weight edges with W-gadgets to form a simple adjacency matrix. Instead, we will build a hypergraph on-top-of the fully-connected graph, wherein each edge becomes a hyper-node, and each W-gadget defines a (directed) hyper-edge. This can be hard to draw on the page, so we will illustrate the hyper-edges induced by the $j_i$-labelled

W-gadgets separately from those of the $k$ and $k'$ W-gadgets,

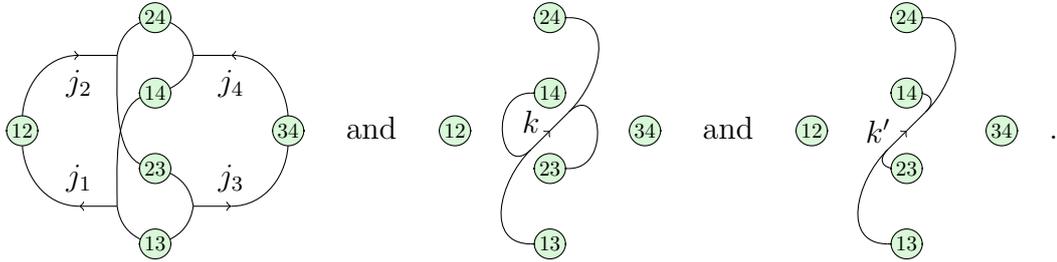

In the above, each hyper-node $v_{ab}$ (written above without the $v$) is labelled by the indices of the two W-nodes ($j_a$ and $j_b$) connected by the underlying edge, and each hyper-edge is labelled by the corresponding W-gadget's label (directed from the gadget's left leg to its right, arbitrarily). For example, the central diagram above draws the hyper-edge $e_k$ (written above without the $e$) corresponding to the $k$-labelled W-gadget, and it goes from $(v_{13}, v_{14})$ to $(v_{23}, v_{24})$.

Now we can give the *incidence matrix* for the hypergraph. Denoting the head and tail of each hyper-edge $e_\ell$ by $H(e_\ell)$ and $T(e_\ell)$ respectively, and denoting the hyper-nodes by $v_{ab}$ to correspond to the underlying edge between $j_a$ and $j_b$, the incidence matrix is given by $I = (m_{(ab)\ell})$, where

$$m_{(ab)\ell} = \begin{cases} -1 & v_{ab} \in T(e_\ell) \\ +1 & v_{ab} \in H(e_\ell) \\ 0 & \text{otherwise} \end{cases} \quad .$$

For our running example, this gives,

| | $e_{j_1}$ | $e_{j_2}$ | $e_{j_3}$ | $e_{j_4}$ | $e_k$ | $e_{k'}$ |
|---|---|---|---|---|---|---|
| $v_{12}$ | 1 | −1 | 0 | 0 | 0 | 0 |
| $v_{13}$ | −1 | 0 | −1 | 0 | −1 | −1 |
| $v_{14}$ | −1 | 0 | 0 | 1 | −1 | 1 |
| $v_{23}$ | 0 | 1 | −1 | 0 | 1 | −1 |
| $v_{24}$ | 0 | 1 | 0 | 1 | 1 | 1 |
| $v_{34}$ | 0 | 0 | 1 | −1 | 0 | 0 |

where rows indicate hyper-nodes (edges in the underlying graph), and columns indicate hyper-edges (W-gadgets). Alternatively, we can write the nodes $j_i$ of the underlying graph on the rows and present, in a cell $m_{i\ell}$, whether a node $j_i$ is present in the head or tail of a hyper-edge $e_\ell$. As an example, take $e_1$ (corresponding to the $j_1$-labelled W-gadget): this hyper edge goes from nodes $(j_1, j_3, j_4)$ on the left to $(j_1, j_2)$ on the right, so we can say that $j_3, j_4 \in T(e_1)$ and $j_2 \in H(e_1)$ and that $j_1$ is cancelled out. So, for the full network,

|          | $e_{j_1}$ | $e_{j_2}$ | $e_{j_3}$ | $e_{j_4}$ | $e_k$ | $e_{k'}$ |
|----------|-----------|-----------|-----------|-----------|-------|----------|
| $v_{j_1}$ | 0 | 1 | −1 | 1 | −1 | 0 |
| $v_{j_2}$ | 1 | 0 | −1 | 1 | 1 | 0 |
| $v_{j_3}$ | −1 | 1 | 0 | −1 | 0 | −1 |
| $v_{j_4}$ | −1 | 1 | 1 | 0 | 0 | 1 |

.

For this form, we will not count nodes more than once (i.e. whether or not a node appears none, once, or *multiple* times in the head or tail, we still only consider whether it does or does not appear in the head or tail). It is not yet clear to us whether counting—essentially weighting this hybrid matrix—would be preferable.

### 5.1.3 The Dimension Problem

Now let us look at the other half of the story; omit the W-gadgets and bring back the wire dimension labels. This omission of gadgets does not change the dimensions, following lemma 53. We will perform the same rewrites, but this time pay attention to what happens to the wire dimensions.

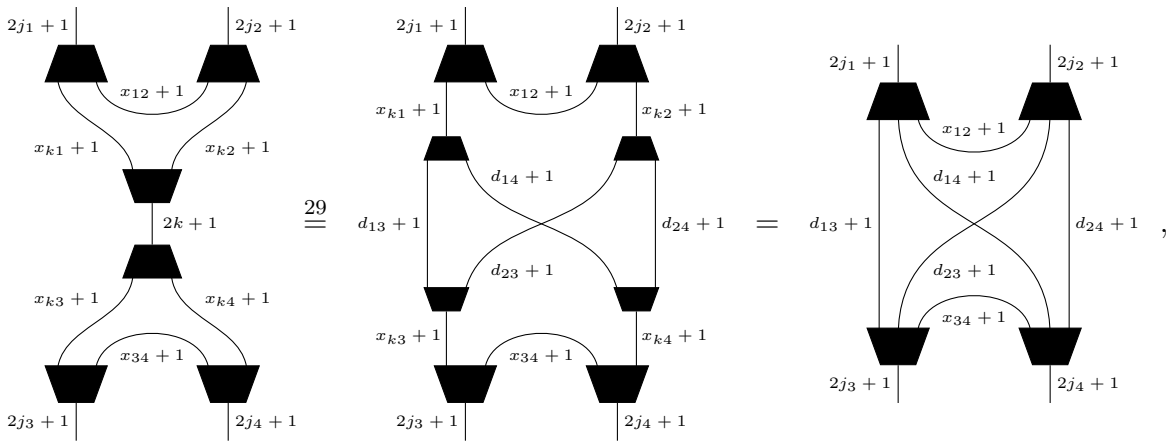

where $d_{ab} = \min\{x_{ka}, x_{kb}\}$. For these rewrites to hold, the following assumptions must be made:

- $x_{k1} \geq \min\{2j_1, d_{13} + d_{14}\} \implies x_{k1} \geq d_{13} + d_{14} = \min\{x_{k1}, x_{k3}\} + \min\{x_{k1}, x_{k4}\}$. Since $x_{ab} \geq 1$ for any $a, b$, this necessarily means that $x_{k3} \leq x_{k1}$ and $x_{k4} \leq x_{k1}$, hence $x_{k1} = k + j_1 - j_2 \geq x_{k3} + x_{k4} = k + j_3 - j_4 + k - j_3 + j_4 = 2k$. This in turn implies that $j_1 - j_2 \geq k$.

- $x_{k2} \geq \min\{2j_2, d_{23} + d_{24}\} \implies x_{k2} \geq d_{23} + d_{24} = \min\{x_{k2}, x_{k3}\} + \min\{x_{k2}, x_{k4}\}$. By the same argument, $x_{k2} \geq x_{k3} + x_{k4}$, which implies $j_2 - j_1 \geq k$.

- $x_{k3} \geq \min\{2j_3, d_{13} + d_{23}\} \implies x_{k3} \geq d_{13} + d_{23} = \min\{x_{k1}, x_{k3}\} + \min\{x_{k2}, x_{k3}\}$. Similarly, $j_3 - j_4 \geq k$.

- $x_{k4} \geq \min\{2j_4, d_{14} + d_{24}\} \implies x_{k4} \geq d_{14} + d_{24} = \min\{x_{k1}, x_{k4}\} + \min\{x_{k2}, x_{k4}\}$. And again, $j_4 - j_3 \geq k$.

From the two assumptions that $j_1 - j_2 \geq k$ and $j_2 - j_1 \geq k$, and further recalling the Clebsch-Gordan conditions for $j_1, j_2, k$, which says that $|j_1 - j_2| \leq k \leq j_1 + j_2$, we must further enforce that $j_1 - j_2 = 0 \implies j_1 = j_2$, and $k = 0$. Likewise, we should also enforce $j_3 = j_4$. Resolving this condition to something more general is a current open problem, since it restricts the spin networks we may consider to only those trivial instances which disconnect internally; that is,

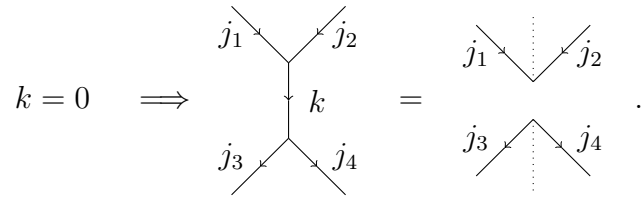

One thing we could do, to characterise this spin network, is write the adjacency matrix of the graph between the $j$-labelled W-nodes. Since this is fully connected, it would be quite a boring adjacency, though luckily the information of interest is the dimensionality of the wires within the graph. Therefore, we will write our second matrix of the normal form as the adjacency matrix of the fully-connected graph between the $j_i$-labelled W-nodes *weighted by the dimensions of the wires between them*; for our spin network above, this gives,

|          | $v_{j_1}$ | $v_{j_2}$ | $v_{j_3}$ | $v_{j_4}$ |
|----------|-----------|-----------|-----------|-----------|
| $v_{j_1}$ | $\cdot$   | $x_{12}$  | $d_{13}$  | $d_{14}$  |
| $v_{j_2}$ | $x_{12}$  | $\cdot$   | $d_{23}$  | $d_{24}$  |
| $v_{j_3}$ | $d_{13}$  | $d_{23}$  | $\cdot$   | $x_{34}$  |
| $v_{j_4}$ | $d_{14}$  | $d_{24}$  | $x_{34}$  | $\cdot$   |

$=$

|          | $v_{j_1}$ | $v_{j_2}$ | $v_{j_3}$ | $v_{j_4}$ |
|----------|-----------|-----------|-----------|-----------|
| $v_{j_1}$ | $\cdot$   | $x_{12}$  | $x_{k3}$  | $x_{k4}$  |
| $v_{j_2}$ | $x_{12}$  | $\cdot$   | $x_{k3}$  | $x_{k4}$  |
| $v_{j_3}$ | $x_{k1}$  | $x_{k2}$  | $\cdot$   | $x_{34}$  |
| $v_{j_4}$ | $x_{k1}$  | $x_{k2}$  | $x_{34}$  | $\cdot$   |

.

There is some freedom with the diagonal—we have left it blank here, but we could justifiably write $2j_i$.

---

### Structure of the Normal Form: Summary

- We have made a reasonable though limiting simplification of our spin networks by substituting singlet states for Bell states. This allows us to reduce to our proposed normal forms, though is not fully general.

- Once we have written a spin network in the W-node form (see Prop. 56), we can then *expand* the symmetriser via a (WW) bi-algebra, push gadgets through into the structure between them, then *absorb* the internal W-nodes into the external ones.

- The process of absorption introduces a dimension problem, which we leave open for future work.

- This leaves us with a single W-node corresponding to each open link, and a 'nest' of W-gadgets occupying the structure between them.

- We can write the incidence matrix of the hypergraph induced by the W-gadgets to yield our first matrix. Then, we can write the adjacency matrix of the fully-connected graph, with each edge weighted by the dimension of its corresponding wire, to yield our second matrix.

- Alternatively, we can hybridise the incidence matrix of the hypergraph by writing the underlying graph's nodes on the rows (the $j_i$-labelled W-nodes) rather than the hypergraph's nodes to present a different matrix with potentially interesting patterns.

## 5.2 Notable Examples

To give a flavour of this matrix-like normal form, we now demonstrate the form for several elementary spin networks of different structures that often appear as sub-networks in larger spin networks. As the adjacency matrix for wire dimension is straightforward to define, we leave this matrix out of these examples, focusing on the far less trivial hypergraph matrices. Call the structure of the previous section 'example A'.

Throughout these examples (as in the previous section), we label open (external) links by spin $j_i$ and internal links by spin(s) $k$ (and $\ell$).

### 5.2.1 Example B

Consider the following spin network and its direct W-node translation (again, omitting wire dimensions for now),

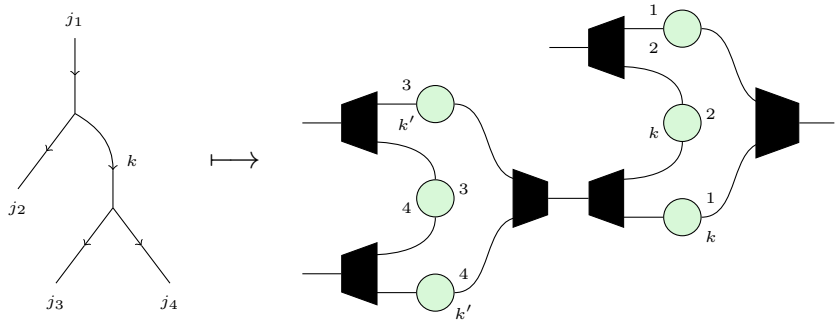

Performing the expansion and absorption steps,

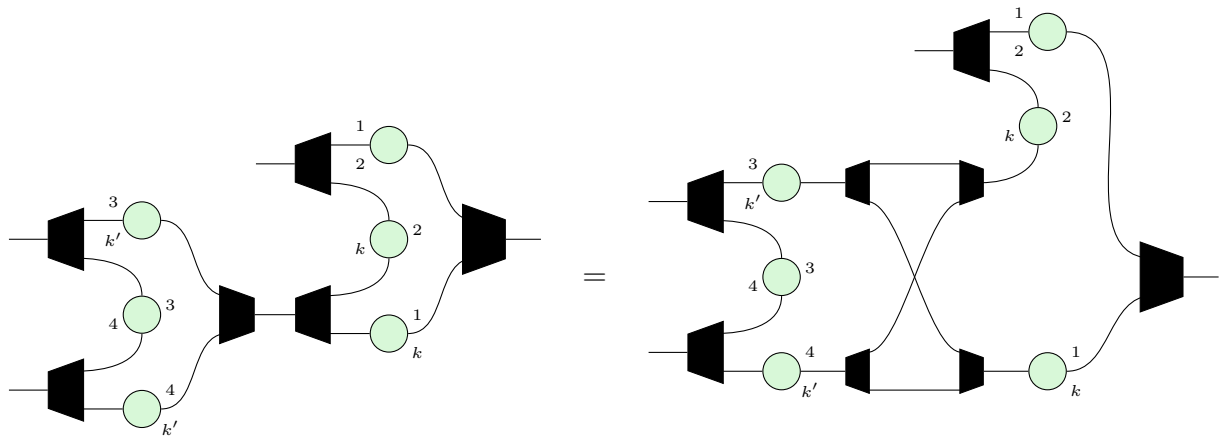

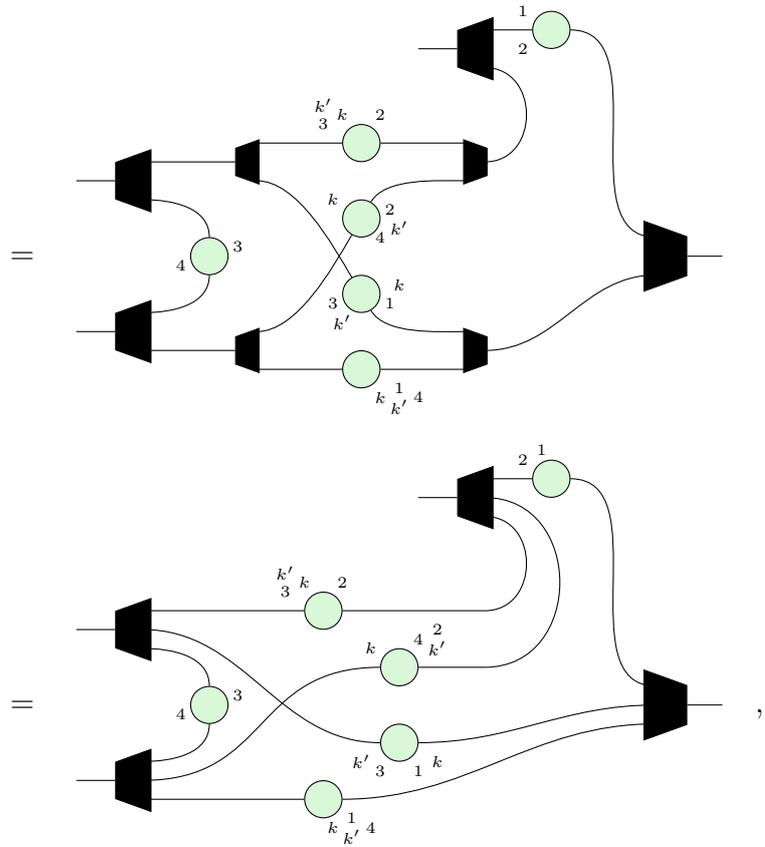

which gives us the following (directed) hypergraph induced by the W-gadget nest on the fully-connected structure between the open $j_i$ links,

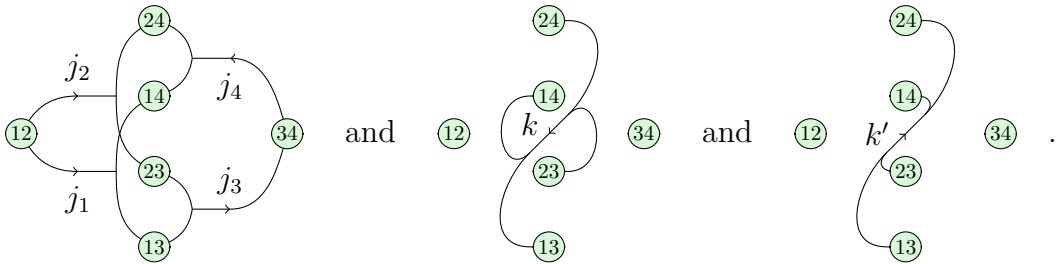

The incidence matrix for this hypergraph is,

|          | $j_1$ | $j_2$ | $j_3$ | $j_4$ | $k$  | $k'$ |
|----------|-------|-------|-------|-------|------|------|
| $v_{12}$ | $-1$  | $-1$  | 0     | 0     | 0    | 0    |
| $v_{13}$ | 1     | 0     | $-1$  | 0     | 1    | $-1$ |
| $v_{14}$ | 1     | 0     | 0     | 1     | 1    | 1    |
| $v_{23}$ | 0     | 1     | $-1$  | 0     | $-1$ | $-1$ |
| $v_{24}$ | 0     | 1     | 0     | 1     | $-1$ | 1    |
| $v_{34}$ | 0     | 0     | 1     | $-1$  | 0    | 0    |

,

or the alternative hybrid matrix,

|          | $j_1$ | $j_2$ | $j_3$ | $j_4$ | $k$ | $k'$ |
|----------|-------|-------|-------|-------|-----|------|
| $v_{j_1}$ | 0     | $-1$  | $-1$  | 1     | 1   | 0    |
| $v_{j_2}$ | $-1$  | 0     | $-1$  | 1     | $-1$ | 0   |
| $v_{j_3}$ | 1     | 1     | 0     | $-1$  | 0   | $-1$ |
| $v_{j_4}$ | 1     | 1     | 1     | 0     | 0   | 1    |

,

## 5.2.2 Example C

Consider the following,

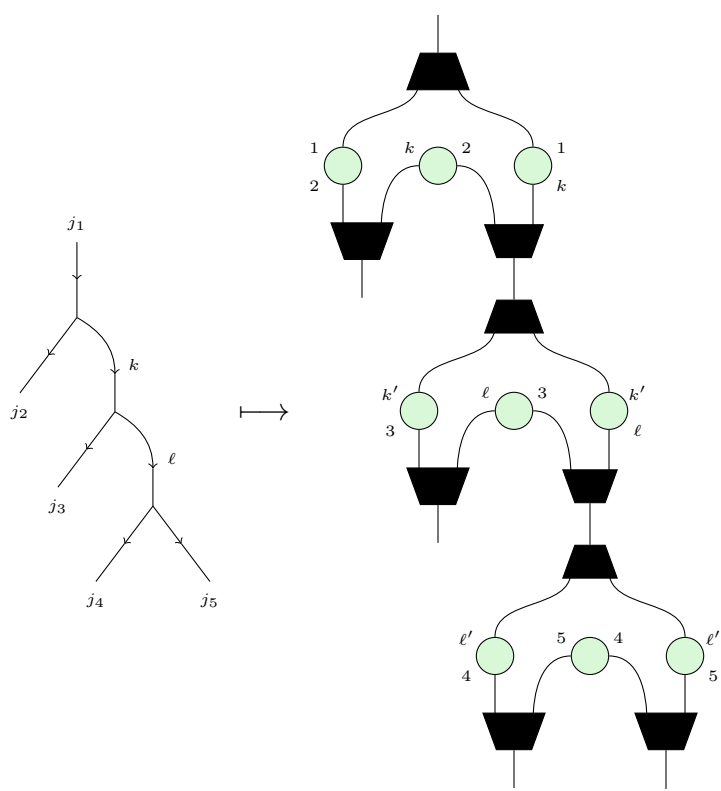

.

Performing expansion and absorption,

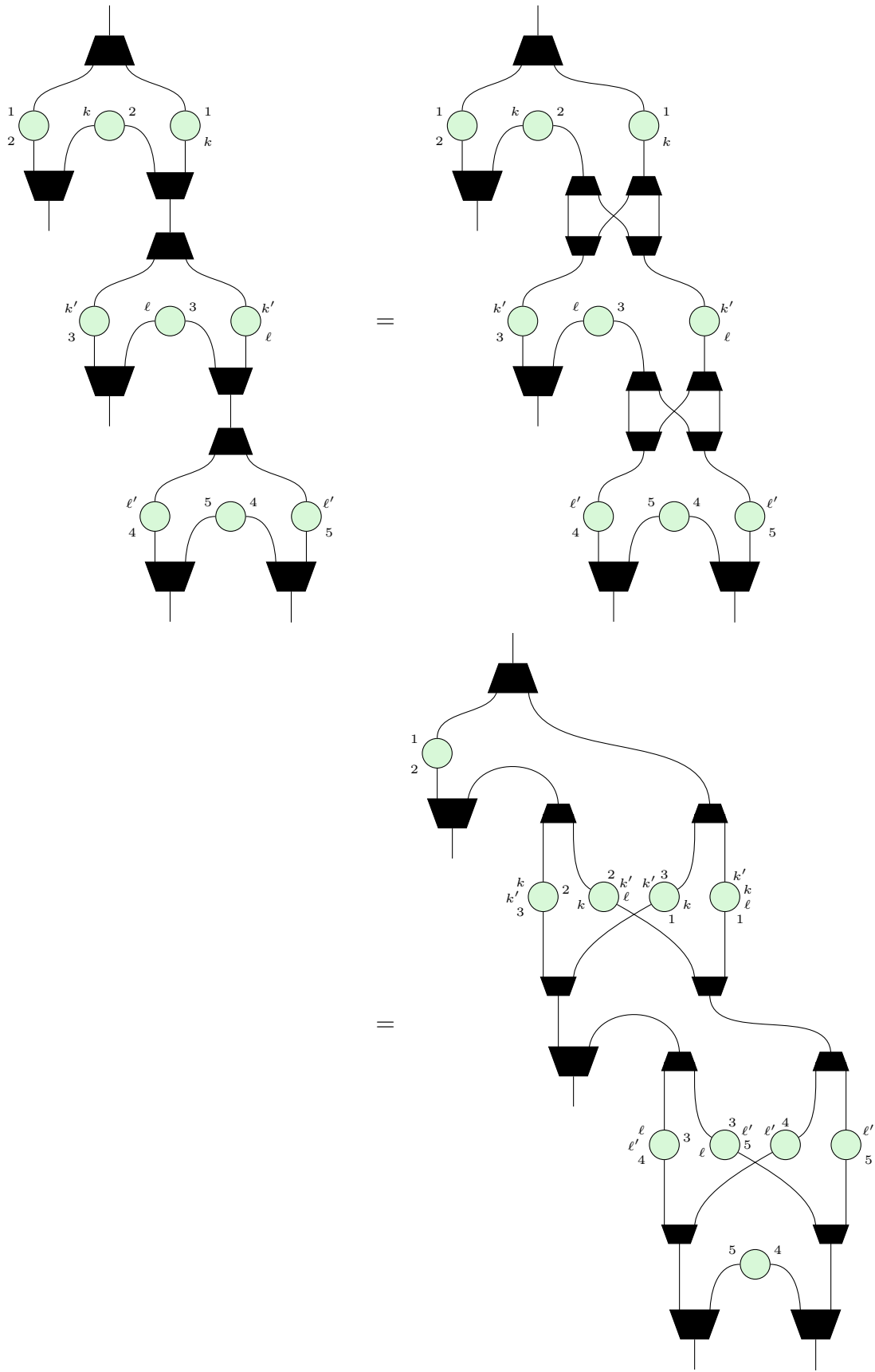

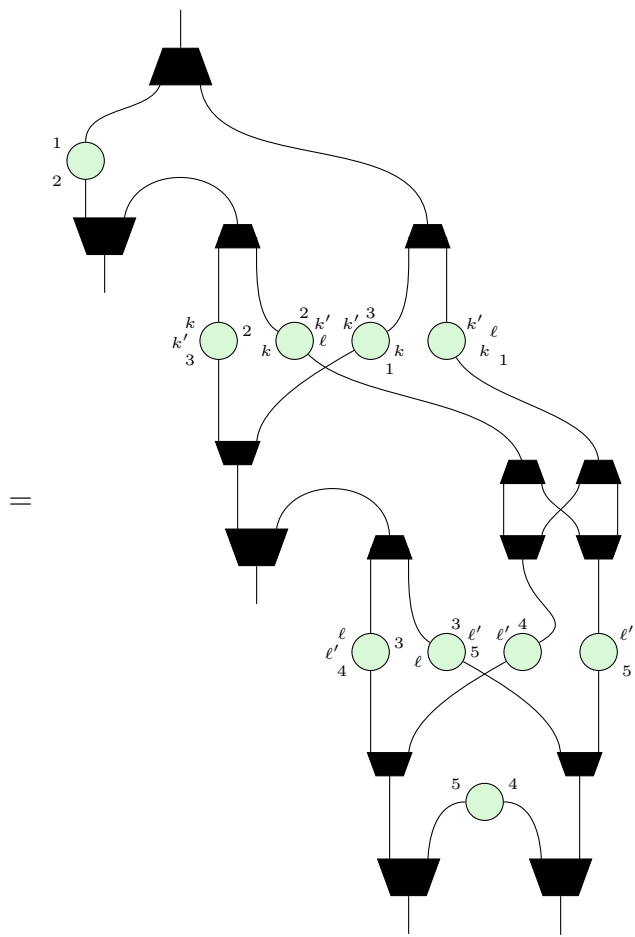

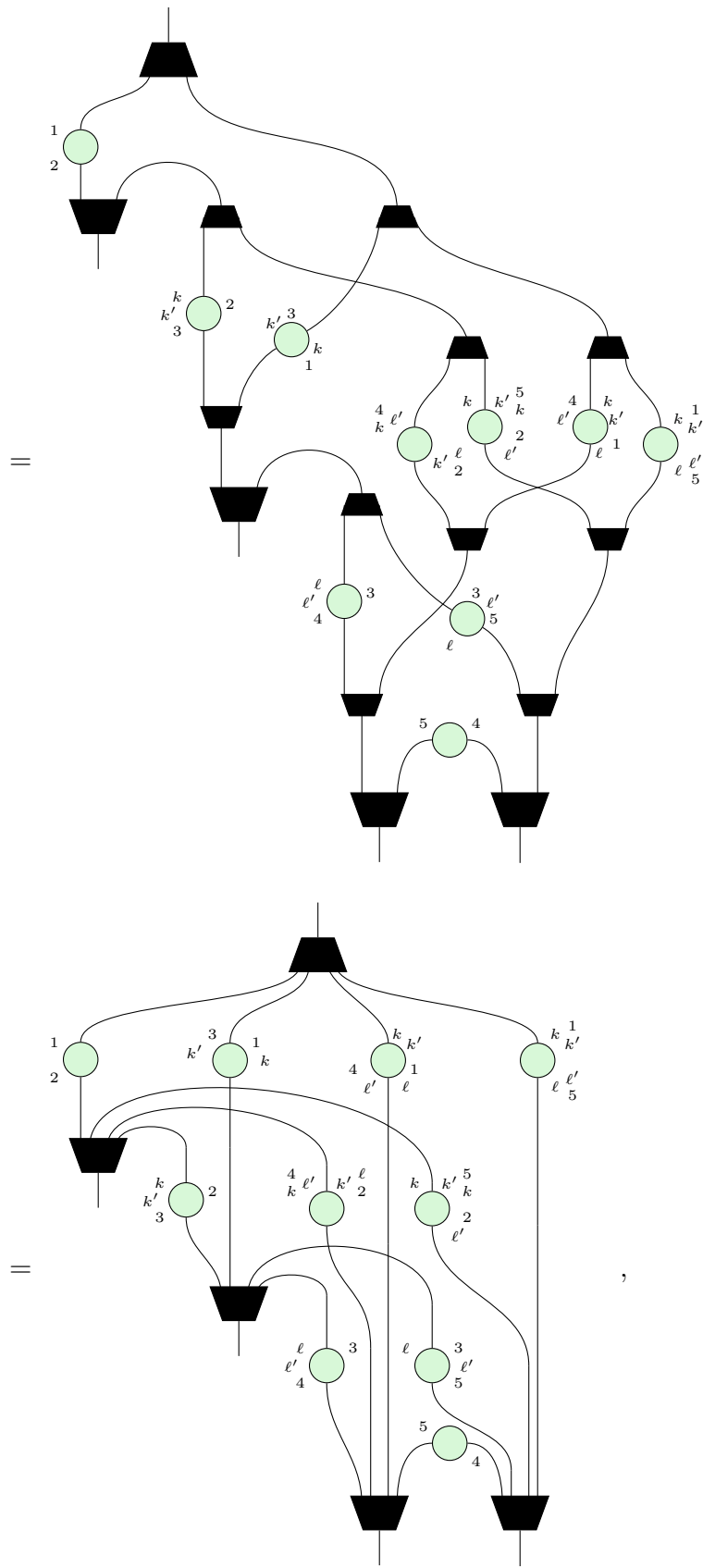

Notice that in the above, since there were two symmetrisation expansions,

we had freedom with which (WW) bi-algebra the $(k', \ell)$-gadget passed into (we arbitrarily chose upward). Ultimately, this is an illusion of free will, as they merge all the same regardless.

To draw the hypergraph, we separate all hyperedges for improved interpretability,

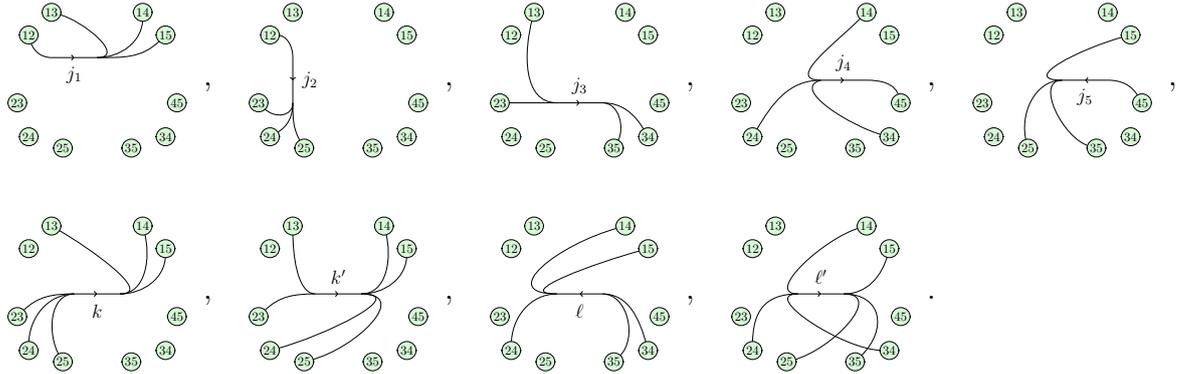

This gives the matrices,

|       | $e_{j_1}$ | $e_{j_2}$ | $e_{j_3}$ | $e_{j_4}$ | $e_{j_5}$ | $e_k$ | $e_k$ | $e_\ell$ | $e_{\ell'}$ |
|-------|-----------|-----------|-----------|-----------|-----------|-------|-------|----------|-------------|
| $v_{12}$ | $-1$ | $-1$ | $0$ | $0$ | $0$ | $0$ | $0$ | $0$ | $0$ |
| $v_{13}$ | $1$ | $0$ | $-1$ | $0$ | $0$ | $1$ | $-1$ | $0$ | $0$ |
| $v_{14}$ | $1$ | $0$ | $0$ | $-1$ | $0$ | $1$ | $1$ | $1$ | $-1$ |
| $v_{15}$ | $1$ | $0$ | $0$ | $0$ | $1$ | $1$ | $1$ | $1$ | $1$ |
| $v_{23}$ | $0$ | $1$ | $-1$ | $0$ | $0$ | $-1$ | $-1$ | $0$ | $0$ |
| $v_{24}$ | $0$ | $1$ | $0$ | $-1$ | $0$ | $-1$ | $1$ | $1$ | $-1$ |
| $v_{25}$ | $0$ | $1$ | $0$ | $0$ | $1$ | $-1$ | $1$ | $0$ | $1$ |
| $v_{34}$ | $0$ | $0$ | $1$ | $-1$ | $0$ | $0$ | $0$ | $-1$ | $-1$ |
| $v_{35}$ | $0$ | $0$ | $1$ | $0$ | $1$ | $0$ | $0$ | $-1$ | $1$ |
| $v_{45}$ | $0$ | $0$ | $0$ | $1$ | $-1$ | $0$ | $0$ | $0$ | $0$ |

,

|       | $e_{j_1}$ | $e_{j_2}$ | $e_{j_3}$ | $e_{j_4}$ | $e_{j_5}$ | $e_k$ | $e_{k'}$ | $e_\ell$ | $e_{\ell'}$ |
|-------|-----------|-----------|-----------|-----------|-----------|-------|----------|----------|-------------|
| $v_{j_1}$ | $0$ | $-1$ | $-1$ | $-1$ | $1$ | $1$ | $0$ | $1$ | $0$ |
| $v_{j_2}$ | $-1$ | $0$ | $-1$ | $-1$ | $1$ | $-1$ | $0$ | $1$ | $0$ |
| $v_{j_3}$ | $1$ | $1$ | $0$ | $-1$ | $1$ | $0$ | $-1$ | $-1$ | $0$ |
| $v_{j_4}$ | $1$ | $1$ | $1$ | $0$ | $-1$ | $0$ | $1$ | $0$ | $-1$ |
| $v_{j_5}$ | $1$ | $1$ | $1$ | $1$ | $0$ | $0$ | $1$ | $0$ | $1$ |

.

*The most cherished goal in physics, as in bad romance novels, is unification.*

— Lee Smolin [96]

# 6
# Discussion and Conclusion

## Contents

In this work, we have derived the generating objects and important relations for spin networks in the Penrose Spin Calculus (PSC) [4], expanding on East, Martin-Dussaud, and Wetering [3] by generalising the qubit ZX-calculus to the finite-dimensional ZX-calculus. In particular, we present the ZX-form for the "surgical removal" of spin network portions with 2- and 3-loops and prove their correctness, then combine this with a proof for the associativity of spin network nodes to give an inductive argument for $n$-loop removal. These loop removal results enable the algorithmic method of Guedes et al. [1] to handle changing graphs, and can thus be used to move their calculations of the action of the Hamiltonian constraint to the PSC. This resolves the issue faced in LQG by providing a graphical calculus that is both *intuitive* and facilitates *accurate* numerics through its capacity to handle (superpositions of) dynamic spin networks.

Throughout our derivations of loop removal, we find a novel presentation of the $6j$-symbol, $\Theta$-graph, and tetrahedral net symbol in the finite-dimensional ZX-calculus to extend the work of Wang et al. [4] for the PSC. These representations give a high-level description of these objects which enables our direct diagrammatic reasoning, unlike the low-level status quo of existing graphical languages for spin networks.

Having established a loop-free normal form for arbitrary spin networks following $n$-loop removal, we investigated the potential for a matrix-like normal form using our novel W-node perspective. Our normal form, while primitive, indicates the possibility to work with spin networks as natively computable objects, beyond their applicability to quantum computation following their translation to the ZX-calculus [as indicated by 3].

Indeed, this work has generally explored connections between the quantum theory of angular momentum and quantum information in mixed dimensions, including pursuits of different useful notations, which is undoubtedly crucial as we design new experiments for tests of quantum gravity. We hope that the results of this work, together with the PSC as a whole [4], can kickstart new research in LQG building on new algorithms and techniques in the wider world of ZX-calculus.

# 6.1 Limitations and Open Problems

Notice that the associativity of spin network nodes introduces a sum. This being a core component of the inductive argument to $n$-loop removal, we thus see sums over spin networks for removal of loops for $n > 3$ nodes. This is not a problem on its own—we expect to handle superpositions of graphs anyway—but it limits our current understanding of the ZX-form for loops beyond the bubble and triple bubble. The scalar belonging to 4-loop removal, for example, will be defined by a sum over $6j$-symbols, but we have not yet understood how the PSC presents this scalar. We leave a general proof of loop removal in the PSC *which addresses general scalars intuitively and correctly* as an open problem, though have indicated the expected form and correctness result in this work.

Our matrix-like normal form derived via W-nodes is still early in its lifetime, and thus presents with several limitations and open problems to consider before it can be put to practical use. The two fundamental open problems are: (1) the handling of singlet states; and (2) the dimension problem. Firstly, re-injecting the singlet states into the form we considered in chapter 5 (e.g. in example A) complicates the *absorption* step as there will be obstacles to the fusion of the W-nodes. In our early workings of this normal form, we noted that the $K_1$-phase X-spider of the singlet does not commute with a W-node, meaning it cannot be moved away to permit absorption. Secondly, addressing the assumptions on wire dimensions in lemma 30 of Poór, Shaikh, and Wang [2] is necessary to extend our normal form to general spin networks without internal disconnects. The dimension problem is most open of the two, and may benefit exploration into dimension-enlarging rules before the absorption step.

## 6.2   Avenues for Future Work

The most immediate next step for this work is to perform the calculations in Guedes et al. [1] in the PSC, now that the ZX-form and correctness of loop removal has been shown in this work. A practical derivation of the action of the Hamiltonian constraint to recover the result of Guedes et al. [1], while not novel, would be a landmark in the study of canonical LQG demonstrated in the ZX-calculus.

Furthermore, entanglement calculations in open spin networks is something we believe will be made visually intuitive in the PSC. For example, the work of Fan, Korepin, and Roychowdhury [97] is very well-suited to the normal forms in this work, as they consider only acyclic spin networks from the outset. Extensions to more complex spin network graphs would also be interesting [a la 98].

More generally, it would be interesting to see our normal forms (and the PSC generally) applied to calculating entanglement connections between arbitrary (open) regions of spin networks [99]. In this context, a ZX-diagrammatic intuition may paint a beautiful visual picture for the connection between gravity, holography, and entanglement, which would be recognised as a powerful result in the formulation for a quantum theory of gravity [100, would be an excellent starting point for this exploration]. At the very least, it would be of great interest to many a physicist to present the emergence of horizon-like surfaces with the increasing entanglement content of the bulk of the spin network *within the framework of the (finite-dimensional) ZX-calculus.*

Besides quantum gravity, the use of $SU(2)$-based contractions and analysis, provided by our derivations in the PSC, may be of practical benefit in an attempt to engage with quantum chemistry [e.g. 101]. Or the domains of particle physics could be tempted into the PSC if we can expand from $SU(2)$ into $SU(3)$; taking products of representations and with $SU(2)$ and $SU(3)$ would allow us to discuss the standard model. The work done here on $SU(2)$ is the first step to expanding beyond the ZX-calculus as a tool for discussing matrices to one for representation theory.

*Cor blimey, that's a hefty read right there.*

— Probably Plato, or someone similar. I might be
paraphrasing.